\documentclass[12pt]{article}
\usepackage{amsmath,cite,epsfig,a4,times,euscript,listings,amssymb}
\usepackage[text-hat-accent]{euler}
\usepackage{graphics}
\usepackage[usenames]{color}
\renewcommand{\baselinestretch}{1.1}
\newcommand{\myTitle}[1]{\begin{center}{\bf\Huge #1}\\[5ex]\end{center}}
\newcommand{\myAuthor}[1]{\begin{center}{\Large #1}\\[2ex]\end{center}}
\newcommand{\myAffiliation}[1]{\\[1ex]{\it\large #1}}

\newcommand{\myDate}{\begin{center}{\large\today}\\[5ex]\end{center}}
\newcommand{\myAbstract}[1]{\begin{center}\renewcommand{\baselinestretch}{1}{\bf Abstract}\\[2ex]\parbox{0.8\linewidth}{\small\hspace{15pt} #1}\end{center}\vspace{\baselineskip}}
\newcommand{\myReport}[1]{\hspace{\fill} #1}
\newcommand{\myPreprint}[1]{}
\newcommand{\myKeywords}[1]{}

\newcommand{\myScript}[1]{\EuScript{#1}}

\newcommand{\slashp}{p\hspace{-6.5pt}/}

\newcommand{\slashr}{r\hspace{-6.0pt}/}
\newcommand{\Appendix}[1]{Appendix~\ref{#1}}   
\newcommand{\Section}[1]{Section~\ref{#1}}

\newcommand{\Figure}[1]{Fig.~\ref{#1}}
\newcommand{\Equation}[1]{Eq.~(\ref{#1})}
\newcommand{\ie}{{\it i.e.}}

\newcommand{\eg}{{\it e.g.}}
\newcommand{\Ord}{\myScript{O}}
\newcommand{\srac}[2]{{\textstyle\frac{#1}{#2}}}
\newcommand{\Tr}{\mathrm{Tr}}
\newcommand{\imag}{\mathrm{i}}

\newcommand{\vep}{{\bar{\epsilon}}}
\newcommand{\vepv}{\epsilon}
\newcommand{\piep}{\pi_{\epsilon}}

\newcommand{\cNLO}{a_\epsilon}
\newcommand{\gQCD}{g_\mathrm{s}}

\newcommand{\ANG}[1]{\langle#1\rangle}
\newcommand{\SQR}[1]{[#1]}

\newcommand{\brktAS}[1]{\langle#1]}

\newcommand{\leftA}[1]{\langle#1|}
\newcommand{\Arght}[1]{|#1\rangle}
\newcommand{\leftS}[1]{[#1|}
\newcommand{\Srght}[1]{|#1]}
\newcommand{\lop}[2]{#1\!\cdot\!#2}
\newcommand{\Nc}{N_\mathrm{c}}
\newcommand{\sgn}{\mathrm{sgn}}
\newcommand{\xP}{x}
\newcommand{\xM}{\bar{x}}
\newcommand{\xiP}{\xi}
\newcommand{\xiM}{\bar{\xi}}

\newcommand{\aM}{a}
\newcommand{\nuSq}{{\nu^2}}
\newcommand{\sTot}{{S}}
\newcommand{\pP}{P}

\newcommand{\pPA}{P_A}
\newcommand{\pM}{\bar{P}}

\newcommand{\alphaS}{\alpha_{\mathrm{s}}}

\newcommand{\rperp}{r_{\scriptscriptstyle\perp}}
\newcommand{\qperp}{q_{\scriptscriptstyle\perp}}
\newcommand{\kperp}{k_{\scriptscriptstyle\perp}}
\newcommand{\Kperp}{K_{\scriptscriptstyle\perp}}
\newcommand{\qbarperp}{\bar{q}_{\scriptscriptstyle\perp}}

\newcommand{\OrdL}[1]{\Ord\big(\Lambda^{#1}\big)}
\newcommand{\Splt}{\EuScript{S}}

\newcommand{\rL}{r}
\newcommand{\qL}{q}
\newcommand{\qB}{{\bar{q}}}
\newcommand{\zr}{z_{\rL}}
\newcommand{\zq}{z_{\qL}}
\newcommand{\zqbar}{z_{\qB}}
\newcommand{\Sqr}{\mathrm{Sqr}}
\newcommand{\Ang}{\mathrm{Ang}}
\newcommand{\Born}{\mathrm{B}}
\newcommand{\Virt}{\mathrm{V}}
\newcommand{\Real}{\mathrm{R}}
\renewcommand{\fam}{\mathrm{fam}}
\newcommand{\unf}{\mathrm{unf}}
\newcommand{\familiar}{familiar}
\newcommand{\Familiar}{Familiar}
\newcommand{\unfamiliar}{unfamiliar}
\newcommand{\Unfamiliar}{Unfamiliar}
\newcommand{\Tree}{\mathrm{tree}}
\newcommand{\Loop}{\mathrm{loop}}
\newcommand{\Freal}{\EuScript{R}}
\newcommand{\aux}{\mathrm{aux}}
\newcommand{\auxq}{\mathrm{aux}\textrm{-}\mathrm{q}}
\newcommand{\auxg}{\mathrm{aux}\textrm{-}\mathrm{g}}
\newcommand{\inlbl}{\mathrm{in}}
\newcommand{\inbar}{\overline{\mathrm{{\imath}n}}}
\newcommand{\Vcoef}{\EuScript{V}}

\newcommand{\dBstar}{d\Born^\star}

\newcommand{\dVunf}{d\Virt^{\star\,\unf}}
\newcommand{\dR}{d\Real}
\newcommand{\dRunf}{\dR^{\star\,\unf}}
\newcommand{\dRfam}{\dR^{\star\,\fam}}
\newcommand{\coll}{\mathrm{coll}}
\newcommand{\collBar}{\overline{\mathrm{co}}\mathrm{ll}}
\newcommand{\dRfamColl}{\dR^{\star\,\fam}_{\coll}}

\newcommand{\JetB}{J_{\Born}}
\newcommand{\JetR}{J_{\Real}}

\newcommand{\CBar}{C_{\inbar}}
\newcommand{\PBar}{\EuScript{P}_{\inbar}}

\newcommand{\mycite}[1]{\cite{#1}}
\newcommand{\Mtree}{\big|\overline{\EuScript{M}}\big|^2}
\newcommand{\MtreeAux}{\big|\overline{\EuScript{M}}^{\aux}\big|^2}
\newcommand{\justMStar}{\big|\overline{\EuScript{M}}^{\,\star}\big|^2}
\newcommand{\justMAuxq}{\big|\overline{\EuScript{M}}^{\,\star}_{\auxq}\big|^2}
\newcommand{\justMAuxg}{\big|\overline{\EuScript{M}}^{\,\star}_{\auxg}\big|^2}
\newcommand{\justMAuxqCor}[2]{\big(\overline{\EuScript{M}}^{\,\star}_{\auxq}\big)^2_{#1#2}}
\newcommand{\justMAuxgCor}[2]{\big(\overline{\EuScript{M}}^{\,\star}_{\auxg}\big)^2_{#1#2}}
\newcommand{\justMStarCor}[2]{\big(\overline{\EuScript{M}}^{\,\star}\big)^2_{#1#2}}

\newcommand{\MtreeStar}{\big|\overline{\EuScript{M}}^{\,\star}\big|^2}
\newcommand{\Amp}{\EuScript{A}}
\newcommand{\ColoredAmp}{\EuScript{M}}
\newcommand{\Caux}{C_{\aux}}
\newcommand{\Cauxq}{C_{\auxq}}
\newcommand{\Cauxg}{C_{\auxg}}
\newcommand{\Paux}{\EuScript{Q}_{\aux}}
\newcommand{\upper}{u}
\newcommand{\pole}{\mathrm{pole}}
\newcommand{\rest}{\mathrm{rest}}

\newcommand{\univ}{\mathrm{univ}}
\newcommand{\Cq}{c_q}
\newcommand{\Cr}{c_r}
\newcommand{\Cqbar}{c_{\bar{q}}}
\newcommand{\xiPin}{\xiP_{\inlbl}}
\newcommand{\xPin}{x_{\inlbl}}
\newcommand{\kin}{k_{\inlbl}}
\newcommand{\kinBar}{k_{\inbar}}
\newcommand{\xMin}{\bar{x}_{\inbar}}
\newcommand{\Lambdak}{\Lambda'}
\newcommand{\PSTstr}{\Sigma^{\star}}
\newcommand{\PSTaux}{\Sigma^{\aux}}
\newcommand{\matrixelement}{squared matrix element}
\newcommand{\rupp}{v}
\newcommand{\kapp}{\kappa}
\newcommand{\kstr}{\kappa^*}
\newcommand{\graph}[3]{\raisebox{-#3ex}{\epsfig{file=#1.pdf,width=#2ex}}}

\newcommand{\tweakcodepar}[3]%
  {\vspace{#1ex}\newline\noindent\hspace*{4.0ex}{\small\tt #3}\vspace{#2ex}\newline\noindent}

\begin{document}

\myReport{IFJPAN-IV-2022-9}
\myPreprint{}\\[2ex]

\myTitle{%
Hybrid $\boldsymbol{k_T}$-factorization and\\[0.5ex] impact factors at NLO
}

\myAuthor{%
Andreas~van~Hameren$^{a}\footnote{hameren@ifj.edu.pl}$,
Leszek Motyka$^{b}$\footnote{leszek.motyka@uj.edu.pl},\\
and Grzegorz Ziarko$^{a}\footnote{grzegorz.ziarko@ifj.edu.pl}$%
\myAffiliation{%
$^a$~Institute of Nuclear Physics Polisch Academy of Sciences,\\
PL-31342 Krak\'ow, Poland\\
$^b$~Institute of Theoretical Physics, Jagiellonian University,\\[0.5ex]
S.~\L{}ojasiewicza 11, 30-348 Kraków, Poland
}
}

\myDate

\myAbstract{%
In the hybrid $k_T$-factorization formula, one initial-state parton momentum is space-like and carries non-vanishing transverse components, while the other is on-shell.
We promote this factorization formula to next-to-leading order. 
Studying the partonic cross section, we identify all soft and collinear divergencies in the real and virtual contribution, and recognize that all non-cancelling ones can be attributed to PDF evolution, evolution kernel, and target impact factors.
In result, we construct a framework that may be used to compute NLO impact factors in general.
In particular, we recover known expressions for inclusive NLO quark-and gluon impact factor corrections. 
}

\myKeywords{QCD}

%\begin{document}
%

\newpage%
\setcounter{tocdepth}{1}
\tableofcontents

\section{Introduction}
The high energy factorization approach to scattering amplitudes has played a seminal role in developments of theory and phenomenology of strong interactions since quantum chromodynamics was discovered 
\cite{%BFKL%
 Lipatov:1976zz%
,Kuraev:1976ge%
,Kuraev:1977fs%
,Balitsky:1978ic%
,Fadin:1989kf%
,Lipatov:1996ts%
},
\cite{%KTFAC%
 Catani:1990xk%
,Catani:1990eg%
,Catani:1994sq%
},
\cite{%NLLBFKL%
 Fadin:1995xg%
,Fadin:1996tb%
,Fadin:1993wh%
,Fadin:1994fj%
,Fadin:2005zj%
,Fadin:1996yv%
,Fadin:1997hr%
,DelDuca:1995ki%
,DelDuca:1998kx%
,DelDuca:1998cx%
,Fadin:1998py%
,Camici:1997ij%
,Ciafaloni:1998gs%
,Ciafaloni:1998iv%
,Fadin:2005zj%
,Balitsky:2007feb%
,Salam:1999cn%
}.
The cornerstone of this factorization, closely related to an even earlier concept of Regge factorization, may be traced back to the Lorentz dilation of time between the projectile and target frames in ultrarelativistic particle collisions.
Due to this effect, the dynamics in the hadronic projectile and in the hadronic target is governed by vastly different time scales and the dynamics approximately factorizes.
From theory side, the high energy factorization approach is based on isolating the leading powers of the total hadron--hadron collision energy squared, $S$, in the scattering amplitudes.
Then, in the perturbative approach to the high energy scattering amplitudes one finds perturbative corrections that are enhanced by powers of logarithms of the energy, $L_S^n$, where $L_S= \ln(S/S_0)$, with $S_0$ being a fixed and moderate scale, and the total collision energy squared $S\gg S_0$ can be arbitrarily large.
The formal all order resummations of these enhanced perturbative corrections are realized within the BFKL scheme 
\cite{%BFKL%
 Lipatov:1976zz%
,Kuraev:1976ge%
,Kuraev:1977fs%
,Balitsky:1978ic%
,Fadin:1989kf%
,Lipatov:1996ts%
},
\cite{%NLLBFKL%
 Fadin:1995xg%
,Fadin:1996tb%
,Fadin:1993wh%
,Fadin:1994fj%
,Fadin:2005zj%
,Fadin:1996yv%
,Fadin:1997hr%
,DelDuca:1995ki%
,DelDuca:1998kx%
,DelDuca:1998cx%
,Fadin:1998py%
,Camici:1997ij%
,Ciafaloni:1998gs%
,Ciafaloni:1998iv%
,Fadin:2005zj%
,Balitsky:2007feb%
,Salam:1999cn%
}.
In general, the Regge factorization principle allows to factorize the dynamics in the projectile $A$ phase space region, in the target $B$ region and in the intermediate (exchange) region.
The scattering amplitude ${\cal A}$ may be written symbolically as  
\[
{ \cal A} = iS \Phi_A \otimes {\cal G}(S) \otimes \Phi_B,
\]
where $\Phi_A$ and $\Phi_B$ are process dependent impact factors, ${\cal G}(S)$ is an universal exchange part, the convolutions denoted by $\otimes$ are performed over transverse momenta, and the other details were omitted as we focus only on the structural features here.

In QCD the BFKL resummation framework was constructed for the gluonic exchange ${\cal G}(S)$ at the leading logarithmic (LL) and the next-to-leading logarithmic accuracy (NLL), where terms $(\alpha_s L_S)^n$ and $\alpha_s^{n+1} L_S^{n}$ are resummed correspondingly.
In recent years within ${\cal N}=4$ supersymmetric Young-Mills theories this approach was further developed to reach the next-to-next-to-leading logarithmic accuracy
\cite{%SYMBFKL%
 Kotikov:2000pm%
,Gromov:2015vua%
,Caron-Huot:2016tzz%
}.
In order to fully use the potential of the BFKL approach in phenomenological applications at the presently highest available NLL accuracy, it is necessary to include the impact factors $\Phi_{A,B}$ at the NLO accuracy.
Such computations are highly non-trivial in the high energy limit.
To this date they have been performed for about a dozen of processes yielding interesting predictions mostly for high energy hadron colliders, like the Large Hadron Collider, RHIC and Tevatron, and for electron--proton colliders: HERA and the Electron--Ion Collider.

The calculations of the impact factors at the NLO started from inclusive quark and gluon impact factors 
\cite{%PARTON-NLOIF%
 Ciafaloni:1998hu%
,Fadin:1999de%
,Fadin:1999df%
},
then the forward jet impact factors were calculated
\cite{%JET-NLOIF%
 Bartels:2001ge%
,Bartels:2002yj%
,Hentschinski:2011tz%
,Caporale:2012ih%
,Chachamis:2012cc%
,Hentschinski:2014esa%
}.
Based on these results applications of the NLL BFKL to collider data was developed, to start with numerous detailed analyses of forward jet cross-sections in hadron colliders 
\cite{%FORJET%
 SabioVera:2006cza%
,Colferai:2010wu%
,Caporale:2011cc%
,Ivanov:2012ms%
,Ducloue:2013hia%
,Ducloue:2013bva%
,Caporale:2014gpa%
}.
Following those pioneering studies the formalism was applied to other high energy scattering and particle production processes.
In particular: for excusive vector meson photoproduction 
\cite{%VMESONS%
 Ivanov:2004pp%
,Ivanov:2005gn%
,Ivanov:2006gt%
,Boussarie:2016bkq%
},
inclusive and diffractive dijets
\cite{%DIJETS%
 Taels:2022tza%
,Caucal:2021ent%
,Boussarie:2019ero%
}
and multi-particle final states 
\cite{%MULTI%
 Boussarie:2016ogo%
,Roy:2019hwr%
}.
The inclusive $\gamma^*$ impact factor was calculated as well 
\cite{%GAMSTARIF%
 Bartels:2002uz%
,Fadin:2002tu%
,Balitsky:2010ze%
,Balitsky:2012bs%
},
that allowed to analyze the DIS processes
\cite{%DIS%
 Beuf:2017bpd%
,Hanninen:2017ddy%
}
and total $\gamma^*\gamma^*$ cross-sections 
\cite{%GAMSTARGAMSTAR%
 Ivanov:2014hpa%
}.
Other examples are the forward hadroproduction of heavy quarks 
\cite{%HEAVYQUARKS%
 Ciafaloni:2000sq%
,Celiberto:2017nyx%
,Bolognino:2019yls%
},
dihadrons
\cite{%DIHADRONS%
 Celiberto:2016hae%
,Celiberto:2017ptm%
},
the Higgs boson
\cite{%HIGGS%
 Celiberto:2020tmb%
,Hentschinski:2020tbi%
,Celiberto:2022fgx%
}
and heavy baryons 
\cite{%LAMBDA%
 Celiberto:2021dzy%
}.
So far the calculations of the NLO impact factors were performed process--by--process in an explicit form.
In the present letter we propose a computational framework that allows to obtain in a straightforward way the NLO impact factors from NLO collinear scattering amplitudes.
%
%To be more specific, our results allow to obtain the NLO transition amplitudes of an incoming parton to $n$ outgoing partons by an exchange of a virtual gluon $g^*$ from a corresponding NLO collinear scattering amplitudes with $n+2$ external lines, and the LO collinear amplitude with $n+3$ external lines.
%%
%We believe that the proposed scheme offers a systematic and efficient way to compute the impact factors NLO for more interesting processes and also to cross check the existing calculations.
%

Complementary to the quest for 
higher logarithic accuracy in BFKL resummation and higher perturbative precision for
impact factors is the application high energy factorization to define and calculate scattering processes involving space-like initial-state gluons and several final-state jet and/or particles.
The idea is to have a factorized formula with parton density functions (PDFs) and a parton level cross section, both however depending on transverse momentum components (the $k_T$) of the initial-state gluons besides the components parallel to the initial-state hadron momenta.
Basically, the formula for heavy quark production established in
\cite{Catani:1990eg,Collins:1991ty}
is generalized by replacing the partonic cross section by the one for other processes, involving jets, massive vector bosons, etc.
The calculations are at tree-level, for which the main hurdle to take is the gauge invariant definition of tree-level amplitudes with space-like initial state partons, which can be achieved with the help of effective vertices from Lipatov's effective action~
\cite{Lipatov:1995pn,Antonov:2004hh}%
, or with the auxiliary parton method~%
\cite{vanHameren:2012if}%
.
Originally, the PDFs were imagined to undergo BFKL evolution, but the factorized formula has also been applied with $k_T$-dependent PDFs obtained from ``unintegration procedures''~
\cite{Kimber:2001sc,Hautmann:2017fcj}%
, and from non-linear evolution~%
\cite{Balitsky:1995ub,Kovchegov:1999ua}%
.

Recently, calculations within this kind of factorization have been preformed at next-to-leading order (NLO) for processes with one particle in the final state~
\cite{Nefedov:2019mrg,Nefedov:2020ecb,Hentschinski:2020tbi}%
.
Moving to NLO inevitably leads to infrared divergences in real and virtual contributions which need to cancel or treated within a renormalization procedure.
In collinear factorization this is rigorously established partly through renormalization of the PDFs.
In this paper we show how the factorization formula with $k_T$ dependence can be promoted to NLO within the auxiliary parton method.
More specifically, we consider the so-called hybrid factorization formula, for which one initial-state parton is light-like, and its associated PDF only depends only on the longitudinal momentum fraction, while the other is space-like and its associated PDF also depends on $k_T$.
We will identify the divergences in real and virtual contributions that do not manifestly cancel, and we will show that they are process independent.
With the process we mean here the one with the initial-state space-like gluon.
The divergences are {\em not} independent of the auxiliary partons, which may a scattering gluon or quark.
We will, however, show that the difference between the choice of auxiliary partons is process independent.
%

%We apply the auxiliary parton approach, which is designed to extract amplitudes involving space-like partons from completely on-shell amplitudes.
%%
%While the method allows to construct amplitudes directly, with the help of eikonal Feynman rules, it can also be applied to extract space-like amplitudes from existing expressions for on-shell amplitudes.
%%
%For tree-level applications so far, the method was treated rather as a ``trick'' to obtain the necessary gauge invariant amplitudes.
%%
%In order to move to NLO, the auxiliary partons must be taken ``mores seriously'', and that the method must be raised to the level of cross sections.
%%
 
%
In a heuristic picture, the auxiliary parton plays a role of a target, with the dynamics that is Regge factorized from the dynamics of the transition in the projectile region.
Such a picture is more or less explicitly present in most of the LO calculations of the impact factors, and at the LO it is straightforward and simple.
A generalization of this method to the NLO amplitudes is more demanding.
%
%In this case it is necessary to have it guaranteed that the Regge factorization holds of the NLO real and virtual corrections in the auxiliary parton region.
%
%Next, a complete NLO description of the auxiliary parton must be performed.
%
We will see, however, that our result fits the interpretation in which the NLO auxiliary parton scattering amplitude and the LL exchange (related to the BFKL kernel) amplitude taken at ${\cal O}(\alpha_s)$ are factorized from the suitable NLO on-shell amplitude, to result with the explicit expression for the NLO impact factor.
Again, it is important to stress that both the NLO auxiliary parton impact factors and the ${\cal O}(\alpha_s)$ LL exchange amplitude are universal.
Hence, the explicit expressions for these contributions may be used to obtain the NLO impact factor for any process in the projectile region.
This method, including the explicit form of the quark and gluon NLO impact factors, is the main result of this paper.

\section{Main result}
The context of our work is the hybrid $k_T$-type factorized cross section formula for hadron scattering, which at lowest order schematically looks like
%
%%%%%%%%%%%%%%%%%%%%%%%%%%%%%%%%%%%%%%%%
\begin{equation}
d\sigma^{(0)} = \int\frac{d\xPin}{\xPin}\,d^2\kperp\frac{d\xMin}{\xMin}\,F(\xPin,|\kperp|)\,f(\xMin)\,d\Born^\star(\xPin,\kperp,\xMin)
~.
\end{equation}
The functions $F(\xPin,|\kperp|)$ and $f(\xMin)$ are parton density functions (PDFs), both depending on a longitudinal momentum fraction, and only one of them depending on transverse momentum.
Of course both typically depend also on an additional scale and on the parton type.
Although this may not be in accordance with the nomenclature in literature, we refer to PDFs as the functions that come naturally with the measures $d\xPin/\xPin$ and $d\xMin/\xMin$. 

We treat this factorization formula as a conjecture, and the main goal of this paper is to show that it can be made to work at NLO.
If the collinear factorization formula were written down as a conjecture, then the analogous goal would be achieved by showing, the of course well-known fact, that all singularities showing up at NLO cancel, or can be absorbed by PDF corrections.
These PDF corrections would match the known evolution of PDFs dictated by QCD, but the sketched exercise could also be seen as a derivation of this evolution.
It is this exercise that we want to perform on the hybrid factorization formula.

Special attention needs to go to the, here Born-level, differential partonic cross section $d\Born^\star$, which is special compared to collinear factorization in the sense that one of its initial-state partons is space-like, and has momentum
%
%%%%%%%%%%%%%%%%%%%%%%%%%%%%%%%%%%%%%%%%
\begin{equation}
\kin^\mu = \xPin\pPA^\mu + \kperp^\mu
\label{Eq:048}
~,
\end{equation}
%%%%%%%%%%%%%%%%%%%%%%%%%%%%%%%%%%%%%%%%
%
where $\pPA$ is the light-like hadron momentum and $\kperp$ is transverse, \ie\ $\lop{\kperp}{\pPA}=0$.
The `$\star$' in $d\Born^\star$ highlights the $\kperp$-dependence.
At the lowest order it is straightforward to define and calculate $d\Born^\star$, and essentially only involves tree-level amplitudes, and final-state momenta that are well-separated by a number of phase space cuts.

The NLO contribution looks, again rather schematically, like
%
%%%%%%%%%%%%%%%%%%%%%%%%%%%%%%%%%%%%%%%%
\begin{align}
d\sigma^{(1)} &
= \int\frac{d\xPin}{\xPin}\,d^2\kperp\frac{d\xMin}{\xMin}\bigg\{
  F(\xPin,|\kperp|)\,f(\xMin)
  \Big[d\Virt^\star(\xPin,\kperp,\xMin)+d\Real^\star(\xPin,\kperp,\xMin)\Big]
\notag\\&\hspace{16ex}
  +\Big[F^{(1)}(\xPin,|\kperp|)\,f(\xMin)+F(\xPin,|\kperp|)\,f^{(1)}(\xMin)\Big]
  d\Born^\star(\xPin,\kperp,\xMin)
\bigg\}
~.
\label{Eq:061}
\end{align}
%%%%%%%%%%%%%%%%%%%%%%%%%%%%%%%%%%%%%%%%
%
Here, $d\Virt^\star$ represents the virtual contribution involving one-loop amplitudes, and $d\Real^\star$ the real contribution involving extra phase space integrals.
Both carry one more power of the strong coupling constant than $d\Born^\star$, but are combined with the same PDFs as at the lowest order.
In the second line of \Equation{Eq:061} the functions $f^{(1)},F^{(1)}$ are higher order corrections, and carry an extra power of the coupling constant.

The loop integrals in $d\Virt^\star$ and the phase space integrals in $d\Real^\star$ cause soft and collinear divergences.
Within collinear factorization, these divergences cancel up to certain collinear divergences, which we denote $\Delta_{\coll}$ and $\Delta_{\collBar}$, that are associated with the variables $\xPin$ and $\xMin$.
The well-known formula for $\Delta_{\collBar}$ is given in \Equation{Eq:064}.
It is absorbed into $f^{(1)}(\xMin)$, or at least the factorization prescription can be interpreted as such.
The same happens with $\Delta_{\coll}$ and the correction to the other collinear PDF that appears instead of $F(\xPin,|\kperp|)$ in the collinear factorization formula.
For the hybrid formula above, we again find $\Delta_{\collBar}$, which can be treated the same way, and we find a different but similar $\Delta^\star_{\coll}$ (\Equation{Eq:065}), which now must be  absorbed into $F^{(1)}(\xPin,|\kperp|)$.

In the following, we will identify extra non-cancelling divergencies in $d\Virt^\star$ and $d\Real^\star$, which we call {\em \unfamiliar} contributions $\Delta_{\unf}$ (\Equation{Eq:047}).
They break a notion of universality that exists at Born level.
They are proportional to the LO cross section, and must be interpreted as corrections to the impact factor associated with the space-like gluon.
So the structure we eventually find is
%
%%%%%%%%%%%%%%%%%%%%%%%%%%%%%%%%%%%%%%%%
\begin{align}
d\sigma^{(1)} &
= \int\frac{d\xPin}{\xPin}\,d^2\kperp\frac{d\xMin}{\xMin}\bigg\{F(\xPin,|\kperp|)\,f(\xMin)
  \Big[d\Virt^\star(\xPin,\kperp,\xMin)+d\Real^\star(\xPin,\kperp,\xMin)\Big]_{\mathrm{cancelling}}
\notag\\&\hspace{14ex}
 +\Big[F^{(1)}(\xPin,|\kperp|)+F(\xPin,|\kperp|)\Delta_{\unf}+\Delta^\star_{\coll}\Big]f(\xMin)\,d\Born^\star(\xPin,\kperp,\xMin)
\notag\\&\hspace{30ex}
   +F(\xPin,|\kperp|)\Big[f^{(1)}(\xMin)+\Delta_{\collBar}\Big]
   d\Born^\star(\xPin,\kperp,\xMin)
\bigg\}
\label{Eq:094}
~.
\end{align}
%%%%%%%%%%%%%%%%%%%%%%%%%%%%%%%%%%%%%%%%
%
The term $d\Real^\star(\xPin,\kperp,\xMin)$ inside the cancelling contribution only involves tree-level amplitudes.
It requires numerical phase space integration, and the IR divergencies can be dealt with using methods known in literature, \eg\ the subtraction method of~\cite{Catani:1996vz}.
Only a special subtraction term related to the space-like gluon is necessary, which will be addressed in this paper.
The term $d\Virt^\star(\xPin,\kperp,\xMin)$ requires one-loop amplitudes.
For processes that are not too complicated, they can rather straightforwardly be obtained from analytic expressions for on-shell one-loop amplitudes with one more external leg, and in this paper we describe the terms appearing in this procedure that that must go to $\Delta_{\unf}d\Born^\star(\xPin,\kperp,\xMin)$.

The procedure mentioned above is the {\em auxiliary parton method} which will be described in this section.
For tree-level amplitudes, the necessity for analytic expressions of on-shell amplitudes can be avoided and the amplitudes with a space-like gluon can be constructed directly using eikonal Feynman rules~\cite{vanHameren:2012if}.
In the case of one-loop amplitudes, an approach to avoid the necessity for analytic expressions of on-shell one-loop amplitudes is described in~\cite{vanHameren:2017hxx}.

\subsection{Notation}
In order to proceed, we need to establish some notation.
We introduce the decomposition of a general four vector $K^\mu$ as
%%%%%%%%%%%%%%%%%%%%%%%%%%%%%%%%%%%%%%%%
\begin{equation}
K^\mu = \xiP_K\pP^\mu + \xiM_K\pM^\mu + \Kperp^\mu 
\end{equation}
%%%%%%%%%%%%%%%%%%%%%%%%%%%%%%%%%%%%%%%%
%
where the momenta $\pP$ and $\pM$ have positive energy and satisfy
%
%%%%%%%%%%%%%%%%%%%%%%%%%%%%%%%%%%%%%%%%
\begin{equation}
\pP^2=\pM^2=0
\quad,\quad
2\lop{\pP}{\pM}=\nuSq>0
\quad,\quad
\lop{\pP}{\Kperp}=\lop{\pM}{\Kperp}=0
\end{equation}
%%%%%%%%%%%%%%%%%%%%%%%%%%%%%%%%%%%%%%%%
%
so
%
%%%%%%%%%%%%%%%%%%%%%%%%%%%%%%%%%%%%%%%%
\begin{equation}
 \xiP_K = \frac{\lop{\pM}{K}}{\lop{\pP}{\pM}}
\;\;,\quad
 \xiM_K = \frac{\lop{\pP}{K}}{\lop{\pP}{\pM}}
~.
\end{equation}
%%%%%%%%%%%%%%%%%%%%%%%%%%%%%%%%%%%%%%%%
%
The momenta $\pP$ and $\pM$ are the directions of the scattering hadron momenta, and we prefer to let them keep a scale rather than to normalize them.
The coordinates $\xiP$ and $\xiM$ are dimensionless and strictly positive for positive-energy momenta.
We need to distinguish between variables $\xiP$ and direction $\pP$ versus variables $\xP$ and momentum $\pPA$ satisfying
%%%%%%%%%%%%%%%%%%%%%%%%%%%%%%%%%%%%%%%%
\begin{equation}
\xiP_K\pP^\mu = \xP_K\pP_A^\mu
~,
\end{equation}
%%%%%%%%%%%%%%%%%%%%%%%%%%%%%%%%%%%%%%%%
as will become clear later.
For the $\xiM$ variables and $\pM$ there is no such issue, and $\pM$ can in fact be considered the hadron momentum itself and we use the variable
%%%%%%%%%%%%%%%%%%%%%%%%%%%%%%%%%%%%%%%%
\begin{equation}
\xM_K = \xiM_K
\end{equation}
%%%%%%%%%%%%%%%%%%%%%%%%%%%%%%%%%%%%%%%%
from now on.

The infinitesimal volume of a light-like momentum $K$ in terms of the decomposition becomes
%
%%%%%%%%%%%%%%%%%%%%%%%%%%%%%%%%%%%%%%%%
\begin{equation}
d^4K\,\delta(K^2)
 = d\xiP_K d\xM_K d^2\Kperp
     \,\frac{1}{2\xiP_K}\,\delta\left(\xM_K-\frac{|\Kperp|^2}{\nuSq\xiP_K}\right)
~.
\label{Eq:001}
\end{equation}
%%%%%%%%%%%%%%%%%%%%%%%%%%%%%%%%%%%%%%%%
%
We will use the same notation for $\Kperp$ as a two-dimensional Euclidean vector, and for its embedding in Minkowski space.
We will use the absolute value notation to make clear that a square is positive.
We introduce a notation for the $n$-particle parton-level differential cross section including a tree-level on-shell \matrixelement\ and the flux factor
%
%%%%%%%%%%%%%%%%%%%%%%%%%%%%%%%%%%%%%%%%
\begin{align}
&d\PSTstr_n\big(\kin,\kinBar\,;\{p_i\}_{i=1}^{n}\big) =
\notag\\&\hspace{8ex}
 \bigg(\prod_{i=1}^{n}d^4p_i\delta(p_i^2-m_i^2)\bigg)\delta^4\bigg(\kin +\kinBar-\sum_{i=1}^np_i\bigg)\,\frac{\MtreeStar\big(\kin,\kinBar\,;\{p_i\}_{i=1}^{n}\big)}{4\xiPin\xMin\lop{\pP}{\pM}}
~.
\label{Eq:002}
\end{align}
%%%%%%%%%%%%%%%%%%%%%%%%%%%%%%%%%%%%%%%% 
%
Let us state immediately that in this write-up,
%
%%%%%%%%%%%%%%%%%%%%%%%%%%%%%%%%%%%%%%%%
\begin{equation}
\textrm{the symbol $\ColoredAmp$ always refers to a {\em tree-level} amplitude.}
\end{equation}
%%%%%%%%%%%%%%%%%%%%%%%%%%%%%%%%%%%%%%%%
%
The arguments in $\PSTstr_n$ and $\MtreeStar$ are separated into initial-state and final-state ones by the semi-colon.
The star highlights that there is a space-like initial state, so
%
%%%%%%%%%%%%%%%%%%%%%%%%%%%%%%%%%%%%%%%%
\begin{equation}
\kin^\mu = \xiPin\pP^\mu+\kperp^\mu
\quad,\quad
\kinBar^\mu = \xMin\pM^\mu
~.
\end{equation}
%%%%%%%%%%%%%%%%%%%%%%%%%%%%%%%%%%%%%%%%
%
If $|\kperp|=0$ and both initial states are light-like, then the star is omitted.
The \matrixelement\ $\MtreeStar$ is assumed to be summed over final-state colors and spins, and averaged over initial-state colors and spins.
One might say that the subscript $n$ in $\PSTstr_n$ is superfluous because its arguments are listed explicitly, just like in $\MtreeStar$, but we keep it for later convenience.
%
%The delta-function expressing momentum conservation can be expressed in more detail as
%%
%%%%%%%%%%%%%%%%%%%%%%%%%%%%%%%%%%%%%%%%%
%\begin{equation}
%\delta^4\bigg(\xiPin\pP+\kperp+\xMin\pM-\sum_{i=1}^np_i\bigg)
%=
%\frac{1}{\lop{\pP}{\pM}}
%\,\delta\bigg(\xiPin - \sum_{i=1}^n\xiP_i\bigg)
%\,\delta\bigg(\xMin - \sum_{i=1}^n\xM_i\bigg)
%\,\delta^2\bigg(\kperp-\sum_{i=1}^np_{{\scriptscriptstyle\perp} i}\bigg)
%~.
%\end{equation}
%%%%%%%%%%%%%%%%%%%%%%%%%%%%%%%%%%%%%%%%%
%%

In order to define a differential Born-level cross section, we also need a ``jet function'' that guarantees that final-state partons cannot become arbitrarily soft or arbitrarily collinear to each other or to the initial-state momenta if this causes a singularity.
Realize that this includes the case when a final-state momentum becomes collinear to $\pP$.
Also \matrixelement{}s with a space-like initial-state can have such a collinear singularity.
We write
%
%%%%%%%%%%%%%%%%%%%%%%%%%%%%%%%%%%%%%%%%
\begin{equation}
\dBstar\big(\kin,\kinBar\,;\{p_i\}_{i=1}^{n}\big)
=
\frac{1}{|\kperp|^2}\, d\PSTstr_n\big(\kin,\kinBar\,;\{p_i\}_{i=1}^{n}\big)
\,\JetB\big(\{p_i\}_{i=1}^{n}\big)
~.
\label{Eq:003}
\end{equation}
%%%%%%%%%%%%%%%%%%%%%%%%%%%%%%%%%%%%%%%%
%
Whereas the ``space-like gluon propagator'' was taken out in the definition of the squared matrix element to have a smooth on-shell limit, it appears to be most convenient to include the factor $1/|\kperp|^2$ in the definition of the partonic cross section.
We include possible non-parton momenta in the argument list of the jet function $\JetB\big(\{p_i\}_{i=1}^{n}\big)$ and simply assumes that it selects only parton momenta.

In order to obtain the real contribution at NLO, the partonic scattering process is augmented with one more final-state parton, and the jet function is advanced in that it incorporates an at most one-step recombination of parton momenta, so it produces both the resolved situation with one more jet than at Born level, and the unresolved situation with the same number of jets.
We denote it
%
%%%%%%%%%%%%%%%%%%%%%%%%%%%%%%%%%%%%%%%%
\begin{equation}
\JetR\big(\{p_i\}_{i=1}^{n+1}\big)
~.
\end{equation}
%%%%%%%%%%%%%%%%%%%%%%%%%%%%%%%%%%%%%%%%
%
We deliberately equip the jet functions with the labels $\Born$ for ``Born'' and $\Real$ for ``real'' rather than $n$ and $n+1$, to highlight that $\JetR$ is not merely a generalization of $\JetB$ regarding phase space cuts.
We mention explicitly that also in the space-like initial-state kinematics, the jet function considers a phase space configuration unresolved if a final-state momentum becomes collinear to $\pP$. 
Given this jet function, we define what we call the {\em \familiar\ real contribution} as
%
%%%%%%%%%%%%%%%%%%%%%%%%%%%%%%%%%%%%%%%%
\begin{align}
\dRfam\big(\kin,\kinBar\,;\{p_i\}_{i=1}^{n+1}\big)
=
    \frac{\cNLO\mu^{2\vepv}}{\piep}\,\frac{1}{|\kperp|^2}
    \,d\PSTstr_{n+1}\big(\kin,\kinBar\,;\{p_i\}_{i=1}^{n+1}\big)
    \,\JetR\big(\{p_i\}_{i=1}^{n+1}\big)
~.
\label{Eq:004}
\end{align}
This is what one naively would write down as the real radiation contribution given the Born formula. 
Of course, actual integration of this differential cross section leads to soft and collinear divergencies, but these can be dealt with using existing methods that can be found in literature of NLO calculations.
Only the collinear singularity associated with the space-like initial-state is different than usual, and we will return to this later.

The overall constant in \Equation{Eq:004} gives the correction factor compared to Born level considering that we are using dimensional regularization, we assume $\gQCD=1$ inside the \matrixelement, and considering the fact that we omitted factors of $2\pi$ in our definition of the phase space in \Equation{Eq:002} compared to usual definitions found in literature.
The symbols in \Equation{Eq:004} represent
%
%%%%%%%%%%%%%%%%%%%%%%%%%%%%%%%%%%%%%%%%
\begin{equation}
\piep = \frac{\pi^{1-\vepv}}{\Gamma(1-\vepv)}
\quad,\quad
\cNLO = \frac{\alphaS}{2\pi}\,\frac{(4\pi)^{\vepv}}{\Gamma(1-\vepv)}
\quad,\quad
\vepv = \frac{4-\mathrm{dim}}{2}
~.
\end{equation}
%%%%%%%%%%%%%%%%%%%%%%%%%%%%%%%%%%%%%%%%
%
In other words, both the Born and the real, and later the virtual, contribution are missing an overall factor
$(2\pi)^{4-3n}\gQCD^{2n}=(2\pi)^{4-3n}(4\pi\alphaS)^{n}$.

\subsection{The auxiliary parton method}
As mentioned in the introduction, the tree-level \matrixelement\ with the space-like initial-state gluon can be defined with the auxiliary parton method proposed in~\mycite{vanHameren:2012if} and most recently applied in~\mycite{Blanco:2020akb}.
We would like to mention that its connection with approaches employing Wilson line operators can, at least at tree-level, be understood from the work in~\cite{Kotko:2014aba}.
Let the desired parton-level process be
%
%%%%%%%%%%%%%%%%%%%%%%%%%%%%%%%%%%%%%%%%
\begin{equation}
g^\star(k_{\inlbl})\;\omega_{\inbar}(k_{\inbar})
\;\to\;
\omega_{1}(p_{1})
\;\omega_{2}(p_{2})
\;\cdots
\;\omega_{n}(p_{n})
\end{equation}
%%%%%%%%%%%%%%%%%%%%%%%%%%%%%%%%%%%%%%%%
%
where the $\omega_i$ are the type of partons or particles involved besides the space-like gluon.
In the auxiliary parton method, this process is obtained from the process
%
%%%%%%%%%%%%%%%%%%%%%%%%%%%%%%%%%%%%%%%%
\begin{equation}
q\big(k_{1}(\Lambda)\big)\;\omega_{\inbar}(k_{\inbar})
\;\to\;
q\big(k_{2}(\Lambda)\big)
\;\omega_{1}(p_{1})
\;\omega_{2}(p_{2})
\;\cdots
\;\omega_{n}(p_{n})
\label{Eq:005}
\end{equation}
%%%%%%%%%%%%%%%%%%%%%%%%%%%%%%%%%%%%%%%%
%
where the momenta $k_{1}(\Lambda)$ and $k_{2}(\Lambda)$ are parametrized with $\Lambda$ in such a way that $k_1^2=k_2^2=0$ for any value of $\Lambda$, and $k_{1}-k_{2}=k_{\inlbl}+\Ord\big(\Lambda^{-1}\big)$.
In fact, in publications so far the parametrization was such that even the $\Ord\big(\Lambda^{-1}\big)$ is absent, but here we will allow this irrelevant term, which allows us to use the parametrization
%
%%%%%%%%%%%%%%%%%%%%%%%%%%%%%%%%%%%%%%%%
\begin{equation}
k_1^\mu = \Lambda\pP^\mu
\quad,\quad
k_2^\mu=p_\Lambda^\mu = (\Lambda-\xiPin)\pP^\mu - \kperp^\mu + \frac{|\kperp|^2}{(\Lambda-\xiPin)\nuSq}\,\pM^\mu
\label{Eq:006}
~.
\end{equation}
%%%%%%%%%%%%%%%%%%%%%%%%%%%%%%%%%%%%%%%%
%
The process of \Equation{Eq:005} with the auxiliary quarks is on-shell, and its \matrixelement\ is well-defined and it is known how to calculate it.
The \matrixelement\ of the desired process with the space-like gluon is obtained by taking
%%%%%%%%%%%%%%%%%%%%%%%%%%%%%%%%%%%%%%%%
\begin{equation}
\Lambda\to\infty
~.
\end{equation}
%%%%%%%%%%%%%%%%%%%%%%%%%%%%%%%%%%%%%%%%
%

In \mycite{vanHameren:2017hxx} and \mycite{Blanco:2020akb} it was noted that instead of an auxiliary scattering quark, also an auxiliary scattering gluon can be used.
At the level of \matrixelement{}s summed over color, one just needs to include a different overall factor to correct for the difference in color representation:
%
%%%%%%%%%%%%%%%%%%%%%%%%%%%%%%%%%%%%%%%%
\begin{equation}
\frac{1}{\gQCD^2\Caux}\,\frac{\xiPin^2|\kperp|^2}{\Lambda^2}
\,\MtreeAux\big(\Lambda\pP,\kinBar\,;p_{\Lambda},\{p_i\}_{i=1}^{n}\big)
\;\overset{\Lambda\to\infty}{\longrightarrow}\;
\MtreeStar\big(\kin,\kinBar\,;\{p_i\}_{i=1}^{n}\big)
\label{Eq:007}
\end{equation}
%%%%%%%%%%%%%%%%%%%%%%%%%%%%%%%%%%%%%%%%
with
%
%%%%%%%%%%%%%%%%%%%%%%%%%%%%%%%%%%%%%%%%
\begin{equation}
\Cauxq = \frac{\Nc^2-1}{\Nc}
\quad,\quad
\Cauxg = 2\Nc
~.
\end{equation}
%%%%%%%%%%%%%%%%%%%%%%%%%%%%%%%%%%%%%%%%
%
%
We call this {\em auxiliary parton universality} and we will see that it does not hold anymore at NLO.
Another property that will appear not to hold anymore is the {\em smooth on-shell limit}
%
%%%%%%%%%%%%%%%%%%%%%%%%%%%%%%%%%%%%%%%%
\begin{equation}
\lim_{z\to0}\int\frac{d^2\kperp}{\pi}\,\delta\big(|\kperp|^2-z\big)\,\MtreeStar\big(\xiPin\pP+\kperp,\kinBar\,;\{p_i\}_{i=1}^{n}\big)
=
\Mtree\big(\xiPin\pP,\kin\,;\{p_i\}_{i=1}^{n}\big)
~,
\end{equation}
%%%%%%%%%%%%%%%%%%%%%%%%%%%%%%%%%%%%%%%%
%
where the right-hand-side is the \matrixelement\ for the process with an on-shell initial-state gluon instead of the space-like gluon.

\subsection{\label{Sec:023}Scaling behavior}
As explicated in~\mycite{vanHameren:2014iua}, rather than a function of $\xiPin$ and $\kperp$, the space-like tree level amplitude must be seen as a function of $k_{\inlbl}$ and the light-like {\em direction} $P$ with the single property that $\lop{k_{\inlbl}}{\pP}=0$.
Interpreted like that, the amplitude satisfies the scaling relation
%
%%%%%%%%%%%%%%%%%%%%%%%%%%%%%%%%%%%%%%%%
\begin{equation}
\Amp_{\Tree}^{\star}\big(\lambda\pP,k_{\inlbl},k_{\inbar};\{p_i\}_{i=1}^n\big) 
= \lambda\Amp_{\Tree}^{\star}\big(\pP,k_{\inlbl},k_{\inbar};\{p_i\}_{i=1}^n\big)
~.
\end{equation}
%%%%%%%%%%%%%%%%%%%%%%%%%%%%%%%%%%%%%%%%
%
The factor $\xiPin^2$ in \Equation{Eq:007} is included exactly in order to match the direction $\pP\to\xiPin\pP$ with the on-shell limit momentum.
This could also have been achieved by defining the auxiliary momenta of \Equation{Eq:006} as $\Lambda\xiPin\pP$ and $(\Lambda-1)\xiPin\pP-\cdots$ etc., but that would be inconvenient for the following. 

We see that the amplitude, and thus the \matrixelement, but in fact also $\dBstar$ do not depend on $\xiPin$ or $\pP$ separately, but only on the combination $\xiPin\pP$.
We can scale 
%%%%%%%%%%%%%%%%%%%%%%%%%%%%%%%%%%%%%%%%
\begin{equation}
\xiPin\to\xPin=\lambda^{-1}\xiPin
\quad\textrm{and}\quad
\pP\to\pPA=\lambda\pP
\label{Eq:067}
\end{equation}
%%%%%%%%%%%%%%%%%%%%%%%%%%%%%%%%%%%%%%%%
%
without changing $\dBstar$.
This ``scale'' $\lambda$ is set by introducing an integration over initial-state variables as
%
%%%%%%%%%%%%%%%%%%%%%%%%%%%%%%%%%%%%%%%%
\begin{equation}
\int d^4\kin\,\EuScript{F}(\kin)\,\dBstar\big(\kin,\kinBar\,;\{p_i\}_{i=1}^n\big)
\end{equation}
%%%%%%%%%%%%%%%%%%%%%%%%%%%%%%%%%%%%%%%%
%
with
%
%%%%%%%%%%%%%%%%%%%%%%%%%%%%%%%%%%%%%%%%
\begin{equation}
\EuScript{F}(\kin) =
 \int \frac{d^2\kperp}{\pi}\,\frac{d\xPin}{\xPin}\,F(\xPin,|\kperp|)\,\delta^4\big(\kin-\xPin\pPA-\kperp\big)
~
\end{equation}
%%%%%%%%%%%%%%%%%%%%%%%%%%%%%%%%%%%%%%%%
%
where $F(\xPin,|\kperp|)$ is the $\kperp$-dependent PDF, or TMD.
This introduces the physical hadron momentum and its fraction.

\subsection{\Familiar\ contribution at NLO}
In \Equation{Eq:004} we presented what we call the {\em \familiar\ real} radiation contribution as what one naively would write down given the Born formula for a partonic cross section with a space-like initial-state gluon, by adding a final-state parton to the process, and allowing a pair of partons to become collinear, and a single parton to become soft.
Obviously, final-state collinear behavior is independent of whether an initial-state parton is on-shell or space-like, and formulas for the singularities are the same.
In \Section{Sec:050} we show that the same is true for the soft behavior, that is the formulas for the singularities look the same as if the initial-state gluon were on-shell.
Only the tree-level \matrixelement\ appearing in those formulas has the space-like gluon.

Similarly, we define the {\em \familiar\ virtual} contribution as the finite part, plus the divergent part that looks exactly as if the space-like gluon were on-shell.
Again, only the tree-level \matrixelement\ appearing in that divergent part has the space-like gluon.
The remaining part, we call {\em \unfamiliar}, we will identify in \Section{Sec:040}.
To be precise, the \unfamiliar\ part also contains a finite contribution proportional to the tree-level \matrixelement.
In \Section{Sec:060} we show that the divergent part can indeed be separated into the \unfamiliar\ part of \Section{Sec:040} plus terms that look as if the space-like gluon were on-shell.

Because the divergent parts of both \familiar\ contributions look exactly as if the space-like gluon were on-shell, their cancellation also happens as if it were on-shell.
Only the initial-state collinear behavior related to the space-like gluon is different, and leads to $\Delta^\star_{\coll}$ in \Equation{Eq:094}, and is presented in \Equation{Eq:065}.

The \familiar\ contributions, finally, are independent of the type of auxiliary partons used.

\subsection{Unfamiliar contributions at NLO}
The {\em \unfamiliar} real contribution occurs in the auxiliary parton method because there is a region in phase space where the radiative gluon should be allowed to participate in the ``consumption'' of the large $\pP$-component $\Lambda$ when $\Lambda\to\infty$.
We calculate the contribution in \Section{Sec:030}, and it turns out to violate the three tree-level properties of auxiliary parton universality, the smooth on-shell limit, and the smooth large $\Lambda$ limit.
The {\em \unfamiliar\ real contribution} is proportional to the Born level result and given by
%
%%%%%%%%%%%%%%%%%%%%%%%%%%%%%%%%%%%%%%%%
\begin{equation}
\dRunf\big(\kin,\kinBar\,;\{p_i\}_{i=1}^n\big)
 =
 \Freal_{\aux}%(\xiPin,\kperp,\mu,\Lambda,\xPlow)
 \,\dBstar\big(\kin,\kinBar\,;\{p_i\}_{i=1}^n\big)
~,
\label{Eq:008}
\end{equation}
%%%%%%%%%%%%%%%%%%%%%%%%%%%%%%%%%%%%%%%%
%
with
%
%%%%%%%%%%%%%%%%%%%%%%%%%%%%%%%%%%%%%%%%
\begin{equation}
\Freal_{\aux} = 
    \cNLO\Nc\bigg(\frac{\mu^2}{|\kperp|^2}\bigg)^{\vepv}
    \Big[\bar{\Freal}_{\univ}^{\pole} + \bar{\Freal}_{\aux}\Big]
+ \Ord\big(\vepv\big) + \Ord\big(\Lambda^{-1}\big)
~,
\end{equation}
%%%%%%%%%%%%%%%%%%%%%%%%%%%%%%%%%%%%%%%%
%%%%%%%%%%%%%%%%%%%%%%%%%%%%%%%%%%%%%%%%
%
where
%
%%%%%%%%%%%%%%%%%%%%%%%%%%%%%%%%%%%%%%%%
\begin{align}
\bar{\Freal}_{\univ}^{\pole} &=
  {-}\frac{2}{\vepv}\ln\frac{\nuSq\Lambda}{|\kperp|^2}
\label{Eq:087}\\
\bar{\Freal}_{\auxq} &= 
\frac{3}{\vepv} -\frac{2\pi^2}{3} + \frac{7}{2}
-\frac{1}{\Nc^2}\,\bigg[\frac{1}{\vepv^2} + \frac{3}{2\vepv} + 4\bigg]
~,
\label{Eq:009}
\\
\bar{\Freal}_{\auxg} &= 
\frac{1}{\vepv^2} 
+ \frac{1}{\vepv}\frac{11}{2}
%\notag\\&\hspace{24ex}
 - \frac{2\pi^2}{3}+ \frac{67}{9}
-\frac{n_f}{\Nc}\bigg[
   \frac{2}{3\vepv}
  +\frac{10}{9}
  -\frac{1}{\Nc^2}\bigg(\frac{1}{3\vepv}-\frac{1}{6}\bigg)
\bigg]
~.
\label{Eq:010}
\end{align}
%%%%%%%%%%%%%%%%%%%%%%%%%%%%%%%%%%%%%%%%
%
%
The auxiliary-gluon case includes the contribution for the real process with an initial-state auxiliary gluon and a quark-antiquark pair in the final state, and $n_f$ is the number of quark flavors included.
We must mention here, that the expression for $\bar{\Freal}_{\univ}^{\pole}$ was obtained by restricting the phase space of the radiative gluon following the procedure presented in~\mycite{Ciafaloni:1998hu}.
This will be explained in \Section{Sec:031}.

The virtual contribution at NLO lives in the same phase space as the Born contribution, but involves the interference between the tree-level amplitude and the one-loop amplitude instead of the tree-level \matrixelement.
The paper~\mycite{vanHameren:2017hxx} addresses the calculation of one-loop amplitudes with a space-like gluon using the auxiliary parton method, and it is observed that they contain $\ln\Lambda$.
They also violate auxiliary parton universality and the smooth on-shell limit, and in \Section{Sec:040}, we will argue that the complete virtual contribution causing these violations is again proportional to the Born contribution, and given by
%
%
%%%%%%%%%%%%%%%%%%%%%%%%%%%%%%%%%%%%%%%%
\begin{equation}
\dVunf\big(\xiPin,\kperp,\xMin\,;\{p_i\}_{i=1}^{n}\big) = \cNLO\Nc\,\mathrm{Re}\big(\Vcoef_{\aux}\big)
\,\dBstar\big(\xiPin,\kperp,\xMin\,;\{p_i\}_{i=1}^{n}\big)
~,
\end{equation}
%%%%%%%%%%%%%%%%%%%%%%%%%%%%%%%%%%%%%%%%
%
with
%
%%%%%%%%%%%%%%%%%%%%%%%%%%%%%%%%%%%%%%%%
\begin{equation}
\Vcoef_{\aux} = 
\bigg(\frac{\mu^2}{|\kperp|^2}\bigg)^{\vepv}
\bigg[\frac{2}{\vepv}\ln\frac{\Lambda}{\xiPin}-\imag\pi + \bar{\Vcoef}_{\aux}\bigg]
+\Ord(\vepv) +\Ord\big(\Lambda^{-1}\big)
~,
\label{Eq:063}
\end{equation}
%%%%%%%%%%%%%%%%%%%%%%%%%%%%%%%%%%%%%%%%
%
and
%
%%%%%%%%%%%%%%%%%%%%%%%%%%%%%%%%%%%%%%%%
\begin{align}
\bar{\Vcoef}_{\auxq} &=
   \frac{1}{\vepv}\,\frac{13}{6}
  + \frac{\pi^2}{3} +\frac{80}{18}
+\frac{1}{\Nc^2}\bigg[\frac{1}{\vepv^2}+\frac{3}{2}\frac{1}{\vepv}+4\bigg]
-\frac{n_f}{\Nc}\bigg[\frac{2}{3}\frac{1}{\vepv} + \frac{10}{9}\bigg]
~,
\label{Eq:012}
\\
\bar{\Vcoef}_{\auxg} &=
-\frac{1}{\vepv^2} 
 + \frac{\pi^2}{3} 
~.
\label{Eq:013}
\end{align}
%%%%%%%%%%%%%%%%%%%%%%%%%%%%%%%%%%%%%%%%
%
The parameter $n_f$ is the number of flavors running in quark loops.
We call this the {\em \unfamiliar\ virtual contribution}.
The rest of the virtual contribution we call {\em \familiar}, and is independent of the auxiliary partons used, has a smooth on-shell limit, and behaves regularly with $\Lambda$.
While the last claim will not exceed the status of conjecture in this publication, the former two will get more solid ground.

\subsection{Interpretation of the \unfamiliar\ contributions}
Combining the \unfamiliar\ contributions and organizing them in a suggestive manner, we can write
%
%%%%%%%%%%%%%%%%%%%%%%%%%%%%%%%%%%%%%%%%
\begin{align}
\dRunf + \dVunf
=
\Delta_{\unf}\,\dBstar
~,
\end{align}
%%%%%%%%%%%%%%%%%%%%%%%%%%%%%%%%%%%%%%%%
%
where
%
%%%%%%%%%%%%%%%%%%%%%%%%%%%%%%%%%%%%%%%%
\begin{align}
\Delta_{\unf}
=
\frac{\cNLO\Nc}{\vepv}
\,\bigg(\frac{\mu^2}{|\kperp|^2}\bigg)^{\vepv}
\bigg[
      \EuScript{I}_{\aux}
    + \EuScript{I}_{\univ}+ \EuScript{I}_{\univ}
    - 2\ln\frac{\nuSq\xiPin}{|\kperp|^2}
 \bigg]
\label{Eq:047}
~,
\end{align}
%%%%%%%%%%%%%%%%%%%%%%%%%%%%%%%%%%%%%%%%
%
with
%
%%%%%%%%%%%%%%%%%%%%%%%%%%%%%%%%%%%%%%%%
\begin{equation}
\EuScript{I}_{\univ} = 
  \frac{11}{6}-\frac{n_f}{3\Nc}
  -\frac{\EuScript{K}}{\Nc}(-\vepv)
\quad\textrm{writing}\quad
\EuScript{K} = \Nc\bigg(\frac{67}{18}-\frac{\pi^2}{6}\bigg)-\frac{5n_f}{9}
~,
\label{Eq:014}
\end{equation}
%%%%%%%%%%%%%%%%%%%%%%%%%%%%%%%%%%%%%%%%
%
and
%
%%%%%%%%%%%%%%%%%%%%%%%%%%%%%%%%%%%%%%%%
\begin{equation}
\EuScript{I}_{\auxq} = \frac{3}{2}-\frac{1}{2}(-\vepv)
\quad,\quad
\EuScript{I}_{\auxg} = 
 \frac{11}{6}+\frac{n_f}{3\Nc^3}+\frac{n_f}{6\Nc^3}(-\vepv)
~.
\label{Eq:015}
\end{equation}
%%%%%%%%%%%%%%%%%%%%%%%%%%%%%%%%%%%%%%%%
%
We can now compare with the results in \mycite{Ciafaloni:1998hu} for the corrections to impact factors.
Realize the difference in the definition of parameter for dimensional regularization $\varepsilon=-\vepv$.
We see that $\EuScript{I}_{\auxq}+\EuScript{I}_{\univ}$ is identical to their equation~(4.9), and $\EuScript{I}_{\auxg}+\EuScript{I}_{\univ}$ is identical to their equation~(5.11).
We introduced the symbol $\EuScript{K}$ for the same quantity defined in their equation~(4.10).
The extra term $\EuScript{I}_{\univ}$ in \Equation{Eq:047} is related to the renormalization of the coupling constant, and appears because our virtual result is not UV-subtracted.
In other words, $\vepv\bar{\Vcoef}_{\aux}-\EuScript{I}_{\univ}$ gives (two times) the virtual expressions between the square brackets in (4.6) and (5.9) in~\mycite{Ciafaloni:1998hu}.

In our approach, the \unfamiliar\ contribution emerges as a consequence of using the auxiliary parton method at the NLO, and of replacing the on-shell gluon parton that enters the collinear factorization framework by an space-like gluon.
It comes from the difference between the emission amplitude of the space-like $t$-channel gluon from the auxiliary parton and the amplitude of collinear splitting of the auxiliary parton into the gluon.
Hence in a natural way the NLO auxiliary parton impact factor enters the \unfamiliar\ contribution.
In addition, however, in the \unfamiliar\ contribution one finds the term proportional to logarithm of the available collision energy, which is not part of the impact factor.
The latter term is just the $\Ord(\alpha_s)$ contribution to the space-like gluon Regge trajectory and, within the high-energy factorization approach, is a part of the evolution kernel.
Notice that, while the individual contributions do, the eventual combination of the real and virtual \unfamiliar\ contribution does not depend on $\Lambda$ anymore, and does not violate the scale invariance mentioned in \Section{Sec:023}.
Also, it does not depend on the type of auxiliary partons.

The form of the \unfamiliar\ contribution is not unique --- there is some scheme dependence left in the finite contribution for $\vepv \to 0$.
It is necessary that the collinear singularity does not appear in the real unfamiliar contribution, as it is already taken into account in the familiar real contribution.
The remaining scheme dependence comes from certain arbitrariness of separation criterion of the phase space for real gluon emission into the impact factor region and the $k_T$-dependent evolution region.
This issue was discussed in detail in \mycite{Ciafaloni:1998kx,Ciafaloni:1998hu}.
Of course, the unique definition of the \unfamiliar\ contribution --- or equivalently --- of the NLO auxiliary parton impact factor sets a scheme which, in turn, constrains the definition of the space-like gluon at NLO and defines the energy scale of the $k_T$-dependent evolution.
The prescription applied in the present paper which corresponds to simply attributing the double IR pole contribution to the \familiar\ part (hence dropping it from the \unfamiliar\ part) is exactly implementing the scheme proposed in \mycite{Ciafaloni:1998kx}, which separates the impact factor region from the evolution region by applying the angular ordering criterion on the real gluon radiation.

\subsection{Non-cancelling \familiar\ contributions}
The \unfamiliar\ contributions are those with the non-smooth on-shell limit, therefore \familiar\ ones have a smooth on-shell limit.
We must specify here that the smoothness really only refers to the $1/\vepv$ divergencies.
The \familiar\ real contribution may still develop finite logarithms of $|\kperp|$, but they are eventually integrable underneath the $\kperp$-integral.

Since the \familiar\ contributions have a smooth on-shell limit, the (non)-cancellation of their divergencies are expected happen just as in the on-shell limit.
In \Section{Sec:060} and \Section{Sec:050} we show that this is indeed the case and that the only non-cancelling ones are of the collinear kind related to the initial-state partons.
At the on-shell $\pM$-side they are of the well-known form, at the space-like $\pP$-side they are slightly different.

Let us repeat the usual treatment of the non-cancelling collinear divergencies.
The tree-level \matrixelement{}s in the \familiar\ real contribution have singularities when the radiative gluon becomes collinear to $\pM$.
Part of the divergent contribution can only be written in a form such that they are proportional to the Born result if the convolution with the PDF is included.
This contribution can then be written as the Born cross section, but with the collinear PDF $f(\xMin)$ replaced by 
\begin{equation}
    -\frac{\cNLO}{\vepv}
\int_{\xMin}^{1}dz\,\PBar^{\mathrm{reg}}(z)\,f\bigg(\frac{\xMin}{z}\bigg)
\label{Eq:059}
\end{equation}
where $\PBar^{\mathrm{reg}}(z)$ is the regulated splitting function coming from the collinear limit of the tree-level \matrixelement.
Remember that in our conventions, $f(\xMin)$ is the function that naturally comes with the measure $d\xMin/\xMin$.
We must mention here that the splitting function the color factor that we later in this work will often take out.
So at this point, it just has its singularity at $z=1$ regulated via the plus-prescription, but has no terms proportional to $\delta(1-z)$.
The plus-prescription takes out the soft-collinear contribution which {\em does} cancel against the virtual contribution.
This term can then be interpreted within the subtractive renormalization of the PDF, at which point the term $\gamma_{\inbar}\delta(1-z)$, coming from the non-cancelling collinear virtual contribution, is added to the splitting function.
The well-known coefficients are
%
%%%%%%%%%%%%%%%%%%%%%%%%%%%%%%%%%%%%%%%%
\begin{equation}
\gamma_{\mathrm{q}} = \gamma_{\bar{\mathrm{q}}} = \frac{3}{2}\frac{\Nc^2-1}{2\Nc}
\quad,\quad
\gamma_{\mathrm{g}} = \frac{11\Nc-2n_f}{6}
~.
\end{equation}
%%%%%%%%%%%%%%%%%%%%%%%%%%%%%%%%%%%%%%%%
%
This has to happen in conjunction with ultraviolet (UV) renormalization, because within dimensional regularization, collinear and UV divergencies typically cancel to a large degree among themselves in one-loop amplitudes, and the collinear divergencies have to be ``put back'' in the form of UV counterterms.
So in short, in the $\overline{\mathrm{MS}}$ scheme
\begin{equation}
\Delta_{\collBar}
=
    -\frac{\cNLO}{\vepv}
\int_{\xMin}^{1}dz\,\big[\PBar^{\mathrm{reg}}(z)+\gamma_{\inbar}\delta(1-z)\big]f\bigg(\frac{\xMin}{z}\bigg)
\label{Eq:064}
\end{equation}
is imagined to be absorbed into the NLO PDF.

The tree-level \matrixelement\ also has singularities when the radiative gluon becomes collinear to $\pP$, and in \Section{Sec:050} we find that the associated non-cancelling contribution equivalent to the above can again be written as the Born cross section, but the $\kperp$-dependent PDF $F(\xPin,|\kperp|)$ replaced by
%
%%%%%%%%%%%%%%%%%%%%%%%%%%%%%%%%%%%%%%%%
\begin{equation}
-\frac{\cNLO}{\vepv}
\int_{\xPin}^1dz\bigg[\frac{2\Nc}{[1-z]_+}+\frac{2\Nc}{z}\bigg]F\bigg(\frac{\xPin}{z},|\kperp|\bigg)
~.
\label{Eq:062}
\end{equation}
%%%%%%%%%%%%%%%%%%%%%%%%%%%%%%%%%%%%%%%%
%
Regarding the non-cancelling \familiar\ virtual contributions, we will show in \Section{Sec:060} that they are identical to the ones in the on-shell limit, so we see that
%
%%%%%%%%%%%%%%%%%%%%%%%%%%%%%%%%%%%%%%%%
\begin{equation}
\Delta^\star_{\coll}
=
-\frac{\cNLO}{\vepv}
\int_{\xPin}^1dz\bigg[\frac{2\Nc}{[1-z]_+}+\frac{2\Nc}{z}+\gamma_{\mathrm{g}}\delta(1-z)\bigg]F\bigg(\frac{\xPin}{z},|\kperp|\bigg)
\label{Eq:065}
\end{equation}
%%%%%%%%%%%%%%%%%%%%%%%%%%%%%%%%%%%%%%%%
%
must be absorbed into the $\kperp$-dependent NLO PDF.

Since the contribution looks very similar to the collinear case, it makes sense to consider a scenario like employed in~\mycite{Nefedov:2020ecb}, and write the function $F$ using a convolution involving a collinear PDF as
%
%%%%%%%%%%%%%%%%%%%%%%%%%%%%%%%%%%%%%%%%
\begin{equation}
F(\xPin,|\kperp|) = \int_{\xPin}^1dz\,\EuScript{C}(z,|\kperp|)\,f\bigg(\frac{\xPin}{z}\bigg)
~.
\end{equation}
%%%%%%%%%%%%%%%%%%%%%%%%%%%%%%%%%%%%%%%%
%
We will not digress on the meaning of the function $\EuScript{C}(z,|\kperp|)$ as it is beyond the scope of this work, but we do observe that
%
%%%%%%%%%%%%%%%%%%%%%%%%%%%%%%%%%%%%%%%%
\begin{align}
&\int_{\xPin}^1dz\bigg[\frac{1}{[1-z]_+}+\frac{1}{z}\bigg]F\bigg(\frac{\xPin}{z},|\kperp|\bigg)
=
\int_{\xPin}^1dz\,\EuScript{C}(z,|\kperp|)
 \int_{\xPin/z}^1dy\bigg[\frac{1}{[1-y]_+}+\frac{1}{y}\bigg]f\bigg(\frac{\xPin}{yz}\bigg)
~,
\end{align}
%%%%%%%%%%%%%%%%%%%%%%%%%%%%%%%%%%%%%%%%
%
and see that this non-cancelling contribution can be interpreted in terms of evolution of the collinear PDF.

\subsection{Interpretation within the high-energy factorization framework}
The definition of the impact factors $\Phi_A$ and $\Phi_B$ and of the unintegrated gluon density $F$ in our work is given in an operational form --- by a procedure to obtain them --- rather than an explicit diagrammatic or operator form.
In the operational definition we use the principle of isolating the factorizing pieces of amplitude with specific dependence on the hadron collision energy $ S = \Lambda\nu^2$.
The basis of the definition is the Regge factorization formula for high energy scattering amplitudes $\myScript M$:  ${\myScript M} \sim S \, \Phi_A \otimes {\myScript G} \otimes \Phi_B$ with impact factors $\Phi_A$, $\Phi_B$  and the Green's function (the evolution factor) ${\myScript G}$.
The main principle of factorizing ${\myScript M}$ into its components in our procedure is the vanishing energy dependence of the impact factors, up to neglected power corrections $\sim 1/S$, and including all the logarithmic effects $\sim \ln S$ in the Green's function.
At the LO the well known procedure with an auxiliary parton leads to a very simple prescription --- the auxiliary parton impact factors are constant functions  of $g^*$ transverse momentum, proportional to the color factor of the parton.
In order to access the unintegrated gluon density $F_A$ (or $F_B$) in $A$ (or $B$) at the LO it is enough to consider the scattering amplitude $AB \to A'B'$ with an auxiliary parton $B$ (or $A$), and factor out the constant impact factor of the auxiliary parton $B$ (or $A$).

Furthermore, in order to access the remaining LO impact factor, say $\Phi_B$, that is a part of $F_B$, one needs to factorize the obtained $F_B \sim {\myScript G} \otimes \Phi_B$ into $\myScript G$ and $\Phi_B$.

In order to perform this, one needs to fix the form of the Green's function $\myScript G$.
The natural separation of $F$ into $\myScript G$ and $\Phi_B$ at LO is usually done by fixing the form of $\myScript G$ by requiring that if $B$ is an auxiliary parton, say a quark $q$, then $F_B = {\myScript G} \otimes \Phi_q$, with $\Phi_q$ being the constant auxiliary quark impact factor.
Having defined $\myScript G$ by this constraint, one may factor it our from $F_B$ for a general scattering state $B\to B'$.

A similar procedure may be performed at the NLO.
Then, in the reasoning described above, we should use the NLO auxiliary partons $A$ and $B$ instead of those computed at LO.
The difference is that the NLO auxiliary parton impact factors carry a nontrivial $k_T$ dependence, and that, in general, at the NLO one finds some scheme dependence that may be related \eg\ to a phase space separation for real gluon radiation into regions of the impact factors and of the Green's function.
In our paper we discuss in detail a selected separation scheme for the auxiliary parton $A$ that is based on the angular ordering constraint  on the real gluon radiation.
This is not a unique choice, but it has an important theoretical feature --- cancellation of collinear divergencies in the impact factor and in the Green's function.

In our general hybrid factorization formula (4) we do not specify the impact factor $B$ because it does not matter for the structure and for the results we obtained.
Expression (4) is kept general and does not require providing a detailed scheme prescription to separate $F_A$ and $\Phi_B$.
However, in order to have an example of possible scheme, one can require that the separation between the impact factor $\Phi_B$ and the unintegrated gluon density in the auxiliary parton (or another kind of projectile scattering $A\to A'$) $F_A$ is performed in the same way as it was for the NLO auxiliary impact factor $\Phi_A$, and the rest of the amplitude proportional to $F_B$.
This is possible because the Regge factorization formula  ${\myScript M} \sim S \, \Phi_A \otimes {\myScript G} \otimes \Phi_B$ is structurally symmetric between the projectile and the target and hence allows for applying the same scheme for separating out the impact factors from the Green's function.

\section{\label{Sec:030}\Unfamiliar\ real contribution}
\begin{figure}
\begin{center}
\epsfig{figure=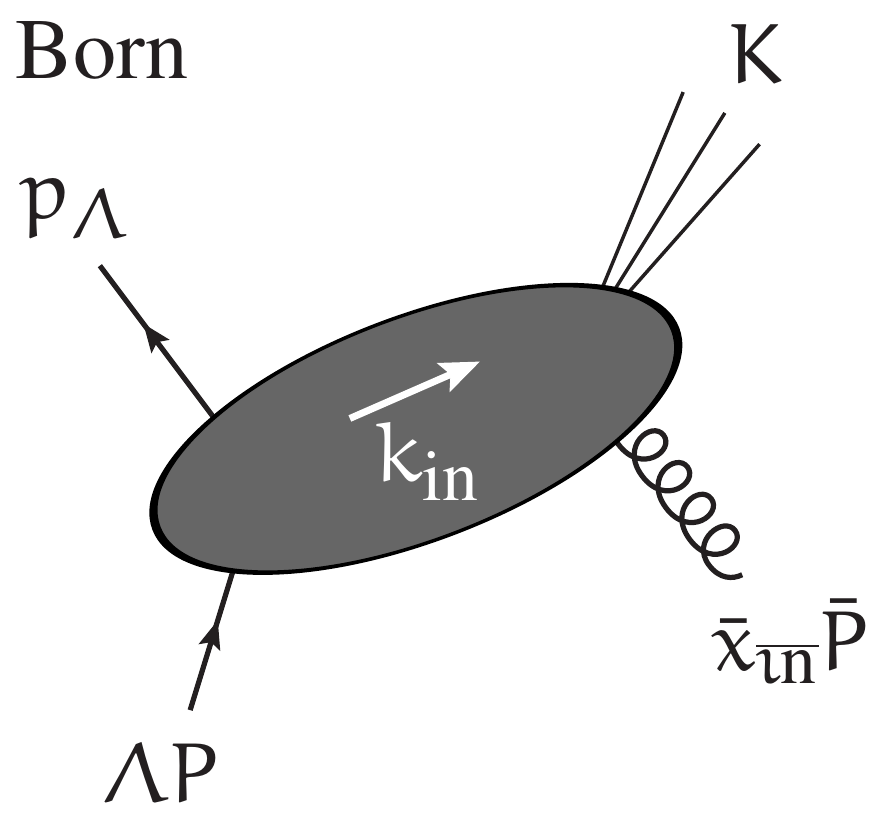,width=22ex}\hspace{6ex}
\raisebox{4ex}{\epsfig{figure=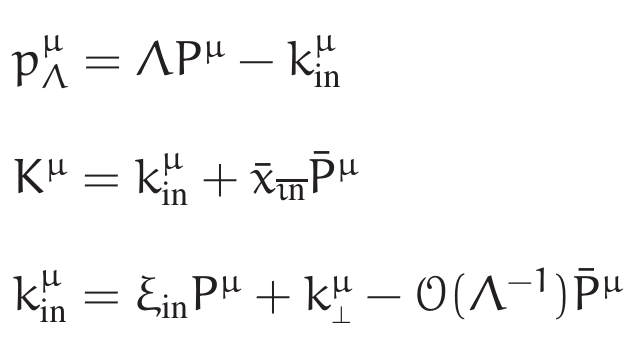,width=28ex}}\\[3ex]
\epsfig{figure=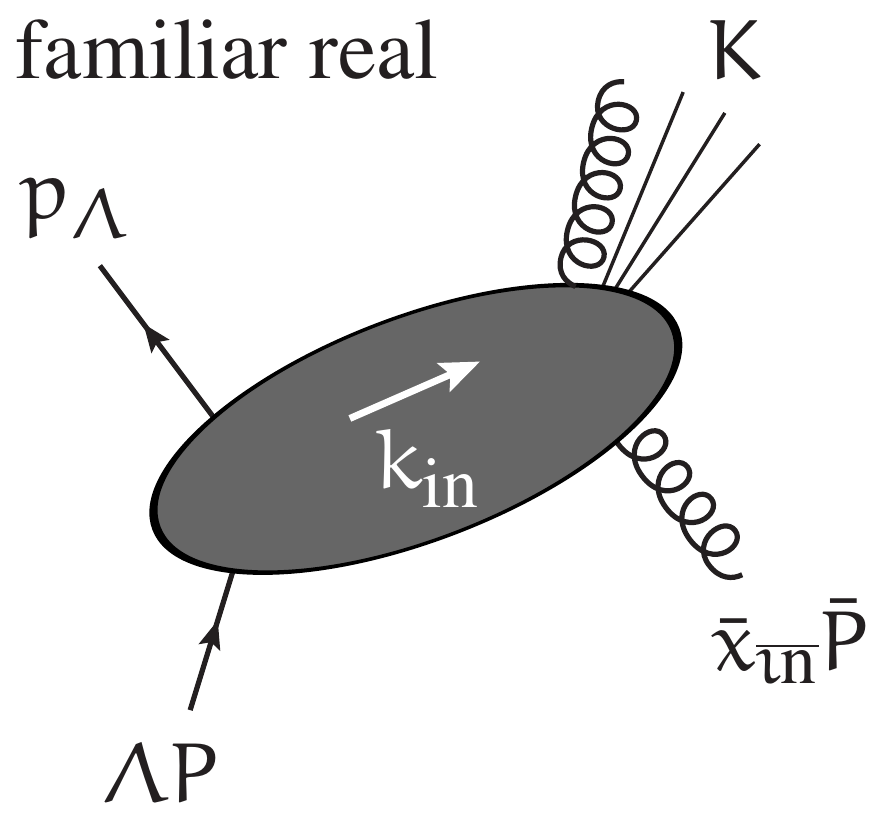,width=22ex}\hspace{11ex}
\epsfig{figure=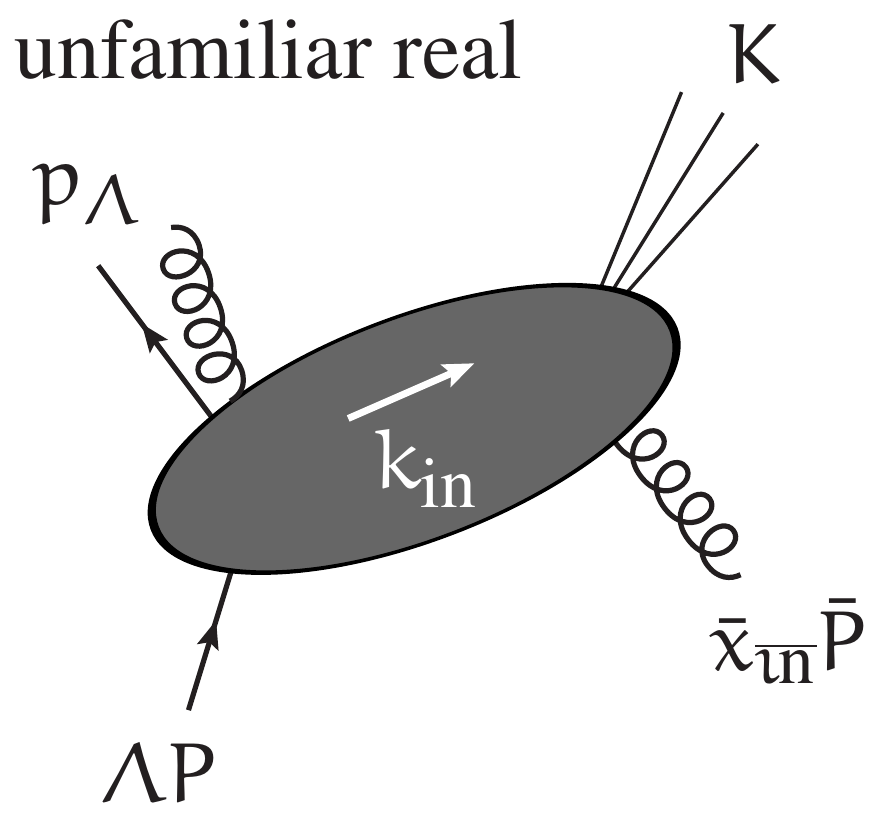,width=22ex}\hspace{0ex}
\caption{\label{Fig:01}Kinematics of the different contributions for the example in which the auxiliary partons are quarks, the collinear initial state parton is a gluon, and the real-radiation parton is also a gluon.}
\end{center}
\end{figure}
Originally, the auxiliary parton method was introduced just as a trick to obtain helicity amplitudes with space-like partons.
In order to go to NLO, we need to lift the method to cross section level including phase space.
First, we need to introduce some more notation.
We will frequently take a few arguments explicitly out of the list $\{p_i\}_{i=1}^n$ in a differential phase space, and give them different symbols so we can write their differential volumes explicitly.
So, for example we can write
%
%%%%%%%%%%%%%%%%%%%%%%%%%%%%%%%%%%%%%%%%
\begin{equation}
d\PSTstr_{n+2}\big(\kin,\kinBar\,;\{p_i\}_{i=1}^{n+2}\big)
=
d^4q\delta(q^2)\,d^4r\delta(r^2)\,\frac{d\PSTstr_{n+2}}{dqdr}\big(\kin,\kinBar\,;q,r,\{p_i\}_{i=1}^{n}\big)
~,
\end{equation}
%%%%%%%%%%%%%%%%%%%%%%%%%%%%%%%%%%%%%%%%
%
implying $q=p_{n+1}$ and $r=p_{n+2}$.
Prepending an expression like on the right-hand side with an integral symbol means we are only integrating over these explicit variables.
The essential observation is now that the \matrixelement\ relation of \Equation{Eq:007} can be formulated in terms of differential cross sections as follows:
%
%%%%%%%%%%%%%%%%%%%%%%%%%%%%%%%%%%%%%%%%
\begin{multline}
\frac{2\xiPin}{\Caux}\int d^4q\delta(q^2)\,\delta(\xiPin+\xiP_q-\Lambda)\,\delta^2(\kperp+\qperp)\,\frac{d\PSTaux_{n+1}}{dq}\big(\Lambda\pP,\kinBar\,;q,\{p_i\}_{i=1}^{n}\big)\,\JetB\big(\{p_i\}_{i=1}^{n}\big)
\\
\overset{\Lambda\to\infty}{\longrightarrow}
\dBstar\big(\kin,\kinBar\,;\{p_i\}_{i=1}^{n}\big)
~.
\label{Eq:016}
\end{multline}
%%%%%%%%%%%%%%%%%%%%%%%%%%%%%%%%%%%%%%%%
%
On the left-hand side there is the on-shell differential cross section for the process with auxiliary partons, with auxiliary initial-state momentum $\Lambda\pP$ and auxiliary final-state momentum $q$ that does {\em not} enter the jet function.
Consequently, it is the only final-state momentum for which the $\pP$ component can become arbitrarily large, ``consuming'' the large initial-state component $\Lambda$.
This is expressed by the delta-function with argument $\xiPin+\xiP_q-\Lambda$.
Realize that the flux factor inside $d\PSTaux_{n+1}$ on the left-hand side has a factor $\Lambda$, playing the role of a denominator factor in \Equation{Eq:007}.
The other one comes from the $1/\xiP_q$ in the differential volume of $q$.
The factor $1/|\kperp|^2$ in the definition of \Equation{Eq:003} was of course chosen to accommodate the factor in \Equation{Eq:007}, and the explicit factor $\xiPin$ together with the flux factor balance the $\xiPin^2$ in \Equation{Eq:007}.
The factor $2$ on the left-hand side above corrects the factor $1/2$ in the differential volume of $q$.

In the case of auxiliary gluons, one could argue that any possible final-state gluon could take the auxiliary role.
This could be imagined to be facilitated by the jet function, choosing any gluon, but only one at a time, to be allowed to have large $\xiP$ variable.
So if the Born process (on the right-hand side of \Equation{Eq:016}) involves $n_g$ final-state gluons, then on the left-hand side there would be $n_g+1$ contributions.
This would effectively increase the symmetry factor $1/(n_g+1)!$ inside $d\PSTaux_{n+1}$ to $1/n_g!$, so we can just as well assume that one specific gluon on the left-hand side is the auxiliary one, and assume that $d\PSTaux_{n+1}$ has a factor $1/n_g!$ from the start.

In order to introduce the \unfamiliar\ real contribution, let us however start with the simple situation of an auxiliary quark-antiquark pair, and a Born level final state that does not involve any gluons.
As mentioned earlier, the \unfamiliar\ contribution appears because we should allow the radiative gluon to take part in the ``consumption'' of the large $\pP$ component $\Lambda$.
For clarity, we use a separate symbol $r$ instead of $p_{n+1}$ for the momentum of the radiative gluon.
In order to let $r$ contribute to the consumption of $\Lambda$ and the provision of $\kperp$, we add $\xiP_r$ and $\rperp$ as arguments to the auxiliary delta functions compared to \Equation{Eq:016}, and define the {\em \unfamiliar\ real contribution} as:
%
%%%%%%%%%%%%%%%%%%%%%%%%%%%%%%%%%%%%%%%%
\begin{align}
&\dRunf\big(\kin,\kinBar\,;\{p_i\}_{i=1}^n\big)
\notag\\&\hspace{2ex}\overset{\Lambda\to\infty}{=}
    \frac{\cNLO\mu^{2\vepv}}{\piep}\frac{2\xiPin}{\Caux}
   \int d^{4-2\vepv}q\delta(q^2)\,d^{4-2\vepv}r\delta(r^2)
    \,\delta(\xiPin+\xiP_q+\xiP_r-\Lambda)
    \,\delta^{2-2\vepv}(\kperp+\qperp+\rperp)
\notag\\&\hspace{38ex}\times
    \,\frac{d\PSTaux_{n+2}}{dqdr}\big(\Lambda\pP,\kinBar\,;r,q,\{p_i\}_{i=1}^{n}\big)
    \,\JetB\big(\{p_i\}_{i=1}^{n}\big)
~.
\label{Eq:044}
\end{align}
%%%%%%%%%%%%%%%%%%%%%%%%%%%%%%%%%%%%%%%%
%
Now, with a similar reasoning as in the Born case, we can argue that the formula can be used also in case of a Born process with final-state gluons, by considering the auxiliary and/or radiative one as special and including a factor $1/n_g!$ in $d\PSTaux_{n+2}$ from the start.
Additionally, when both are gluons, the auxiliary and the radiative one are distinguished by $\xiP_r<\xiP_q$, as we will see later.

While $\xiP_r$ is allowed to become large in \Equation{Eq:044}, it is also allowed to become arbitrarily small, a phase space region we want to reserve for the \familiar\ contribution.
Thus we need to subtract something from the \unfamiliar\ contribution as defined in \Equation{Eq:044}, which we will address \Section{Sec:031}.

In order to proceed, we write the momenta $q$ and $r$ explicitly in the Sudakov decomposition and denote
%
%%%%%%%%%%%%%%%%%%%%%%%%%%%%%%%%%%%%%%%%
\begin{equation}
\Lambdak = \Lambda-\xiPin
\end{equation}
%%%%%%%%%%%%%%%%%%%%%%%%%%%%%%%%%%%%%%%%
%
so we can scale the variables $\xiP_q=\Lambdak\xP_q$ and $\xiP_r=\Lambdak\xP_r$ to get
%
%%%%%%%%%%%%%%%%%%%%%%%%%%%%%%%%%%%%%%%%
\begin{align}
&\dRunf\big(\kin,\kinBar\,;\{p_i\}_{i=1}^n\big)
\notag\\&\hspace{2ex}\overset{\Lambda\to\infty}{=}
  \frac{\cNLO\mu^{2\vepv}}{\piep}\frac{2\xiPin}{\Caux}
  \int_0^1\frac{d\xP_q}{2\xP_q}\int_0^{\xMin}d\xM_q\int d^{2-2\vepv}\qperp
  \delta\left(\xM_q-\frac{|\qperp|^2}{\nuSq\xP_q\Lambdak}\right)
  \notag\\&\hspace{12ex}\times
  \int_{0}^1\frac{d\xP_r}{2\xP_r}\int_0^{\xMin}d\xM_r\int d^{2-2\vepv}\rperp
  \delta\left(\xM_r-\frac{|\rperp|^2}{\nuSq\xP_r\Lambdak}\right)
\label{Eq:017}
  \\&\hspace{12ex}\times
  \Lambdak^{-1}\delta(\xP_q+\xP_r-1)\,\delta^{2-2\vepv}(\kperp+\qperp+\rperp)
  \notag\\&\hspace{3ex}\times
  \,\frac{d\PSTaux_{n+2}}{dqdr}\big((\Lambdak+\xiPin)\pP,\xMin
                            \,;\,\xP_r\Lambdak\pP+\rperp+\xM_r\pM
                            \,,\,\xP_q\Lambdak\pP+\qperp+\xM_q\pM
                            \,,\{p_i\}_{i=1}^{n}\big)
    \,\JetB\big(\{p_i\}_{i=1}^{n}\big) 
\notag~.
\end{align}
%%%%%%%%%%%%%%%%%%%%%%%%%%%%%%%%%%%%%%%%
%
Inside $d\PSTaux_{n+2}$ there is now a tree-level \matrixelement\ with three partons with large coefficients for $\pP$.
Realize that $\xM_q$ and $\xM_r$ are suppressed with $\Lambda$ due to the on-shell conditions.
We calculate this triple-$\Lambda$ limit in \Appendix{App:050} and find to be 
%
%%%%%%%%%%%%%%%%%%%%%%%%%%%%%%%%%%%%%%%%
\begin{multline}
\frac{1}{\Caux}
\,\MtreeAux\big((\Lambdak+\xiPin)\pP,\kinBar\,;\xP_r\Lambdak\pP+\rperp+\xM_r\pM
                   \,,\,\xP_q\Lambdak\pP+\qperp+\xM_q\pM
                   \,,\,\{p_i\}_{i=1}^n\big)
\\
\overset{\Lambdak\to\infty}{\longrightarrow}\;
\Paux(\xP_q,\qperp,\xP_r,\rperp)
\,\frac{\Lambdak^2\MtreeStar\big(\xiPin\pP-\qperp-\rperp,\kinBar\,;\{p_i\}_{i=1}^n\big)}{\xiPin^2|\qperp+\rperp|^2}
~,
\end{multline}
%%%%%%%%%%%%%%%%%%%%%%%%%%%%%%%%%%%%%%%%
%
with
%
%%%%%%%%%%%%%%%%%%%%%%%%%%%%%%%%%%%%%%%%
\begin{multline}
\Paux(\xP_q,\qperp,\xP_r,\rperp)
= 
 \xP_q\xP_r\,\EuScript{P}_{\aux}(\xP_q,\xP_r)\,|\qperp+\rperp|^2
\\
\times\left[
  \frac{\Cqbar}{|\qperp|^2|\rperp|^2}
+ \frac{1}{\xP_r|\qperp|^2+\xP_q|\rperp|^2-\xP_q\xP_r|\qperp+\rperp|^2}
  \bigg(\frac{\Cr\,\xP_r^2}{|\rperp|^2}+\frac{\Cq\,\xP_q^2}{|\qperp|^2}\bigg)
\right]
~.
\label{Eq:018}
\end{multline}
%%%%%%%%%%%%%%%%%%%%%%%%%%%
%
The auxiliary parton dependent quantities are
%
%%%%%%%%%%%%%%%%%%%%%%%%%%%%%%%%%%%%%%%%
\begin{align}
\textrm{auxiliary quarks:}&\quad
\EuScript{P}_{\aux}(\xP_q,\xP_r) = \frac{1+\xP_q^2-\vepv\xP_r^2}{\xP_r}
\quad,\quad \Cq=\Cqbar=\Nc\;\;,\;\;\Cr=\frac{-1}{\Nc}
\label{Eq:089}
~,\\
\textrm{auxiliary gluons:}&\quad
\EuScript{P}_{\aux}(\xP_q,\xP_r) = \frac{1+\xP_q^4+\xP_r^4}{\xP_q\xP_r}
\quad,\quad \Cq=\Cqbar=\Cr=\Nc
\label{Eq:090}
~,\\
\textrm{gluon to quarks:}&\quad
\EuScript{P}_{\aux}(\xP_q,\xP_r) = 1 - \frac{2\xP_q\xP_r}{1-\vepv}
\quad,\quad \Cq=\Cr=\frac{1}{2}\;\;,\;\;\Cqbar=\frac{-1}{2\Nc^2}
~.
\end{align}
%%%%%%%%%%%%%%%%%%%%%%%%%%%%%%%%%%%%%%%%
%
The ``gluon to quarks'' refers to the real process with an initial-state auxiliary gluon and a quark-antiquark pair in the final state. 
We keep the labels $q$ and $r$ for the final-state partons in this case. 

In the following, we prefer to work with a parameter for dimensional regularization that is imagined to be positive, and we introduce
%
%%%%%%%%%%%%%%%%%%%%%%%%%%%%%%%%%%%%%%%%
\begin{equation}
\vep = -2\vepv
~.
\end{equation}
%%%%%%%%%%%%%%%%%%%%%%%%%%%%%%%%%%%%%%%%
%
Then, at this stage, we find (remember the flux factor)
%
%%%%%%%%%%%%%%%%%%%%%%%%%%%%%%%%%%%%%%%%
\begin{align}
&\dRunf\big(\kin,\kinBar\,;\{p_i\}_{i=1}^n\big)
  \notag\\&\hspace{2ex}\overset{\Lambda\to\infty}{\to}
  \frac{2\cNLO}{\piep\mu^\vep}
  \int_0^1\frac{d\xP_q}{2\xP_q}\int_0^{\xMin}d\xM_q\int d^{2+\vep}\qperp
  \delta\left(\xM_q-\frac{|\qperp|^2}{\nuSq\xP_q\Lambdak}\right)
  \notag\\&\hspace{8ex}\times
  \int_{0}^1\frac{d\xP_r}{2\xP_r}\int_0^{\xMin}d\xM_r\int d^{2+\vep}\rperp
  \delta\left(\xM_r-\frac{|\rperp|^2}{\nuSq\xP_r\Lambdak}\right)
%  \notag\\&\hspace{16ex}\times
  \Lambdak^{-1}\delta(\xP_q+\xP_r-1)\,\delta^{2+\vep}(\kperp+\qperp+\rperp)
  \notag\\&\hspace{16ex}\times
   \Paux(\xP_q,\qperp,\xP_r,\rperp)
  \,\frac{\xiPin^2}{\Lambdak}
  \,\frac{\Lambdak^2d\PSTstr_{n}\big(\xiPin\pP+\kperp,\xMin;\{p_i\}_{i=1}^n\big)}
         {\xiPin^2|\kperp|^2}
    \,\JetB\big(\{p_i\}_{i=1}^{n}\big) 
~.
\end{align}
%%%%%%%%%%%%%%%%%%%%%%%%%%%%%%%%%%%%%%%%
%%
%So finally, we find \Equation{Eq:008} with
%%
%%%%%%%%%%%%%%%%%%%%%%%%%%%%%%%%%%%%%%%%%
%\begin{multline}
%\Freal_{\aux}(\xiPin,\kperp,\mu,\Lambda,\xMin)
%= 
%  \frac{2\cNLO}{\piep\mu^\vep}
%  \int_0^1\frac{d\xP_q}{2\xP_q}\int_0^{\xMmax}\!\!d\xM_q\int d^{2+\vep}\qperp
%  \delta\left(\xM_q-\frac{|\qperp|^2}{\nuSq\xP_q\Lambdak}\right)
%\\\times
%  \int_0^1\frac{d\xP_r}{2\xP_r}\int_0^{\xMin}\!\!d\xM_r\int d^{2+\vep}\rperp
%  \delta\left(\xM_r-\frac{|\rperp|^2}{\nuSq\xP_r\Lambdak}\right)
%\\\times
%  \delta^{2+\vep}(\kperp+\qperp+\rperp)
%  \,\delta(\xP_q+\xP_r-1)\,\Paux(\xP_q,\qperp,\xP_r,\rperp)
%~.
%\end{multline}
%%%%%%%%%%%%%%%%%%%%%%%%%%%%%%%%%%%%%%%%%
%
After performing the $\xM_q,\xM_r$ integrals, we find \Equation{Eq:008} with
%%%%%%%%%%%%%%%%%%%%%%%%%%%%%%%%%%%%%%%%
\begin{align}
\Freal_{\aux}(\xiPin,\kperp,\mu,\Lambda,\xMin)
&=
  \frac{2\cNLO}{\piep\mu^\vep}
  \int_0^1\frac{d\xP_q}{2\xP_q}\int d^{2+\vep}\qperp
  \,\theta\big(0<|\qperp|^2<\nuSq\xMin\xP_q\Lambdak\big)
\notag\\&\hspace{4ex}\times
  \int_{0}^1\frac{d\xP_r}{2\xP_r}\int d^{2+\vep}\rperp
  \,\theta\big(0<|\rperp|^2<\nuSq\xMin\xP_r\Lambdak\big)
\label{Eq:080}
\\&\hspace{8ex}\times
  \delta^{2+\vep}(\kperp+\qperp+\rperp)
  \,\delta(\xP_q+\xP_r-1)\,\Paux(\xP_q,\qperp,\xP_r,\rperp)
~.
\notag
\end{align}
%%%%%%%%%%%%%%%%%%%%%%%%%%%%%%%%%%%%%%%%
%
Understanding that $\xP_q+\xP_r=1$ and $\kperp+\qperp+\rperp=0$, we can write
%
%%%%%%%%%%%%%%%%%%%%%%%%%%%%%%%%%%%%%%%%
\begin{multline}
\Paux(\xP_q,\qperp,\xP_r,\rperp)
= 
 \xP_q\xP_r\EuScript{P}_{\aux}(\xP_q,\xP_r)\,|\kperp|^2\bigg(
  \frac{\Cqbar}{|\rperp|^2|\rperp+\kperp|^2}
+ \frac{\Cq\,(1-\xP_r)^2}{|\rperp+\kperp|^2|\rperp+\xP_r\kperp|^2}
\\
+ \frac{\Cr\,\xP_r^2}{|\rperp|^2|\rperp+\xP_r\kperp|^2}
\bigg)
~,
\end{multline}
%%%%%%%%%%%%%%%%%%%%%%%%%%%%%%%%%%%%%%%%
%
and we see that the explicit $1/\xP_q$ and $1/\xP_r$ in the integral cancel, and that the only possible singularities with respect to $\xP_q,\xP_r$ are in $\EuScript{P}_{\aux}(\xP_q,\xP_r)$.
In case of auxiliary quarks, there is only a singularity with respect to $\xP_r$, and in case of auxiliary gluons, we demand that $\xP_r<\xP_q$.
Remember that we distinguish between the gluons and must not include a factor $2$.
Because of this, and the fact that the $\qperp,\rperp$ integrals are UV finite, we can safely take $\Lambda\to\infty$ inside the $\theta\big(0<|\qperp|^2<\nuSq\xMin\xP_q\Lambdak\big)$, and eliminate the $\qperp$ integral and the $\xP_q$ integral.
Thus we find
%
%%%%%%%%%%%%%%%%%%%%%%%%%%%%%%%%%%%%%%%%
\begin{multline}
\Freal_{\aux}
= 
  \frac{\cNLO}{2\piep\mu^\vep}
  \int_{0}^{\upper}dx
  \,\EuScript{P}_{\aux}(1-x,x)
  \int d^{2+\vep}\rperp\,\theta\big(0<|\rperp|^2<\nuSq x\xMin\Lambdak\big)
\\\times
|\kperp|^2
\bigg(
  \frac{\Cqbar}{|\rperp|^2|\rperp+\kperp|^2}
+ \frac{\Cq\,(1-x)^2}{|\rperp+\kperp|^2|\rperp+x\kperp|^2}
+ \frac{\Cr\,x^2}{|\rperp|^2|\rperp+x\kperp|^2}
\bigg)
~.
\label{Eq:019}
\end{multline}
%%%%%%%%%%%%%%%%%%%%%%%%%%%%%%%%%%%%%%%%
%
In order for the formulas to accomodate for the phase space partition for the auxiliary gluon case, we write an upper integration limit ``$\upper$'' for the $\xP_r$-integration, with
%
%%%%%%%%%%%%%%%%%%%%%%%%%%%%%%%%%%%%%%%%
\begin{equation}
\textrm{$\upper=1$ for the auxiliary quark case, and $\upper=1/2$ for the auxiliary-gluon case.}
\end{equation}
%%%%%%%%%%%%%%%%%%%%%%%%%%%%%%%%%%%%%%%%
%
Now let us take out the pole in $\EuScript{P}_{\aux}$ explicitly, and write
%
%%%%%%%%%%%%%%%%%%%%%%%%%%%%%%%%%%%%%%%%
\begin{equation}
\EuScript{P}_{\aux}(1-x,x)
= \frac{\EuScript{P}_{\aux}^{\pole}}{x} + \EuScript{P}_{\aux}^{\rest}(x)
~.
\label{Eq:045}
\end{equation}
%%%%%%%%%%%%%%%%%%%%%%%%%%%%%%%%%%%%%%%%
%
For the terms with $\EuScript{P}_{\aux}^{\rest}$, we can safely take $\Lambda\to\infty$ also in $\theta\big(0<|\rperp|^2<\nuSq x\xMin\Lambdak\big)$.
Then we can use \Equation{Eq:052}, and find that the part with $\EuScript{P}_{\aux}^{\rest}$ is given by
%
%%%%%%%%%%%%%%%%%%%%%%%%%%%%%%%%%%%%%%%%
\begin{align}
\Freal_{\aux}^{\rest} =
\frac{2\cNLO}{\vep}\bigg(\frac{|\kperp|}{\mu}\bigg)^\vep
\int_{0}^{\upper}dx\,\EuScript{P}_{\aux}^{\rest}(x)
\big(\Cqbar + \Cq(1-x)^\vep + \Cr x^\vep\big)
~,
\label{Eq:020}
\end{align}
%%%%%%%%%%%%%%%%%%%%%%%%%%%%%%%%%%%%%%%%
%
which we will calculate later.
Regarding the terms with $\EuScript{P}_{\aux}^{\pole}/x$ in \Equation{Eq:019}, we first of all give them seperate symbols, as
%
%%%%%%%%%%%%%%%%%%%%%%%%%%%%%%%%%%%%%%%%
\begin{equation}
\Freal_{\aux}^{\pole}
=
\Freal_{\aux}^{\pole,\qB} + \Freal_{\aux}^{\pole,\qL} + \Freal_{\aux}^{\pole,\rL}
~,
\end{equation}
%%%%%%%%%%%%%%%%%%%%%%%%%%%%%%%%%%%%%%%%
and we observe that for the last one, with $\Cr$, we can scale $\rperp$ with $x$ to get
%
%%%%%%%%%%%%%%%%%%%%%%%%%%%%%%%%%%%%%%%%
\begin{equation}
\Freal_{\aux}^{\pole,\rL} =
  \frac{\cNLO}{2\piep\mu^\vep}\,\Cr
  \int_{0}^{\upper}dx
  \,\frac{\EuScript{P}_{\aux}^{\pole}}{x}\,x^\vep
  \int d^{2+\vep}\rperp\,\theta\big(0<x|\rperp|^2<\nuSq\xMin\Lambdak\big)\,
\frac{|\kperp|^2}{|\rperp|^2|\rperp+\kperp|^2}
~.
\end{equation}
%%%%%%%%%%%%%%%%%%%%%%%%%%%%%%%%%%%%%%%%
%
We see that the singularity in $x$ is protected by $\vep$, and that $x\to0$ does not conflict with $\Lambda\to\infty$ in the integration boundary, so also here we can take the latter limit, and find
%
%%%%%%%%%%%%%%%%%%%%%%%%%%%%%%%%%%%%%%%%
\begin{align}
\Freal_{\aux}^{\pole,\rL} =
\frac{2\cNLO}{\vep}\bigg(\frac{|\kperp|}{\mu}\bigg)^\vep
\,\Cr\,\frac{\EuScript{P}_{\aux}^{\pole}}{\vep}\,\upper^\vep
~.
\end{align}
%%%%%%%%%%%%%%%%%%%%%%%%%%%%%%%%%%%%%%%%
%
The other terms in, we reorganize as
%
%%%%%%%%%%%%%%%%%%%%%%%%%%%%%%%%%%%%%%%%
\begin{equation}
\Freal_{\aux}^{\pole,\qB} +\Freal_{\aux}^{\pole,\qL}
 = \Freal_{\aux}^{\pole,0} + \Freal_{\aux}^{\pole,1}
~,
\end{equation}
%%%%%%%%%%%%%%%%%%%%%%%%%%%%%%%%%%%%%%%%
%
where
%
%%%%%%%%%%%%%%%%%%%%%%%%%%%%%%%%%%%%%%%%
\begin{equation}
\Freal_{\aux}^{\pole,0}
= \frac{\cNLO}{2\piep\mu^\vep}
  \,|\kperp|^2\,\big[\Cqbar+\Cq\big]
  \int_{0}^{\upper}dx
  \,\frac{\EuScript{P}_{\aux}^{\pole}}{x}
  \int d^{2+\vep}\rperp\,\frac{\theta\big(0<|\rperp|^2<\nuSq x\xMin\Lambdak\big)}{|\rperp|^2|\rperp+\kperp|^2}
\label{Eq:066}
~,
\end{equation}
%%%%%%%%%%%%%%%%%%%%%%%%%%%%%%%%%%%%%%%%
%
and
%
%%%%%%%%%%%%%%%%%%%%%%%%%%%%%%%%%%%%%%%%
\begin{multline}
\Freal_{\aux}^{\pole,1}
= \frac{\cNLO}{2\piep\mu^\vep}
  \,|\kperp|^2\,\Cq
  \int_{0}^{\upper}dx
  \,\frac{\EuScript{P}_{\aux}^{\pole}}{x}
  \int d^{2+\vep}\rperp\,\theta\big(0<|\rperp|^2<\nuSq x\xMin\Lambdak\big)
\\\times
\bigg(
  \frac{(1-x)^2}{|\rperp+\kperp|^2|\rperp+x\kperp|^2}
- \frac{1}{|\rperp|^2|\rperp+\kperp|^2}
\bigg)
~.
\end{multline}
%%%%%%%%%%%%%%
%
We see that the integrand of $\Freal_{\aux}^{\pole,1}$ does not have a singularity in $x$, and we can take $\Lambda\to\infty$ in the integration boundary.
Thus we find
%
%%%%%%%%%%%%%%%%%%%%%%%%%%%%%%%%%%%%%%%%
\begin{align}
\Freal_{\aux}^{\pole,1}
&= 
  \frac{2\cNLO}{\vep}\bigg(\frac{|\kperp|}{\mu}\bigg)^\vep
  \,\Cq\,
  \int_{0}^{\upper}dx
  \,\frac{\EuScript{P}_{\aux}^{\pole}}{x}\big[(1-x)^\vep - 1\big]
\notag\\&\hspace{0ex}=
  2\cNLO\bigg(\frac{|\kperp|}{\mu}\bigg)^\vep
  \,\Cq\,\EuScript{P}_{\aux}^{\pole}\,
  \big[-\mathrm{Li}_2(\upper)+\Ord(\vep)\big]
~.
\end{align}
%%%%%%%%%%%%%%%%%%%%%%%%%%%%%%%%%%%%%%%%
%
For $\Freal_{\aux}^{\pole,0}$ finally, we use \Equation{Eq:054} to find
%
%%%%%%%%%%%%%%%%%%%%%%%%%%%%%%%%%%%%%%%%
\begin{equation}
\Freal_{\aux}^{\pole,0}
  =
    2\cNLO\bigg(\frac{|\kperp|}{\mu}\bigg)^\vep
    \big(\Cqbar+\Cq\big)\,\EuScript{P}_{\aux}^{\pole}
    \bigg[\frac{1}{\vep^2}
    -\frac{1}{\vep}\ln\frac{|\kperp|^2}{\upper \nuSq\xMin\Lambdak}
    +\frac{1}{4}\mathrm{Li}_2\bigg(\frac{|\kperp|^2}{\upper \nuSq\xMin\Lambdak}\bigg)
    \bigg]
\label{Eq:068}
~.
\end{equation}
%%%%%%%%%%%%%%%%%%%%%%%%%%%%%%%%%%%%%%%%
%
It turns out that for the pole contribution really
%
%%%%%%%%%%%%%%%%%%%%%%%%%%%%%%%%%%%%%%%%
\begin{equation}
\Cqbar=\Cq=\Nc
\quad\textrm{and}\quad
\EuScript{P}_{\aux}^{\pole}=2
\end{equation}
%%%%%%%%%%%%%%%%%%%%%%%%%%%%%%%%%%%%%%%%
%
both for auxiliary quarks and for auxiliary gluons.
We would like to extract the remaining auxiliary parton dependence from $\Freal_{\aux}^{\pole,0}$ caused by the dependence on $\upper$, and we can write
%
%%%%%%%%%%%%%%%%%%%%%%%%%%%%%%%%%%%%%%%%
\begin{equation}
\Freal_{\aux}^{\pole,0}
=\Freal_{\univ}^{\pole}
   +8\cNLO\Nc\bigg(\frac{|\kperp|}{\mu}\bigg)^\vep\bigg[
   \frac{1}{\vep}\ln\upper 
    +\frac{1}{4}\mathrm{Li}_2\bigg(\frac{|\kperp|^2}{\upper \nuSq\xMin\Lambdak}\bigg)
    -\frac{1}{4}\mathrm{Li}_2\bigg(\frac{|\kperp|^2}{\nuSq\xMin\Lambdak}\bigg)
   \bigg]
\end{equation}
%%%%%%%%%%%%%%%%%%%%%%%%%%%%%%%%%%%%%%%%
%
with
%
%%%%%%%%%%%%%%%%%%%%%%%%%%%%%%%%%%%%%%%%
\begin{align}
\Freal_{\univ}^{\pole}
  &=
    8\cNLO\Nc\bigg(\frac{|\kperp|}{\mu}\bigg)^\vep
    \bigg[\frac{1}{\vep^2}
    -\frac{1}{\vep}\ln\frac{|\kperp|^2}{\nuSq\xMin\Lambdak}
    +\frac{1}{4}\mathrm{Li}_2\bigg(\frac{|\kperp|^2}{\nuSq\xMin\Lambdak}\bigg)
    \bigg]
\label{Eq:081}
\\&=
\frac{2\cNLO\Nc}{\piep\mu^\vep}
  \,|\kperp|^2
  \int_{0}^{1}\frac{dx}{x}
  \int d^{2+\vep}\rperp\,\frac{1}{|\rperp|^2|\rperp+\kperp|^2}
            \,\theta\big(0<|\rperp|^2<\nuSq x\xMin\Lambdak\big)
\label{Eq:085}
~.
\end{align}
%%%%%%%%%%%%%%%%%%%%%%%%%%%%%%%%%%%%%%%%
%
So summarizing, we eventually find
%
%%%%%%%%%%%%%%%%%%%%%%%%%%%%%%%%%%%%%%%%
\begin{equation}
\Freal_{\aux} = \Freal_{\univ}^{\pole} + \Freal_{\aux}^{\rest} + \Freal_{\aux}^{\pole,\rest}
\end{equation}
where $\Freal_{\univ}^{\pole}$ is given just above, $\Freal_{\aux}^{\rest}$ is given in \Equation{Eq:020} and will be calculated further in the sections below, and
\begin{equation}
\Freal_{\aux}^{\pole,\rest}
= 
    4\cNLO\bigg(\frac{|\kperp|}{\mu}\bigg)^\vep
\bigg\{
    \Nc
    \bigg[
     \frac{2}{\vep}\ln\upper
    -\mathrm{Li}_2(\upper)
%    +\frac{1}{4}\mathrm{Li}_2\bigg(\frac{|\kperp|^2}{\upper \nuSq\xMin\Lambdak}\bigg)
%    -\frac{1}{4}\mathrm{Li}_2\bigg(\frac{|\kperp|^2}{\nuSq\xMin\Lambdak}\bigg)
    \bigg]
    +\Cr\frac{\upper^\vep}{\vep^2}
\bigg\}
~.
\label{Eq:021}
\end{equation}
%%%%%%%%%%%%%%%%%%%%%%%%%%%%%%%%%%%%%%%%
%
where we put to zero the $\mathrm{Li}_2$-functions with arguments that vanish for $\Lambda\to\infty$.

\subsection{\label{Sec:031}Subtraction of double-counted contributions}
The $1/\vep^2$ in $\Freal_{\univ}^{\pole}$ of \Equation{Eq:081} comes from the integration region where the radiative gluon becomes both soft and collinear to $\pP$.
This region is naturally also present in the \familiar\ contribution, and we would like to keep it there, and correct for the double counting here.

First of all, one may argue that the integration limit for the radiative gluon really should be limited to the half plane for which $\xP_r>\xM_r$.
This would change $\xMin\xP_r$ to $\xP_r^2$ in the integration limit in \Equation{Eq:080}.
It is not difficult to see that this would only influence the value of $\Freal_{\univ}^{\pole}$, giving it an overall factor of $1/2$ and replacing the occurance of $\xMin$ with $1$:
%
%%%%%%%%%%%%%%%%%%%%%%%%%%%%%%%%%%%%%%%%
\begin{equation}
\Freal_{\univ}^{\pole}
  \to
    4\cNLO\Nc\bigg(\frac{|\kperp|}{\mu}\bigg)^\vep
    \bigg[\frac{1}{\vep^2}
    -\frac{1}{\vep}\ln\frac{|\kperp|^2}{\nuSq\Lambdak}
    +\frac{1}{4}\mathrm{Li}_2\bigg(\frac{|\kperp|^2}{\nuSq\Lambdak}\bigg)
    \bigg]
\label{Eq:093}
~.
\end{equation}
%%%%%%%%%%%%%%%%%%%%%%%%%%%%%%%%%%%%%%%%
%
This criterion simply divides equally the phase space of the real gluon emission to regions corresponding to the projectile and the target.

In order to remove the $1/\vep^2$ contribution completely, we can follow the prescription of \mycite{Ciafaloni:1998hu} and restrict the integration region further based on a requirement of angular ordering.
It eventually means that $\Freal_{\univ}^{\pole}$ of \Equation{Eq:085} is replaced with
%
%%%%%%%%%%%%%%%%%%%%%%%%%%%%%%%%%%%%%%%%
\begin{align}
\Freal_{\univ}^{\pole}
&\rightarrow
\frac{2\cNLO\Nc}{\piep\mu^{\vep}}
  \,|\kperp|^2
  \int_{0}^{1}\frac{dx}{x}
  \int d^{2+\vep}\rperp\,\frac{1}{|\rperp|^2|\rperp+\kperp|^2}
  \,\theta\bigg(\frac{|\rperp|}{\nu\sqrt{\Lambda}}<x<\frac{|\rperp|}{|\rperp+\kperp|}\bigg)
\notag\\&=
\frac{2\cNLO\Nc}{\piep\mu^{\vep}}
  \,|\kperp|^2
 \int d^{2+\vep}\rperp\,\frac{\theta\big(|\rperp|<\nu\sqrt{\Lambda}\big)}{|\rperp|^2|\rperp+\kperp|^2}
  \,\ln\frac{\nu\sqrt{\Lambda}}{\max\big(|\rperp|,|\rperp+\kperp|\big)}
\notag\\&=
  \frac{4\cNLO\Nc}{\vep}
  \bigg(\frac{|\kperp|}{\mu}\bigg)^{\vep}
  \ln\frac{\nuSq\Lambda}{|\kperp|^2}
  + \Ord(\vep)
\label{Eq:086}
~.
\end{align}
%%%%%%%%%%%%%%%%%%%%%%%%%%%%%%%%%%%%%%%%
%
Notice that the $\theta$-function forbids the radiative gluon to become collinear to the auxiliary parton, and only allows it to become soft.
When the radiative gluon becomes collinear to the auxiliary parton, then the \familiar\ and \unfamiliar\ pictures of \Figure{Fig:01} look identical, and this is exactly the situation that is sensitive to double counting.

The $\theta$-function in the above equation assigns the phase space  which is attributed to the real part of the $k_T$-dependent evolution step to the \familiar\ real contribution.
In a natural way, the remaining part of the phase-space goes into the \unfamiliar\ real contribution  --- which we identify with the NLO auxiliary parton impact factor.
Removing the ``double counting'' is exactly equivalent to the splitting the full phase space into these two complementing regions.
Hence, the argument of the $\theta$-function specifies the scheme in which the auxiliary impact factor and the evolution kernel are treated.
This separation criterion may still be varied, provided that the soft-collinear pole vanishes from the \unfamiliar\ part, which corresponds to a residual freedom of choice of the energy scale, as pointed out in \cite{Ciafaloni:1998kx,Ciafaloni:1998hu}.
%

%%%%%%%%%%%%%%%%%%%%%%%%%%%%%%%%%%%%%%%%%%%%%%%%%%%%%%%%%%%%%%%%%%%%%%%%%%%%%%%%%%%%
%%%%%%%%%%%%%%%%%%%%%%%%%%%%%%%%%%%%%%%%%%%%%%%%%%%%%%%%%%%%%%%%%%%%%%%%%%%%%%%%%%%%
%%%%%%%%%%%%%%%%%%%%%%%%%%%%%%%%%%%%%%%%%%%%%%%%%%%%%%%%%%%%%%%%%%%%%%%%%%%%%%%%%%%%
%%%%%%%%%%%%%%%%%%%%%%%%%%%%%%%%%%%%%%%%%%%%%%%%%%%%%%%%%%%%%%%%%%%%%%%%%%%%%%%%%%%%
\subsection{Auxiliary quarks}
The collinear splitting function is given by
%
%%%%%%%%%%%%%%%%%%%%%%%%%%%%%%%%%%%%%%%%
\begin{equation}
\EuScript{P}_{qg}(1-x,x) = \frac{1+(1-x)^2+\frac{1}{2}\delta\vep x^2}{x}
= \frac{2}{x} + x - 2 + \srac{1}{2}\delta\vep x
~,
\end{equation}
%%%%%%%%%%%%%%%%%%%%%%%%%%%%%%%%%%%%%%%%
%
with $\delta=1$ (remember that $\vep=-2\vepv$).
So we find, putting $\upper=1$ in \Equation{Eq:020},
%
%%%%%%%%%%%%%%%%%%%%%%%%%%%%%%%%%%%%%%%%
\begin{equation}
\Freal_{\auxq}^{\rest} = 
    \frac{2\cNLO}{\vep}\bigg(\frac{|\kperp|}{\mu}\bigg)^\vep
\bigg\{
 \big[\Cqbar+\Cq+\Cr\big]\bigg(- \frac{3}{2} +\frac{\delta}{4}\,\vep\bigg)
+\Cq\,\frac{5}{4}\,\vep
+\Cr\,\frac{7}{4}\,\vep + \Ord\big(\vep^2\big)\bigg\}
~.
\end{equation}
%%%%%%%%%%%%%%%%%%%%%%%%%%%%%%%%%%%%%%%%
%
Putting the coefficients
%
%%%%%%%%%%%%%%%%%%%%%%%%%%%%%%%%%%%%%%%%
\begin{equation}
\Cqbar = \Cq = \Nc
\quad,\quad
\Cr = -1/\Nc
~,
\end{equation}
%%%%%%%%%%%%%%%%%%%%%%%%%%%%%%%%%%%%%%%%
%
and forgetting about $\Ord\big(\vep^2\big)$, we get
%
%%%%%%%%%%%%%%%%%%%%%%%%%%%%%%%%%%%%%%%%
\begin{equation}
\Freal_{\auxq}^{\rest} = 
    2\cNLO\Nc\bigg(\frac{|\kperp|}{\mu}\bigg)^\vep
\bigg\{
 - \frac{3}{\vep} + \frac{5+2\delta}{4}
-\frac{1}{\Nc^2}\,\bigg[- \frac{3}{2\vep} + \frac{7+\delta}{4}\bigg]\bigg\}
~.
\end{equation}
%%%%%%%%%%%%%%%%%%%%%%%%%%%%%%%%%%%%%%%%
%
Adding $\Freal_{\univ}^{\pole}$ and $\Freal_{\aux}^{\pole,\rest}$ from \Equation{Eq:021} with $\upper=1$, we get
%
%%%%%%%%%%%%%%%%%%%%%%%%%%%%%%%%%%%%%%%%
\begin{equation}
\Freal_{\auxq} = \Freal_{\univ}^{\pole} +
    2\cNLO\Nc\bigg(\frac{|\kperp|}{\mu}\bigg)^\vep
\bigg\{
 {}-{}\frac{3}{\vep}
 -\frac{\pi^2}{3} + \frac{5+2\delta}{4}
-\frac{1}{\Nc^2}\,\bigg[\frac{2}{\vep^2}- \frac{3}{2\vep} + \frac{7+\delta}{4}\bigg]
\bigg\}
~,
\label{Eq:022}
\end{equation}
%%%%%%%%%%%%%%%%%%%%%%%%%%%%%%%%%%%%%%%%
%
and putting $\vep=-2\vepv$ and $\delta=1$, we find \Equation{Eq:009}.
%

%%%%%%%%%%%%%%%%%%%%%%%%%%%%%%%%%%%%%%%%%%%%%%%%%%%%%%%%%%%%%%%%%%%%%%%%%%%%%%%%%%%%
%%%%%%%%%%%%%%%%%%%%%%%%%%%%%%%%%%%%%%%%%%%%%%%%%%%%%%%%%%%%%%%%%%%%%%%%%%%%%%%%%%%%
\subsection{Auxiliary gluons}
Now the collinear splitting function is given by
%
%%%%%%%%%%%%%%%%%%%%%%%%%%%%%%%%%%%%%%%%
\begin{equation}
\EuScript{P}_{gg}(1-x,x) = \frac{1+x^4+(1-x)^4}{x(1-x)}
= \frac{2}{x} + 2\left(\frac{1}{1-x} - 2 + x - x^2\right)
~.
\end{equation}
%%%%%%%%%%%%%%%%%%%%%%%%%%%%%%%%%%%%%%%%
%
Putting  $\upper=1/2$ in \Equation{Eq:020}, we find
%
%%%%%%%%%%%%%%%%%%%%%%%%%%%%%%%%%%%%%%%%
\begin{multline}
\Freal_{\auxg}^{\rest} = 
    \frac{2\cNLO}{\vep}\bigg(\frac{|\kperp|}{\mu}\bigg)^\vep
 \bigg\{
 \big[\Cqbar+\Cq+\Cr\big]\bigg(2\ln2-\frac{11}{6}\bigg)
\\
+\Cq\bigg(\frac{131}{72}-\frac{132}{72}\ln2-\ln^22\bigg)\vep
+\Cr\bigg(\frac{137}{72}+\frac{132}{72}\ln2-\ln^22-\frac{\pi^2}{6}\bigg)\vep + \Ord\big(\vep^2\big)\bigg\}
~.
\end{multline}
%%%%%%%%%%%%%%%%%%%%%%%%%%%%%%%%%%%%%%%%
%
Setting $\Cqbar=\Cq=\Cr=\Nc$ and forgetting about the $\Ord\big(\vep^2\big)$, we get
%
%%%%%%%%%%%%%%%%%%%%%%%%%%%%%%%%%%%%%%%%
\begin{equation}
\Freal_{\auxg}^{\rest} = 
    2\cNLO\Nc\bigg(\frac{|\kperp|}{\mu}\bigg)^\vep
\bigg\{
\frac{3}{\vep}\bigg(2\ln2-\frac{11}{6}\bigg)
+ \frac{67}{18} - \frac{\pi^2}{6} - 2\ln^22
\bigg\}
~.
\end{equation}
%%%%%%%%%%%%%%%%%%%%%%%%%%%%%%%%%%%%%%%%
%
Adding $\Freal_{\univ}^{\pole}$ and $\Freal_{\aux}^{\pole,\rest}$ from \Equation{Eq:021} with $\upper=1/2$ we get
%
%
%%%%%%%%%%%%%%%%%%%%%%%%%%%%%%%%%%%%%%%%
\begin{align}
\Freal_{\auxg} 
&= 
%     2\cNLO\Nc\bigg(\frac{|\kperp|}{\mu}\bigg)^\vep
% \bigg\{
% \frac{2^{1-\vep}}{\vep^2} 
% + \frac{1}{\vep}\bigg[6\ln2-\frac{11}{2}-4\ln\frac{2\xPlow}{\Lambda}\bigg]
%     -\mathrm{Li}_2\bigg(\frac{|\kperp|^2}{\nuSq\xMin\xPlow}\bigg)
% \notag\\&\hspace{28ex}
%  + \frac{67}{18} - \frac{\pi^2}{6} - 2\ln^22
%  - \frac{\pi^2}{6} + \ln^22
% \bigg\}
% \notag\\&=
  \Freal_{\univ}^{\pole}
+
    2\cNLO\Nc\bigg(\frac{|\kperp|}{\mu}\bigg)^\vep
\bigg\{
\frac{2}{\vep^2} 
- \frac{1}{\vep}\,\frac{11}{2}
 - \frac{\pi^2}{3}+ \frac{67}{18}
\bigg\}
+\Ord(\vep)
~,
\end{align}
%%%%%%%%%%%%%%%%%%%%%%%%%%%%%%%%%%%%%%%%
%
and putting $\vep=-2\vepv$, we get
%
%%%%%%%%%%%%%%%%%%%%%%%%%%%%%%%%%%%%%%%%
\begin{equation}
\Freal_{\auxg} = 
  \Freal_{\univ}^{\pole}
+
    \cNLO\Nc\bigg(\frac{\mu^2}{|\kperp|^2}\bigg)^{\vepv}
\bigg\{
\frac{1}{\vepv^2} 
+ \frac{1}{\vepv}\,\frac{11}{2}
%\notag\\&\hspace{24ex}
 - \frac{2\pi^2}{3}+ \frac{67}{9}
\bigg\}
+\Ord(\vepv)
~.
\label{Eq:023}
\end{equation}
%%%%%%%%%%%%%%%%%%%%%%%%%%%%%%%%%%%%%%%%
%

%%%%%%%%%%%%%%%%%%%%%%%%%%%%%%%%%%%%%%%%%%%%%%%%%%%%%%%%%%%%%%%%%%%%%%%%%%%%%%%%%%%%
%%%%%%%%%%%%%%%%%%%%%%%%%%%%%%%%%%%%%%%%%%%%%%%%%%%%%%%%%%%%%%%%%%%%%%%%%%%%%%%%%%%%
\subsection{Initial-state gluon and final-state quark-antiquark pair}
Now the collinear splitting function is given by
%
%%%%%%%%%%%%%%%%%%%%%%%%%%%%%%%%%%%%%%%%
\begin{equation}
\EuScript{P}_{gg}(1-x,x) = 1 - \frac{2x(1-x)}{1+\frac{1}{2}\delta\vep}
= 1 -2x + 2x^2 + (x-x^2)\delta\vep + \Ord(\vep^2)
~.
\end{equation}
%%%%%%%%%%%%%%%%%%%%%%%%%%%%%%%%%%%%%%%%
%
Putting  $\upper=1$ in \Equation{Eq:020}, we find
%
%%%%%%%%%%%%%%%%%%%%%%%%%%%%%%%%%%%%%%%%
\begin{equation}
\Freal^{\rest} = 
    \frac{2\cNLO}{\vep}\bigg(\frac{|\kperp|}{\mu}\bigg)^\vep
\bigg\{
 \big[\Cqbar+\Cq+\Cr\big]\bigg(\frac{2}{3}+\frac{\delta}{6}\vep\bigg)
+\big[\Cq+\Cr\big]\frac{-13}{18}\vep
+ \Ord(\vep^2)\bigg\}
~.
\end{equation}
%%%%%%%%%%%%%%%%%%%%%%%%%%%%%%%%%%%%%%%%
%
We keep the labels $\rL,\qL$ for final-state variables, now the quark-antiquark pair, so we can still use \Equation{Eq:018} instead of \Equation{Eq:043}.
Thus, we need to set $\Cqbar=-1/(2\Nc^2)$ and $\Cq=\Cr=1/2$.
Forgetting about $\Ord\big(\vep^2\big)$, we then find
%
%%%%%%%%%%%%%%%%%%%%%%%%%%%%%%%%%%%%%%%%
\begin{equation}
\Freal_{\auxg}^{\qL\qB} =
\Freal^{\rest} = 
    2\cNLO\bigg(\frac{|\kperp|}{\mu}\bigg)^\vep
\bigg\{
 \frac{2}{3\vep}
-\frac{13}{18} + \frac{\delta}{6}
-\frac{1}{\Nc^2}\bigg(\frac{1}{3\vep}+\frac{\delta}{12}\bigg)
\bigg\}
~.
\end{equation}
%%%%%%%%%%%%%%%%%%%%%%%%%%%%%%%%%%%%%%%%
%
Putting $\vep=-2\vepv$, we get
%
%%%%%%%%%%%%%%%%%%%%%%%%%%%%%%%%%%%%%%%%
\begin{equation}
\Freal_{\auxg}^{\qL\qB} = 
    \cNLO\bigg(\frac{\mu^2}{|\kperp|^2}\bigg)^{\vepv}
\bigg\{
-\frac{2}{3\vepv}
-\frac{13}{9} + \frac{\delta}{3}
+\frac{1}{\Nc^2}\bigg(\frac{1}{3\vepv}-\frac{\delta}{6}\bigg)
\bigg\}
~.
\label{Eq:024}
\end{equation}
%%%%%%%%%%%%%%%%%%%%%%%%%%%%%%%%%%%%%%%%

\section{\label{Sec:040}\Unfamiliar\ virtual contribution}
The \unfamiliar\ real contributions have the general form
%
%%%%%%%%%%%%%%%%%%%%%%%%%%%%%%%%%%%%%%%%
\begin{equation}
\dRunf = 
\cNLO\Nc\bigg(\frac{\mu^2}{|\kperp|^2}\bigg)^{\vepv}
\bigg\{
\frac{A}{\vepv^2}
+\frac{B}{\vepv}
+C
\bigg\}\dBstar
~,
\label{Eq:025}
\end{equation}
%%%%%%%%%%%%%%%%%%%%%%%%%%%%%%%%%%%%%%%%
%
where the coefficients $A,B,C$ depend on the auxiliary parton type.
So besides violating parton universality, they also clearly violate the smooth on-shell limit.
We may expect that there are also virtual contributions that violate these tree-level properties and in fact we find that they also violate the smooth limit $\Lambda\to\infty$, while having the general form of \Equation{Eq:025}.
In the following we derive the relevant virtual contributions, which we again call {\em \unfamiliar}.
{\em We conjecture that there are no other $\Lambda$-dependent virtual contributions than the one we identify here}.

Regarding phase space, the virtual contribution is the same as the Born contribution.
Instead of the tree-level \matrixelement, it however contains the interference between the tree-level amplitude and the one-loop amplitude.
The non-smooth on-shell limit suggests to have a closer look at it in one-loop amplitudes, in particular the limit of $|\kperp|\to0$ before $\Lambda\to\infty$.
For, maybe large, but fixed value of $\Lambda$, this limit corresponds to the collinear limit for the auxiliary parton momenta: with the momenta of \Equation{Eq:006}, we have
%
%%%%%%%%%%%%%%%%%%%%%%%%%%%%%%%%%%%%%%%%
\begin{equation}
2\lop{k_1}{k_2} =2\lop{(-\Lambda\pP)}{p_\Lambda} = \frac{-\Lambda}{\Lambda-\xiPin}\,|\kperp|^2 \to0
\quad\textrm{for}\quad |\kperp|\to0
~.
\end{equation}
%%%%%%%%%%%%%%%%%%%%%%%%%%%%%%%%%%%%%%%%
%
We let the symbol $k_1$ refer to a negative energy momentum here to be consistent with the literature in the following.
The collinear limit of one-loop amplitudes has been studied in~\mycite{Bern:1995ix,Bern:1999ry}.
There, color decompositions of amplitudes are employed, and it is noted that collinear limits can be studied on so-called {\em primitive amplitudes}.
These do not explicitly depend on color, and correspond to a fixed ordering of the external legs.

Let us however start with tree-level.
In particular, we will show now that the two limits $\Lambda\to\infty$ and $|\kperp|\to0$ commute.
Following the formalism of \mycite{Bern:1999ry}, we have at tree-level the collinear factorization formula
%
%%%%%%%%%%%%%%%%%%%%%%%%%%%%%%%%%%%%%%%%
\begin{equation}
\Amp_{\Tree}(1^{h_1},2^{h_2},3,\ldots,n)
\overset{1||2}{\longrightarrow}
\sum_{h=\pm}\mathrm{Split}^{\Tree}_{-h}(1^{h_1},2^{h_2})\,
   \Amp_{\Tree}(K^h,3,\ldots,n)
\end{equation}
%%%%%%%%%%%%%%%%%%%%%%%%%%%%%%%%%%%%%%%%
%
where $\Amp_{\Tree}$ is a primitive amplitude (which at tree-level is identical to a so called {\em partial amplitude} and therefore directly associated with a color structure), and the adjacent (incoming) momenta $k_1,k_2$ become collinear with $k_1+k_2\to K$.
%
%The arguments of the function $\mathrm{Split}^{\Tree}_{-h}$ are before the limit.
%
The superscripts $h$ indicate helicities.
The splittings are typically represented as a function of the splitting variable $z$ in
%
%%%%%%%%%%%%%%%%%%%%%%%%%%%%%%%%%%%%%%%%
\begin{equation}
k_1 \to zK \quad,\quad k_2 \to (1-z)K
~.
\end{equation}
%%%%%%%%%%%%%%%%%%%%%%%%%%%%%%%%%%%%%%%%
%
For our situation, we have $K = -\xiPin\pP$ and $z=\Lambda/\xiPin$, so our splitting variable is not bounded by $1$.
Therefore, it is more convenient to use the representation of the splitting functions in \Appendix{App:020} in terms of more general momentum fractions $x$ and $y$ such that
%
%%%%%%%%%%%%%%%%%%%%%%%%%%%%%%%%%%%%%%%%
\begin{equation}
k_1 \to xK \quad,\quad k_2 \to yK
~.
\end{equation}
%%%%%%%%%%%%%%%%%%%%%%%%%%%%%%%%%%%%%%%%
%
Inserting into those formulas
%
%%%%%%%%%%%%%%%%%%%%%%%%%%%%%%%%%%%%%%%%
\begin{equation}
x\leftarrow-\Lambda
\quad\textrm{and}\quad
y\leftarrow(\Lambda-\xiPin)
~,
\end{equation}
%%%%%%%%%%%%%%%%%%%%%%%%%%%%%%%%%%%%%%%%
%
on can easily see that only the ones for which the auxiliary helicities are opposite survive for large $\Lambda$, and that they become $\Lambda/(\xiPin\kapp)$ and $\Lambda/(\xiPin\kstr)$.
The quantities $\kapp,\kstr$ are proportional to $|\kperp|$, see \Appendix{App:010}.
Thus, we find
%
%%%%%%%%%%%%%%%%%%%%%%%%%%%%%%%%%%%%%%%%
\begin{align}
\Amp_{\Tree}\big((-\Lambda\pP)^{h_1},p_\Lambda^{h_2},3,\ldots,n\big)
&\overset{|\kperp|\to0}{\longrightarrow}
 \frac{\Lambda}{\xiPin\kapp}\,\Amp_{\Tree}\big((-\xiPin\pP)^+,3,\ldots,n\big)
\notag\\&\hspace{3ex}
+\frac{\Lambda}{\xiPin\kstr}\,\Amp_{\Tree}\big((-\xiPin\pP)^-,3,\ldots,n\big)
\;+\;\Ord\big(\Lambda^{0}\big)
~,
\end{align}
%%%%%%%%%%%%%%%%%%%%%%%%%%%%%%%%%%%%%%%%
%
which indeed is $\Lambda/(\xiPin|\kperp|)$ times the on-shell limit $|\kperp|\to0$ of the space-like amplitude \mbox{$\Amp_{\Tree}^\star\big({-}\xiPin\pP-\kperp,3,\ldots,n\big)$}, see for example Section~5 in~\mycite{Blanco:2020akb}.

The one-loop factorization formula is a bit more complicated, and following \mycite{Bern:1999ry} it is given by
%
%%%%%%%%%%%%%%%%%%%%%%%%%%%%%%%%%%%%%%%%
\begin{multline}
\Amp_{\Loop}(1^{h_1},2^{h_2},3,\ldots,n)
\overset{1||2}{\longrightarrow}
\sum_{h=\pm}\bigg\{
\mathrm{Split}^{\Tree}_{-h}(1^{h_1},2^{h_2})\,
   \Amp_{\Loop}(K^h,3,\ldots,n)
\\
+\mathrm{Split}^{\Loop}_{-h}(1^{h_1},2^{h_2})\,
   \Amp_{\Tree}(K^h,3,\ldots,n)
\bigg\}
~.
\label{Eq:026}
\end{multline}
%%%%%%%%%%%%%%%%%%%%%%%%%%%%%%%%%%%%%%%%
%
Remember that we are looking for contributions of the type of \Equation{Eq:025}, which are clearly absent in the first term within curly brackets in \Equation{Eq:026}: they are not in the tree-level splitting function, and the loop amplitude is on-shell and independent of $\kperp$.
Thus our interest goes to the second contributions.
The one-loop splitting functions are given by
%
%%%%%%%%%%%%%%%%%%%%%%%%%%%%%%%%%%%%%%%%
\begin{equation}
\mathrm{Split}^{\Loop}_{-h}(1^{h_1},2^{h_2})
= \gQCD^2c_\Gamma\times\mathrm{Split}^{\Tree}_{-h}(1^{h_1},2^{h_2})
 \times \Vcoef(-h,1^{h_1},2^{h_2})
~,
\end{equation}
%%%%%%%%%%%%%%%%%%%%%%%%%%%%%%%%%%%%%%%%
%
with
%
%%%%%%%%%%%%%%%%%%%%%%%%%%%%%%%%%%%%%%%%
\begin{equation}
\gQCD^2 c_\Gamma = \frac{\gQCD^2}{(4\pi)^{2-\vepv}}\,\frac{\Gamma(1+\vepv)\Gamma^2(1-\vepv)}{\Gamma(1-2\vepv)}
=
\frac{\cNLO}{2} + \Ord\big(\vepv^3\big)
~.
\end{equation}
%%%%%%%%%%%%%%%%%%%%%%%%%%%%%%%%%%%%%%%%
%
The opposite-helicity $\Vcoef$ functions can be found in \mycite{Bern:1999ry} to be
%
%%%%%%%%%%%%%%%%%%%%%%%%%%%%%%%%%%%%%%%%
\begin{multline}
\Vcoef_{g\to \bar{q}_1q_2}(h,1^\pm,2^\mp) =
\bigg(\frac{\mu^2}{-s_{12}}\bigg)^{\vepv}
\bigg\{
%  \frac{1}{\vepv^2}\big[ 2-z^{-\vepv} -(1-z)^{-\vepv} \big]
%  + \ln z\ln(1-z)
   \frac{1}{\vepv}\bigg(\ln[z(1-z)]+ \frac{13}{6}\bigg)
  - \frac{1}{2}\ln^2\frac{z}{1-z}
  - \frac{\pi^2}{6} +\frac{83}{18} - \frac{\delta_R}{6}
\\
+\frac{1}{\Nc^2}\bigg[\frac{1}{\vepv^2}+\frac{3}{2}\frac{1}{\vepv}+\frac{7+\delta_R}{2}\bigg]
-\frac{n_f}{\Nc}\bigg[\frac{2}{3}\frac{1}{\vepv} + \frac{10}{9}\bigg]
\bigg\}
+\Ord(\vepv)
\label{Eq:027}
\end{multline}
%%%%%%%%%%%%%%%%%%%%%%%%%%%%%%%%%%%%%%%%
%
with $\delta_R=1$ in the 't~Hooft-Veltman scheme of dimensional regulariztion and $\delta_R=0$ for dimensional reduction, and
%
%%%%%%%%%%%%%%%%%%%%%%%%%%%%%%%%%%%%%%%%
\begin{equation}
\Vcoef_{g\to g_1g_2}(h,1^\pm,2^\mp) =
\bigg(\frac{\mu^2}{-s_{12}}\bigg)^{\vepv}
\bigg\{
%\frac{1}{\vepv^2}\big[ -z^{-\vepv}(1-z)^{-\vepv} \big]
%+ 2\ln z\ln(1-z)
-\frac{1}{\vepv^2} + \frac{1}{\vepv}\ln[z(1-z)] - \frac{1}{2}\ln^2\frac{z}{1-z}
 - \frac{\pi^2}{6} 
\bigg\}
+\Ord(\vepv)
~.
\label{Eq:028}
\end{equation}
%%%%%%%%%%%%%%%%%%%%%%%%%%%%%%%%%%%%%%%%
%
The invariant $s_{12}=(k_1+k_2)^2$ is to be taken before the collinear limit of course.
Since these functions do not actually depend on the values of the helicities, and we saw that at tree-level the collinear limit commutes with the large $\Lambda$ limit, we are tempted to identify the \unfamiliar\ part of the one-loop amplitude as
%
%%%%%%%%%%%%%%%%%%%%%%%%%%%%%%%%%%%%%%%%
\begin{equation}
\Amp_{\Loop}^{\star\,\unf}\big({-}\xiPin\pP-\kperp,3,\ldots,n\big)
=
\frac{\cNLO}{2}\,\Vcoef_{\aux}\,\Amp_{\Tree}^{\star}\big({-}\xiPin\pP-\kperp,3,\ldots,n\big)
\label{Eq:029}
\end{equation}
%%%%%%%%%%%%%%%%%%%%%%%%%%%%%%%%%%%%%%%%
%
where we obtain the expressions for $\Vcoef_{\aux}$ in \Equation{Eq:012} and \Equation{Eq:013} by just inserting $z\leftarrow\Lambda/\xiPin$ into \Equation{Eq:027} and \Equation{Eq:028}.
Realize that $\Lambda/\xiPin>1$ {\em does} generate imaginary parts in the logarithms.

Simply replacing the on-shell tree-level amplitude with the space-like one to arrive at \Equation{Eq:029} seems a bit sketchy and we will give it more solid ground in \Section{Sec:041}.
First however, we want to address color.
In the beginning of Section~4.5 in \mycite{Bern:1999ry} it is pointed out that {\em it is only the leading partial amplitude, $A_{5;1}$ which has the same color structure as the tree-level amplitude, that receives splitting contributions from} $\mathrm{Split}^{\Loop}$.
Therefore, color can be treated the same as at tree-level, and the formula for the \unfamiliar\ contribution can directly be lifted to virtual contribution at cross section level.
This includes a factor $\Nc$ coming with the leading partial amplitudes (see Equation~(2.5-6) and (2.10) in \mycite{Bern:1999ry}), and the factor of $2$ for the virtual contribution being the interference between tree-level and one-loop.
Thus we find
%
%%%%%%%%%%%%%%%%%%%%%%%%%%%%%%%%%%%%%%%%
\begin{equation}
\dVunf = \cNLO\Nc\,\mathrm{Re}\big(\Vcoef_{\aux}\big)
\,\dBstar
~.
\end{equation}
%%%%%%%%%%%%%%%%%%%%%%%%%%%%%%%%%%%%%%%%
%

\subsection{\label{Sec:041}One-loop amplitudes in the auxiliary parton approach}
In this section, we justify using the space-like tree-level amplitude itself instead of its on-shell limit in \Equation{Eq:029}.
This requires a repetition of some technical aspects regarding the calculation of one-loop amplitudes.

The application of the auxiliary parton approach in one-loop amplitudes has been studied in \mycite{vanHameren:2017hxx}, in particular in the context of integrand-based methods to calculate these~\mycite{Ossola:2006us,Giele:2008ve}.
One-loop amplitudes can be decomposed in terms of scalar integrals, and for NLO calculations it suffices to include integrals with at most $4$ propagator denominators, so-called box integrals of the type
%
%%%%%%%%%%%%%%%%%%%%%%%%%%%%%%%%%%%%%%%%
\begin{equation}
\graph{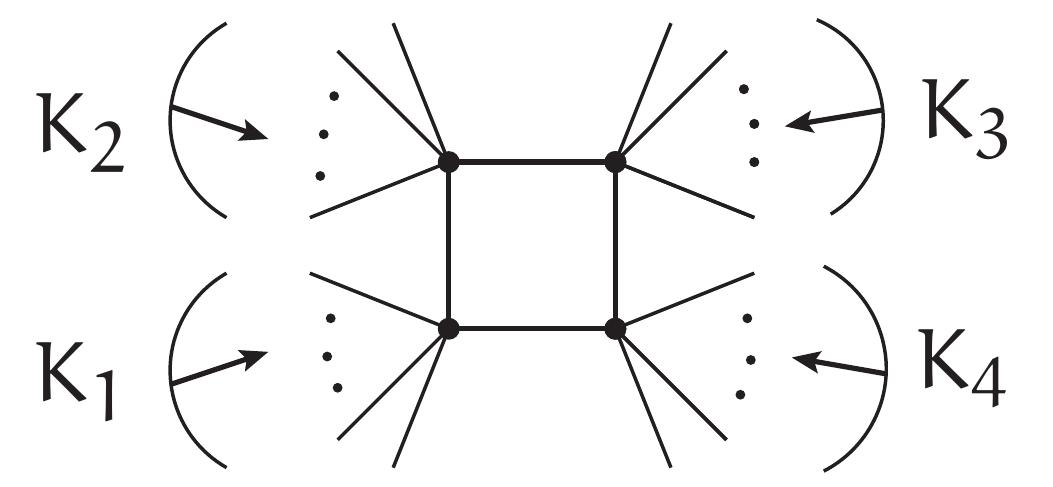}{24}{5}=
\frac{(2\pi\mu)^{2\vepv}}{\imag(2\pi)^{4}c_\Gamma}\int\frac{d^{4-2\vepv}\ell}{\ell^2(\ell+K_1)^2(\ell+K_1+K_2)^2(\ell+K_1+K_2+K_3)^2}
\end{equation}
%%%%%%%%%%%%%%%%%%%%%%%%%%%%%%%%%%%%%%%%
%
where $K_{1,2,3}$ are sums of external momenta, and momentum conservation implies $K_4=-K_1-K_2-K_3$.
Realize that these integrals are invariant under shifts $\ell\to\ell+K$ of the integration momentum.
Fewer denominator factors lead to triangle, bubble, and tadpole integrals.
\begin{figure}
\begin{center}
\epsfig{figure=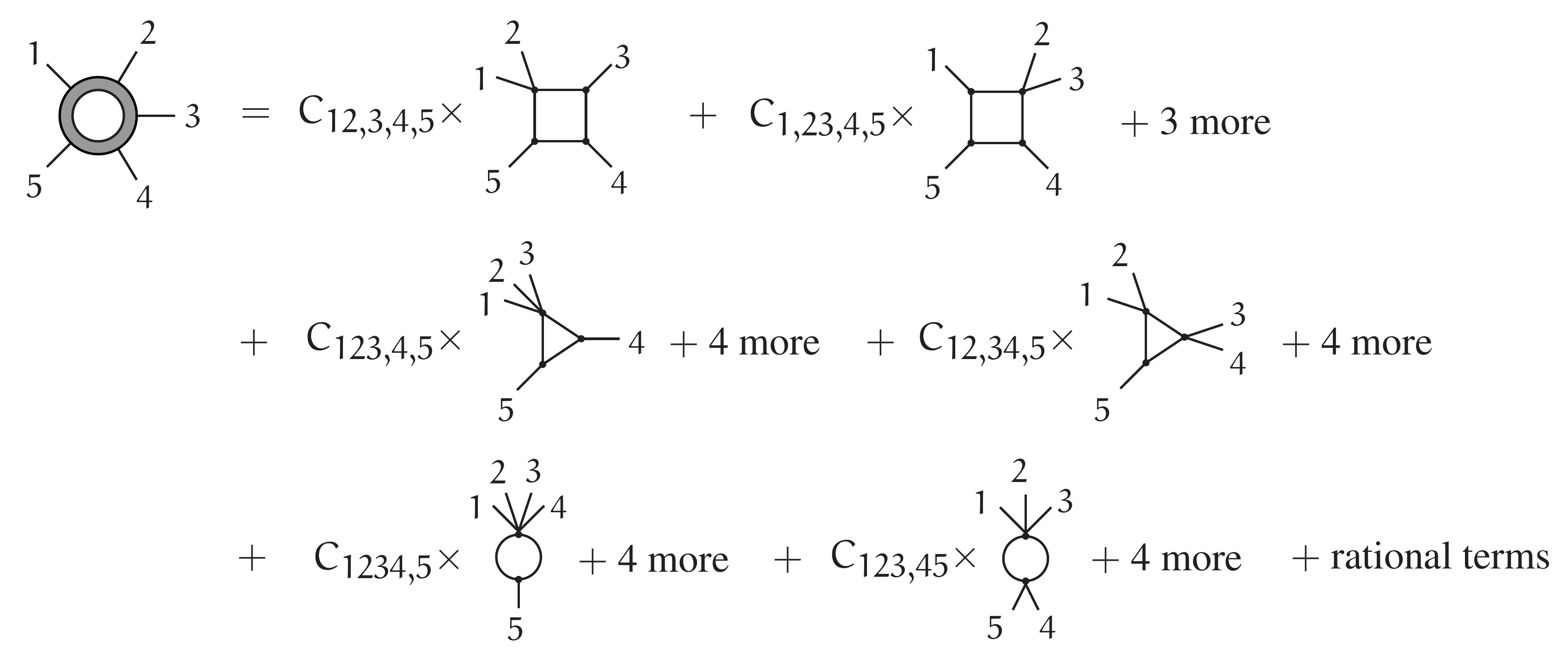,width=\linewidth}
\caption{\label{Fig:03}Example of the decomposition of a planar one-loop amplitude with $5$ external legs into scalar functions times their coefficients. One-point scalar functions are not included, and the decomposition only holds to and including $\Ord(\vepv^0)$. The rational terms are remainders that do not contain logarithms of the invariants of the external momenta, but cannot be captured within the scalar function decomposition. Essentially, to each possible cyclicly ordered partition of $(1,2,3,4,5)$ into $4,3$ or $2$ groups belongs a term.}
\end{center}
\end{figure}
These integrals evaluate, within dimensional regularization, to (poly)loga\-rithms of ratios of invariants from the external momenta, while the coefficients for these integrals are ``tree-level objects'', that is ratios of contractions of momenta and polarization vectors.
The list of possible scalar integrals can simply be deduced from the possible kinematical channels implied by the amplitude, and the integrand-based methods extract the coefficients from the one-loop integrand.
The latter is the function of the loop momentum $\ell$ that needs to be integrated over in order to arrive at the one-loop amplitude:
%
%%%%%%%%%%%%%%%%%%%%%%%%%%%%%%%%%%%%%%%%
\begin{equation}
\graph{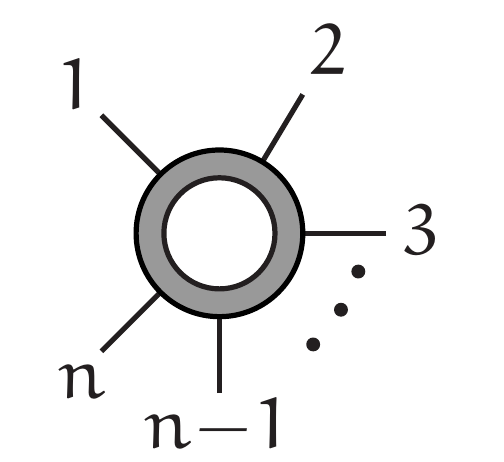}{13}{5}
=
\int d^{4-2\vepv}\ell\,\frac{\EuScript{N}(\ell)}{\ell^2(\ell+p_1)^2(\ell+p_1+p_2)^2\cdots(\ell+p_1+p_2+\cdots+p_{n-1})}
~.
\end{equation}
%%%%%%%%%%%%%%%%%%%%%%%%%%%%%%%%%%%%%%%%
%
We wrote the integrand here as a numerator polynomial $\EuScript{N}(\ell)$ divided by all possible $\ell$-dependend denominator factors for a planar one-loop amplitude.
The integrand-based methods use carefully chosen explicit values of $\ell$ to extract numerical values of the coefficients for the scalar integrals from the integrand.
These methods silently assume that the integrand is calculated in the Feynman gauge in order to avoid other than quadratic denominator factors.

\begin{figure}
\begin{center}
\epsfig{figure=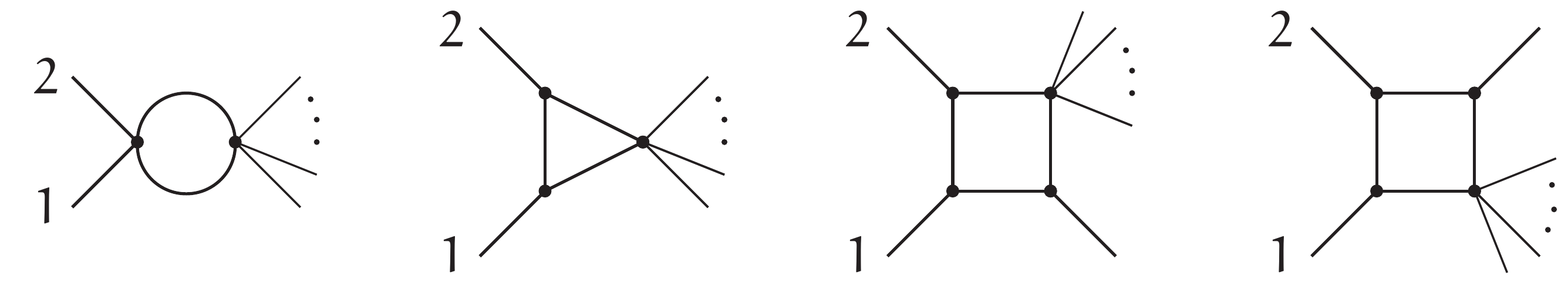,width=68ex}
\caption{\label{Fig:02}Scalar one-loop functions contributing to the limit when $1$ and $2$ become collinear.}
\end{center}
\end{figure}

In \mycite{Bern:1995ix,Bern:1999ry}, $\mathrm{Split}^{\Loop}$ was calculated by realizing that only contributions from scalar integrals of the type in \Figure{Fig:02} need to be taken into account, that is besides the bubble only scalar integrals with the two external lines that are becoming collinear attached to the loop on their own.
The box integrals must have at most one off-shell leg.
Contributions from box integrals with two off-shell legs that satisfy this criterion do not survive the collinear limit.
The fact that the box integrals contribute hints at the main difficulty the authors in the cited work resolved, namely that not only graphs with the factorized form of a one-loop part times a tree-level blob contribute.
For these factorized graphs, the prescription to simply replace the limiting on-shell amplitude by the off-shell amplitude in \Equation{Eq:029} is obviously correct.
In the following we argue that it is also correct for the collection of other graphs.

At first sight, is seems natural to combine the integrand-based methods with the auxiliary parton method by carefully keeping the dependence on $\Lambda$ when evaluating the scalar functions and possibly encounter logarithims of $\Lambda$ in those, while taking the limit $\Lambda\to\infty$ on the integrand, or rather taking the leading power, and extracting the coefficients after this limit.
The integrand will, like the tree-level amplitude, be proportional to $\Lambda$, and this factor is removed by the (square root of the) prescription of \Equation{Eq:007}.
Realize that this integrand, just like tree-level amplitudes, does not depend on the type of auxiliary partons.
In~\mycite{vanHameren:2017hxx} however, it was observed that in particular the triangle integral
%
%%%%%%%%%%%%%%%%%%%%%%%%%%%%%%%%%%%%%%%%
\begin{equation}
\graph{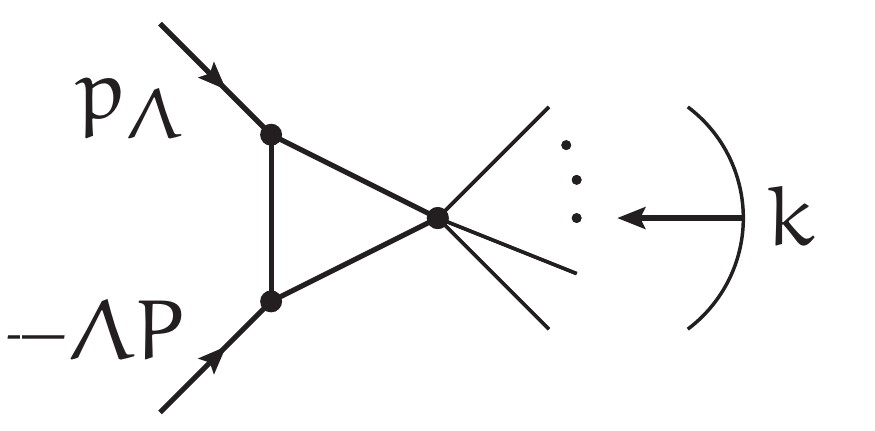}{20}{4}=
\frac{(2\pi\mu)^{2\vepv}}{\imag(2\pi)^{4}c_\Gamma}\int d^{4-2\vepv}\ell\,\frac{\Lambda}{\ell^2(\ell-\Lambda\pP)^2(\ell-k)^2}
= \frac{-\Lambda}{\vepv^2|\kperp|^2}\bigg(\frac{\mu^2}{|\kperp|^2}\bigg)^{\vepv}
\end{equation}
%%%%%%%%%%%%%%%%%%%%%%%%%%%%%%%%%%%%%%%%
%
obstructs this naive approach, because it violates the natural scaling with $\Lambda$: while its integrand behaves as $\Ord\big(\Lambda^0\big)$, the integral is proportional to $\Lambda$.
We apply the convention for scalar functions to include a factor $\Lambda$ in the numerator of the integrand for every denominator factor that behaves as $\Ord(\Lambda)$, so $\Lambda$ effectively acts as a regulator of linear denominator factors.
As a result of the ``anomalous'' behavior of this scalar integral, its correct coefficient cannot be found by taking the leading power in $\Lambda$ on the one-loop integrand, and the subleading terms need to be included.
This then leads to a contribution that depends on the type of auxiliary partons.

Notice that the triangle above is exactly the one appearing in the collinear, or here on-shell, limit.
The box integrals are, up to $\Ord\big(\Lambda^{-1}\big)$ and $\Ord\big(\vepv\big)$ given by
%%%%%%%%%%%%%%%%%%%%%%%%%%%%%%%%%%%%%%%%
\begin{equation}
\graph{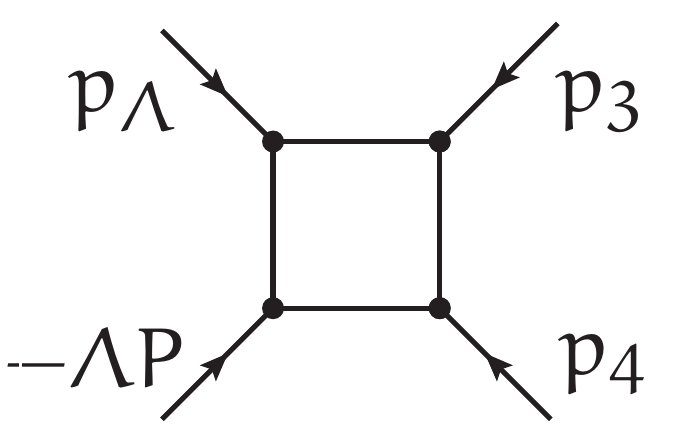}{15}{4}=
  \frac{1}{2\lop{\pP}{p_4}|\kperp|^2}\bigg\{
  \bigg(\frac{\mu^2}{|\kperp|^2}\bigg)^{\vepv}
  \bigg[\frac{4}{\vepv^2}-\frac{2}{\vepv}\ln\frac{2\Lambda\lop{\pP}{p_4}}{|\kperp|^2}\bigg]
  -\pi^2\bigg\}
~,
\end{equation}
\vspace{-3ex}
\begin{multline}
\graph{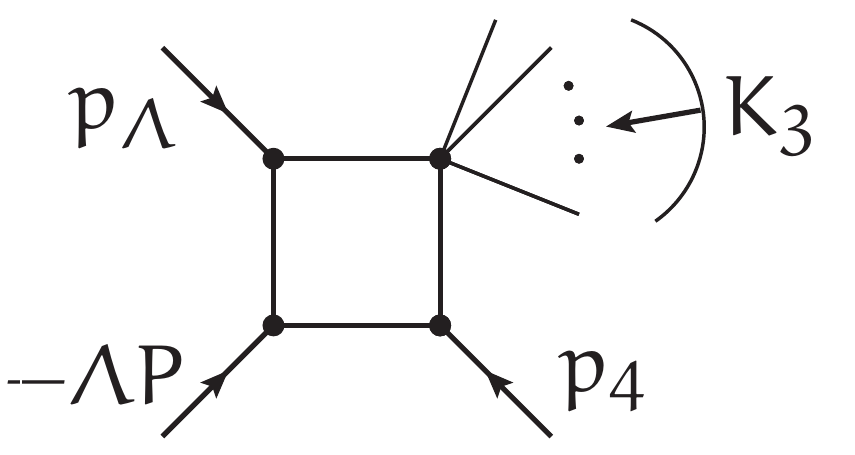}{19.5}{4}=
  \frac{1}{2\lop{\pP}{p_4}|\kperp|^2}\bigg\{
  \bigg(\frac{\mu^2}{|\kperp|^2}\bigg)^{\vepv}
  \bigg[\frac{2}{\vepv^2}-\frac{2}{\vepv}\ln\frac{2\Lambda\lop{\pP}{p_4}}{-K_3^2}\bigg]
\\
  +2\mathrm{Li}_2\bigg(1+\frac{|\kperp|^2}{K_3^2}\bigg)-\frac{2\pi^2}{3}\bigg\}
~.
\end{multline}
We use the convention that momenta indicated by the letter $p$ are light-like.
Also, $\lop{\pP}{p_4}$ must be imagined to have a small negative imaginary part, while $K_3^2$ has a small positive imaginary part.
The formulas have the necessary factor $\big(\mu^2/|\kperp|^2\big)^{\vepv}$.
In fact, the only other scalar function that does, besides the bubble, is
\begin{multline}
\graph{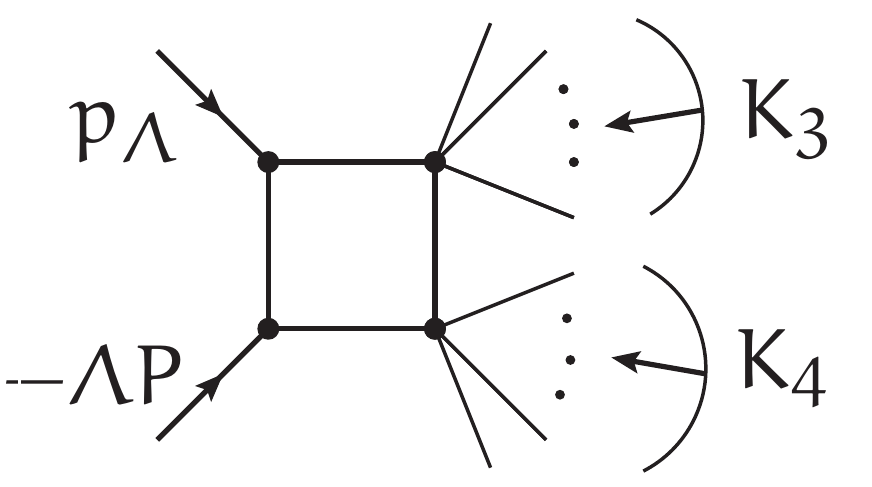}{20}{5}=
  \frac{1}{2\lop{\pP}{K_4}|\kperp|^2}\bigg\{
  \bigg(\frac{\mu^2}{|\kperp|^2}\bigg)^{\vepv}
  \bigg[\frac{1}{\vepv^2}-\frac{2}{\vepv}\ln\frac{\Lambda\lop{\pP}{K_4}}{\mu^2}
        -\frac{1}{\vepv}\ln\frac{\mu^4}{K_3^2K_4^2}\bigg]
\\
  -\frac{1}{2}\ln^2\frac{K_3^2}{K_4^2} -\frac{2\pi^2}{3}\bigg\}
~.
\end{multline}
%%%%%%%%%%%%%%%%%%%%%%%%%%%%%%%%%%%%%%%%
%
As mentioned before, the contribution to the one-loop amplitude of this function times its coefficient is irrelevant in the limit $|\kperp|\to0$, but we will need it in the following.

As can be seen, the box functions do scale correctly with $\Lambda$ and their coefficients can be extracted from the one-loop integrand after the limit $\Lambda\to\infty$.
Consequently, the result will not depend on the type of auxiliary partons, but we still need to include part of it in what we call \unfamiliar\ contribution, both because they do contribute to the non-smooth on-shell limit, and because they carry $\ln\Lambda$.
Now it is important to note that for any of these functions (and realize that $K_3,K_4$ are sums of external momenta and there is a function for each combination) the relevant part regarding $\ln\Lambda$ is almost the same, namely
%
%%%%%%%%%%%%%%%%%%%%%%%%%%%%%%%%%%%%%%%%
\begin{equation}
  \frac{-1}{2\lop{\pP}{K_4}|\kperp|^2}
  \bigg(\frac{\mu^2}{|\kperp|^2}\bigg)^{\vepv}
  \frac{2\ln\Lambda}{\vepv}
~.
\label{Eq:030}
\end{equation}
%%%%%%%%%%%%%%%%%%%%%%%%%%%%%%%%%%%%%%%%
%
Only the factor $1/\lop{\pP}{K_4}$ (or $1/\lop{\pP}{p_4}$) varies.

The calculation of the coefficients of these box functions has been addressed in Section~5.1 of \mycite{vanHameren:2017hxx}.
Thanks to momentum conservation they can be uniquely labeled by the momentum $K_4$.
The pair of specific choices of the loop momentum to extract the coefficients are $\ell_{1},\ell_{2}$, satisfying the conditions that
%
%%%%%%%%%%%%%%%%%%%%%%%%%%%%%%%%%%%%%%%%
\begin{equation}
\ell_{1,2}^2=\lop{\pP}{\ell_{1,2}}=(\ell_{1,2}-k)^2=(\ell_{1,2}-K_4)^2=0
~.
\end{equation}
%%%%%%%%%%%%%%%%%%%%%%%%%%%%%%%%%%%%%%%%
%
Here, the momentum $k=\xiPin\pP+\kperp$ has positive energy, so it differs by a minus sign with \mycite{vanHameren:2017hxx}.
The coefficients are given by
%
%%%%%%%%%%%%%%%%%%%%%%%%%%%%%%%%%%%%%%%%
\begin{equation}
C(K_4)=\frac{1}{2}\big[\mathrm{Res}(K_4;\ell_1)+\mathrm{Res}(K_4;\ell_2)\big]
~,
\end{equation}
%%%%%%%%%%%%%%%%%%%%%%%%%%%%%%%%%%%%%%%%
%
with
%%%%%%%%%%%%%%%%%%%%%%%%%%%%%%%%%%%%%%%%%
\begin{equation}
\mathrm{Res}(K_4;\ell_{1,2}) = 
-\Lambda\,|\kperp|^2\sum_{h=-,+}\graph{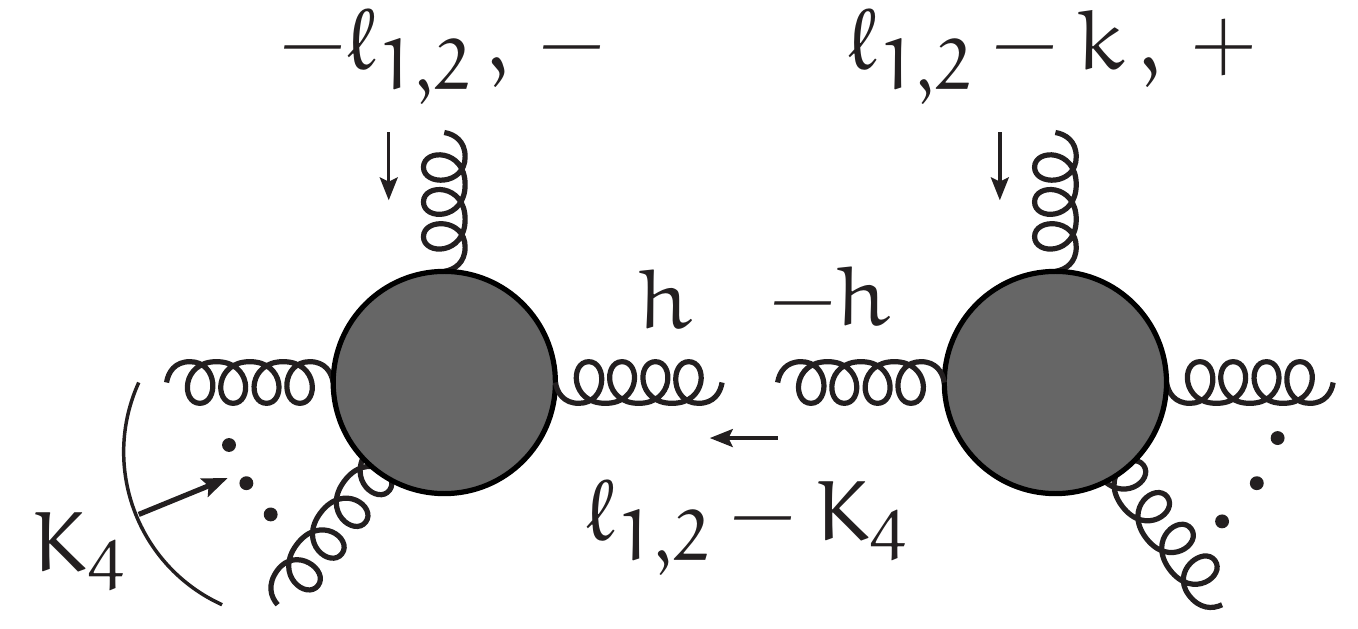}{32}{6}
~.
\label{Eq:031}
\end{equation}
%%%%%%%%%%%%%%%%%%%%%%%%%%%%%%%%%%%%%%%%
%
The blobs are on-shell tree-level amplitudes, and the sum is over the helicities of the ``internal'' gluon.
The expression in \mycite{vanHameren:2017hxx} has $\Lambda^2$ but here we follow the convention to include one factor $\Lambda$ in the box function.
The expression above then scales with $\Lambda$ as expected for a tree-level object.

The reader may have noticed that the above looks like a term in the Britto-Cachazo-Feng (BCF) decomposition of the tree-level amplitude~\mycite{Britto:2004ap,Britto:2005fq}.
Decomposing the transverse momentum $\kperp$ into polarization vectors
%
%%%%%%%%%%%%%%%%%%%%%%%%%%%%%%%%%%%%%%%%
\begin{equation}
k^\mu = \xiPin\pP^\mu + \kperp^\mu
\quad,\quad
\kperp^\mu = -\frac{\kapp}{\sqrt{2}}\,e_{-}^\mu
             -\frac{\kstr}{\sqrt{2}}\,e_{+}^\mu
\end{equation}
%%%%%%%%%%%%%%%%%%%%%%%%%%%%%%%%%%%%%%%%
%
the specific loop momenta are given by
%
%%%%%%%%%%%%%%%%%%%%%%%%%%%%%%%%%%%%%%%%
\begin{equation}
\ell_1^\mu = z_1\pP^\mu - \frac{\kapp}{\sqrt{2}}\,e_{-}^\mu
\quad,\quad
\ell_2^\mu = z_2\pP^\mu - \frac{\kstr}{\sqrt{2}}\,e_{+}^\mu
\end{equation}
%%%%%%%%%%%%%%%%%%%%%%%%%%%%%%%%%%%%%%%%
%
where $z_1,z_2$ are such that $(\ell_{1,2}-K_4)^2=0$.
So we see that $\pP$ plays the role of the ``shift vector'' to accommodate the on-shellness of the internal line.
In other words, we have
%
%%%%%%%%%%%%%%%%%%%%%%%%%%%%%%%%%%%%%%%%
\begin{equation}
\sum_{K_4}
\frac{\mathrm{Res}(K_4;\ell_{1})}{\big(\frac{\kapp}{\sqrt{2}}\,e_{-}+K_4\big)^2}
=
-\Lambda\,|\kperp|^2
\,\Amp_{\Tree}\bigg(\bigg(\frac{\kapp}{\sqrt{2}}\,e_{-}\bigg)^-,\bigg(\frac{\kstr}{\sqrt{2}}\,e_{+}-\xiPin\pP\bigg)^+,3,\ldots,n\bigg) %,p_3^{h_3},\ldots,p_n^{h_n}\bigg)
\label{Eq:032}
\end{equation}
%%%%%%%%%%%%%%%%%%%%%%%%%%%%%%%%%%%%%%%% 
%
where the sum is over all possibilities of
%
%%%%%%%%%%%%%%%%%%%%%%%%%%%%%%%%%%%%%%%%
\begin{equation}
K_4=p_n \quad,\quad K_4=p_n+p_{n-1} \quad,\ldots.
\end{equation}
%%%%%%%%%%%%%%%%%%%%%%%%%%%%%%%%%%%%%%%%
%
With $\ell_2$ the amplitude with $e_+$ is obtained instead.
Realize that these are the momenta, and that $\pP$ plays the role of ``polarization vector''.

We are however not interested in this result.
Rather, we are interested in the sum on the left-hand side of \Equation{Eq:032}, but instead with that denominator, we want the terms to be multiplied with (\ref{Eq:030}).
And indeed, it turns out that
%
%%%%%%%%%%%%%%%%%%%%%%%%%%%%%%%%%%%%%%%%
\begin{equation}
\sum_{K_4}
\frac{\mathrm{Res}(K_4;\ell_{1})}{2\lop{P}{K_4}|\kperp|^2}
=
-\Lambda
\,\Amp_{\Tree}^\star\big({-}\xiPin\pP-\kperp,3,\ldots,n\big) %p_3^{h_3},\ldots,p_n^{h_n}\big)
~.
\end{equation}
%%%%%%%%%%%%%%%%%%%%%%%%%%%%%%%%%%%%%%%% 
%
So far, we were not able to prove this remarkable result, but we were able to confirm it numerically for amplitudes with up to a several number of external gluons.
Incidentally, the result offers an alternative way to compute tree-level amplitudes with a space-like gluon.
Realize that this formula is not to be interpreted recursively.
The blobs in \Equation{Eq:031} are ``normal'' on-shell tree-level amplitudes.

Both $\ell_1$ and $\ell_2$ give the same result, so within the integrand-based framework, the so-called ``tilde coefficients'' of the box function vanish, and the triangle coefficient can be calculated directly from the ``triple cut''.
These then relatively straightforwardly lead to the space-like tree-level amplitude again.
So eventually, these findings support the prescription to replace the on-shell limit with the space-like tree-level amplitude leading to \Equation{Eq:029}.

\section{\label{Sec:060}Divergencies in the \familiar\ virtual contribution}
The divergent structure of on-shell one-loop QCD amplitudes is universal and well known~\mycite{Kunszt:1994np}.
As is common when presenting this structure, we consider the UV-subtracted contribution.
It allows us to write down the complete divergent contribution, that is the sum of \familiar\ and \unfamiliar. 
We will, however, be able to distinguish the two.

If we stick to the terms relevant to auxiliary partons, we have for the divergent virtual contribution in case of auxiliary quarks, explicitly denoted $q$ and $\bar{q}$,  the following expression times $\cNLO$:
%
%%%%%%%%%%%%%%%%%%%%%%%%%%%%%%%%%%%%%%%%
\begin{align}
&-2\frac{C_F}{\vepv^2}\,\justMAuxq
\notag\\&\hspace{0ex}
+ \frac{2}{\vepv}\bigg\{\ln\!\bigg(\frac{\mu^2}{|\kperp|^2}\bigg)\justMAuxqCor{q}{\bar{q}}
+ \sum_{i\neq q,\bar{q}}\bigg[
  \ln\!\bigg(\frac{\mu^2}{2\Lambda\lop{P}{p_i}}\bigg)\justMAuxqCor{i}{q}
+ \ln\!\bigg(\frac{\mu^2}{-2\Lambda\lop{P}{p_i}}\bigg)\justMAuxqCor{i}{\bar{q}}\bigg]\bigg\}
\notag\\&\hspace{0ex}
-2\frac{3C_F}{2\vepv}\,\justMAuxq
~.
\end{align}
%%%%%%%%%%%%%%%%%%%%%%%%%%%%%%%%%%%%%%%%
%
The \matrixelement{}s involve a space-like gluon, but carry the auxiliary label because they are still in the auxiliary quark color representation.
The large-$\Lambda$ limit and the division by $\Caux$ are assumed to be already performed.
Eventually, the real part must be taken for the virtual contribution, but we want the keep the imaginary part explicit here for later comparison.
The first line contains the two soft-collinear terms associated with the auxiliary quarks.
The second line contains the other soft terms associated with those.
The sums are over external partons.
Realize that in literature these sums are typically written over double indices, say $i$ and $j$, with $i\neq j$ rather than $i>j$, explaining the factor $2$ above.
Allowing for a slight abuse of notation incorrectly suggesting positivity, $\justMAuxqCor{i}{j}$ is the color-correlated \matrixelement\ involving external partons $i$ and $j$.
The last line represents the two collinear terms associated with the auxiliary quarks.
In \Appendix{App:060} we calculate
%
%%%%%%%%%%%%%%%%%%%%%%%%%%%%%%%%%%%%%%%%
\begin{equation}
\justMAuxq = \justMStar
\quad,\quad
\justMAuxqCor{q}{\bar{q}} = \frac{1}{2\Nc}\justMStar
\label{Eq:091}
~,
\end{equation}
%%%%%%%%%%%%%%%%%%%%%%%%%%%%%%%%%%%%%%%%
%
and
%
%%%%%%%%%%%%%%%%%%%%%%%%%%%%%%%%%%%%%%%%
\begin{equation}
\justMAuxqCor{i}{q}+\justMAuxqCor{i}{\bar{q}} = \justMStarCor{i}{\star}
~.
\end{equation}
%%%%%%%%%%%%%%%%%%%%%%%%%%%%%%%%%%%%%%%%
%
where the latter is the correlator involving the space-like gluon, independent of the auxiliary color representation.
Using color conservation, we have
%
%%%%%%%%%%%%%%%%%%%%%%%%%%%%%%%%%%%%%%%%
\begin{equation}
\sum_{i\neq q,\bar{q}}\justMAuxqCor{i}{\bar{q}} = -\justMAuxqCor{q}{\bar{q}} -\justMAuxqCor{\bar{q}}{\bar{q}}
=-\bigg(\frac{1}{2\Nc}+C_F\bigg)\justMStar
=-\frac{\Nc}{2}\,\justMStar
\end{equation}
%%%%%%%%%%%%%%%%%%%%%%%%%%%%%%%%%%%%%%%%
%
and the same for $q\leftrightarrow\bar{q}$.
So we have the divergent virtual contribution
%
%%%%%%%%%%%%%%%%%%%%%%%%%%%%%%%%%%%%%%%%
\begin{align}
&-2\frac{C_F}{\vepv^2}\,\justMStar
\notag\\&\hspace{0ex}
+ \frac{1}{\Nc\vepv}\ln\!\bigg(\frac{\mu^2}{|\kperp|^2}\bigg)\justMStar
+ \frac{\Nc}{\vepv}\bigg[2\ln\!\bigg(\frac{\Lambda}{\xiPin}\bigg)-\imag\pi\bigg]\justMStar
+ \frac{2}{\vepv}\sum_{i\neq\star}\ln\!\bigg(\frac{\mu^2}{2\xiPin\lop{P}{p_i}}\bigg)\justMStarCor{i}{\star}
\notag\\&\hspace{0ex}
-3\frac{C_F}{\vepv}\,\justMStar
~.
\end{align}
%%%%%%%%%%%%%%%%%%%%%%%%%%%%%%%%%%%%%%%%
%
We introduced $\xiPin$ here to make sure that the combination $\xiPin\pP$ appears rather than $\pP$ on its own.
After some reorganization, we can write
%
%
%%%%%%%%%%%%%%%%%%%%%%%%%%%%%%%%%%%%%%%%
\begin{align}
&-\frac{C_A}{\vepv^2}\,\justMStar
+ \frac{2}{\vepv}\sum_{i\neq\star}\ln\!\bigg(\frac{\mu^2}{2\xiPin\lop{P}{p_i}}\bigg)\justMStarCor{i}{\star}
-\frac{11\Nc-2n_f}{6\vepv}\,\justMStar
\notag\\&\hspace{0ex}
+\frac{\Nc}{\vepv}\bigg\{
  \frac{1}{\Nc^2}\bigg[\frac{1}{\vepv}
  +\ln\!\bigg(\frac{\mu^2}{|\kperp|^2}\bigg)\bigg]
  + 2\ln\!\bigg(\frac{\Lambda}{\xiPin}\bigg)-\imag\pi
  +\frac{1}{3} + \frac{3}{2\Nc^2} - \frac{n_f}{3\Nc}
\bigg\}\justMStar
~.
\end{align}
%%%%%%%%%%%%%%%%%%%%%%%%%%%%%%%%%%%%%%%%
%
The first line is what we call the divergent \familiar\ virtual contribution related to the space-like gluon.
It looks exactly as if the space-like gluon were at the on-shell limit, but with the space-like tree-level \matrixelement.
The second line is the \unfamiliar\ contribution, and corresponds to the divergent terms in \Equation{Eq:063} and \Equation{Eq:012}.
Realize that those results are UV un-subtracted, while the results above are, so we need to add $(11\Nc-2n_f)/(6\vepv)\justMStar$ back to the result here in order to match%
%(\eg\ $1/3+11/6=13/6$ etc.)
.

In case of auxiliary gluons, labelled $A$ and $B$, we have $\cNLO$ times
%
%%%%%%%%%%%%%%%%%%%%%%%%%%%%%%%%%%%%%%%%
\begin{align}
&-2\frac{C_A}{\vepv^2}\,\justMAuxg
\notag\\&\hspace{0ex}
+ \frac{2}{\vepv}\bigg\{\ln\!\bigg(\frac{\mu^2}{|\kperp|^2}\bigg)\justMAuxgCor{A}{B}
+ \sum_{i\neq A,B}\bigg[
  \ln\!\bigg(\frac{\mu^2}{2\Lambda\lop{P}{p_i}}\bigg)\justMAuxgCor{i}{A}
+ \ln\!\bigg(\frac{\mu^2}{-2\Lambda\lop{P}{p_i}}\bigg)\justMAuxgCor{i}{B}\bigg]\bigg\}
\notag\\&\hspace{0ex}
-2\frac{11\Nc-2n_f}{6\vepv}\,\justMAuxg
~.
\end{align}
%%%%%%%%%%%%%%%%%%%%%%%%%%%%%%%%%%%%%%%%
%
In \Appendix{App:060} we calculate
%
%%%%%%%%%%%%%%%%%%%%%%%%%%%%%%%%%%%%%%%%
\begin{equation}
\justMAuxg = \justMStar
\quad,\quad
\justMAuxgCor{A}{B} = -\frac{\Nc}{2}\,\justMStar
~,
\label{Eq:092}
\end{equation}
%%%%%%%%%%%%%%%%%%%%%%%%%%%%%%%%%%%%%%%%
%
and
%
%%%%%%%%%%%%%%%%%%%%%%%%%%%%%%%%%%%%%%%%
\begin{equation}
\justMAuxgCor{i}{A}+\justMAuxgCor{i}{B} = \justMStarCor{i}{\star}
~.
\end{equation}
%%%%%%%%%%%%%%%%%%%%%%%%%%%%%%%%%%%%%%%%
%
Realize again that we assume the \matrixelement{}s to be divided by $\Caux$ already. 
Using color conservation, we now get
%
%%%%%%%%%%%%%%%%%%%%%%%%%%%%%%%%%%%%%%%%
\begin{equation}
\sum_{i\neq A,B}\justMAuxqCor{i}{B} 
=-\bigg({-}\frac{\Nc}{2}+C_A\bigg)\justMStar
=-\frac{\Nc}{2}\,\justMStar
~,
\end{equation}
%%%%%%%%%%%%%%%%%%%%%%%%%%%%%%%%%%%%%%%%
%
and the divergent contribution becomes
%
%
%%%%%%%%%%%%%%%%%%%%%%%%%%%%%%%%%%%%%%%%
\begin{align}
&-2\frac{C_A}{\vepv^2}\,\justMStar
\notag\\&\hspace{0ex}
- \frac{\Nc}{\vepv}\ln\!\bigg(\frac{\mu^2}{|\kperp|^2}\bigg)\justMStar
+ \frac{\Nc}{\vepv}\bigg[2\ln\!\bigg(\frac{\Lambda}{\xiPin}\bigg)-\imag\pi\bigg]\justMStar
+ \frac{2}{\vepv}\sum_{i\neq\star}\ln\!\bigg(\frac{\mu^2}{2\xiPin\lop{P}{p_i}}\bigg)\justMStarCor{i}{\star}
\notag\\&\hspace{0ex}
-\frac{11\Nc-2n_f}{3\vepv}\,\justMStar
~,
\end{align}
%%%%%%%%%%%%%%%%%%%%%%%%%%%%%%%%%%%%%%%%
%
and after reorganization
%
%%%%%%%%%%%%%%%%%%%%%%%%%%%%%%%%%%%%%%%%
\begin{align}
&-\frac{C_A}{\vepv^2}\,\justMStar
+ \frac{2}{\vepv}\sum_{i\neq\star}\ln\!\bigg(\frac{\mu^2}{2\xiPin\lop{P}{p_i}}\bigg)\justMStarCor{i}{\star}
-\frac{11\Nc-2n_f}{6\vepv}\,\justMStar
\notag\\&\hspace{0ex}
-\frac{\Nc}{\vepv}\bigg\{
 \frac{1}{\vepv}\,
+\ln\!\bigg(\frac{\mu^2}{|\kperp|^2}\bigg)
 - 2\ln\!\bigg(\frac{\Lambda}{\xiPin}\bigg)+\imag\pi
\bigg\}\justMStar
-\frac{11\Nc-2n_f}{6\vepv}\,\justMStar
~.
\end{align}
%%%%%%%%%%%%%%%%%%%%%%%%%%%%%%%%%%%%%%%%
%
We see the same structure as with the auxiliary quarks, with the second line now matching the divergent part of \Equation{Eq:063} and \Equation{Eq:013} after adding $(11\Nc-2n_f)/(6\vepv)\justMStar$.

\section{\label{Sec:050}Divergencies in the \familiar\ real contribution}
The \familiar\ real contribution becomes divergent when a radiative gluon becomes soft or collinear to another parton.
The soft limit of a gluon in QCD \matrixelement{}s is universal and well-known~\cite{Bassetto:1983mvz} (see also Section~4.2 in~\mycite{Catani:1996vz}).
%
%Furthermore, this limit commutes with the large-$\Lambda$ limit.
%
Terms related to the auxiliary partons, let us take quarks, when a final-state gluon with momentum $r$ becomes soft and then taking the large $\Lambda$ limit are given by
%
%%%%%%%%%%%%%%%%%%%%%%%%%%%%%%%%%%%%%%%%
\begin{equation}
  -2\sum_{i\neq q,\bar{q}}\Bigg\{
    \frac{(\lop{\pP}{p_i})}{(\lop{\pP}{r})(\lop{r}{p_i})}\,\justMAuxqCor{i}{q}
   +\frac{(\lop{\pP}{p_i})}{(\lop{\pP}{r})(\lop{r}{p_i})}\,\justMAuxqCor{i}{\bar{q}}
  \Bigg\}
-2\,\frac{(\Lambda\lop{\pP}{p_\Lambda})}{(\lop{\Lambda\pP}{r})(\lop{r}{p_\Lambda})}\,\justMAuxqCor{q}{\bar{q}}
\label{Eq:088}
\end{equation}
%%%%%%%%%%%%%%%%%%%%%%%%%%%%%%%%%%%%%%%%
%
where $q,\bar{q}$ refer to the auxiliary quark pair.
The auxiliary parton momenta are given in \Equation{Eq:006}.
While omitting an overall constant, we do include the characteristic minus-sign, and the factor $2$ coming from the double sum over all external partons.
The terms in the sum over $i$ are exactly what one also gets from taking the soft limit on the amplitude after the large-$\Lambda$ limit constructed with eikonal Feynman rules.
Using the relations for the color-correlated \matrixelement{}s from the previous section again, we find
%
%%%%%%%%%%%%%%%%%%%%%%%%%%%%%%%%%%%%%%%%
\begin{equation}
%\frac{\cNLO\mu^{2\vepv}}{\piep}\sum_{i\neq\star}
    \frac{(\lop{\pP}{p_i})}{(\lop{\pP}{r})(\lop{r}{p_i})}\,\justMAuxqCor{i}{q}
   +\frac{(\lop{\pP}{p_i})}{(\lop{\pP}{r})(\lop{r}{p_i})}\,\justMAuxqCor{i}{\bar{q}}
=
\,\frac{(\lop{\xiPin\pP}{p_i})}{(\lop{\xiPin\pP}{r})(\lop{r}{p_i})}\,\justMStarCor{i}{\star}
~,
\end{equation}
%%%%%%%%%%%%%%%%%%%%%%%%%%%%%%%%%%%%%%%%
%
that is again the pattern that the space-like gluon should be treated as if it were at the on-shell limit, but with the space-like LO \matrixelement.
Since $(\Lambda\lop{\pP}{p_\Lambda})=-\srac{1}{2}|\kperp|^2+\Ord\big(\Lambda^{-1}\big)$, the last term in \Equation{Eq:088} obviouly vanishes for $\Lambda\to\infty$, but we kept it here to show that a similar looking term survives in the \unfamiliar\ contribution.
This can be understood from \Equation{Eq:042}, in which the $\Cr$-term has the double-eikonal form with the auxiliary partons as ``spectators'' like the vanishing term above.
Notice that the values of $\Cr$ in \Equation{Eq:089} and \Equation{Eq:090} match \Equation{Eq:091} and \Equation{Eq:092}.

Besides the other usual soft and collinear singularities, there is also a singularity when a radiative gluon becomes collinear to the momentum $\pP$. 
\Appendix{App:030} we derive the limit
%
%%%%%%%%%%%%%%%%%%%%%%%%%%%%%%%%%%%%%%%%
\begin{multline}
\justMStar\big(\xiPin\pP+\kperp,\kinBar\,;r,\{p_i\}_{i=1}^n\big)
\\
\overset{r\to \xiP_r\pP}{\longrightarrow}
\frac{2\Nc}{\lop{\pP}{r}}\,\frac{\xiPin^2}{\xiP_r(\xiPin-\xiP_r)^2}
\,\justMStar\big((\xiPin-\xiP_r)\pP+\kperp,\kinBar\,;\{p_i\}_{i=1}^n\big)
~.
\label{Eq:034}
\end{multline}
%%%%%%%%%%%%%%%%%%%%%%%%%%%%%%%%%%%%%%%%
%
For comparison, the usual collinear limit, for example with the parton on the on-shell side, looks like
%
%%%%%%%%%%%%%%%%%%%%%%%%%%%%%%%%%%%%%%%
\begin{equation}
\justMStar\big(\kin,\xMin\pM;r,\{p_i\}_{i=1}^n\big)
\overset{r\to \xM_r\pM}{\longrightarrow}
\frac{\CBar}{\lop{\pM}{r}}\,\frac{\PBar(1-\xM_r/\xMin)}{\xMin-\xM_r}
\,\justMStar\big(\kin,(\xMin-\xM_r)\pM;\{p_i\}_{i=1}^n\big)
~,
\end{equation}
%%%%%%%%%%%%%%%%%%%%%%%%%%%%%%%%%%%%%%%%
%
where $\PBar(z)$ is the non-regularized splitting function associated with the on-shell initial-state parton, with the color factor $\CBar$ taken out explicitly.
The singular behavior leads to the non-cancelling collinear divergence of \Equation{Eq:064} that is absorbed into the NLO correction to the collinear PDF following the well-know factorization prescription.
Here we derive \Equation{Eq:065} for the space-like case.
Let us just mention already that this collinear limit {\em does} commute with the large-$\Lambda$ limit.

We can be a bit more ambitious in the definition of the collinear limit, and instead of \Equation{Eq:034}, we can write down a formula that does not compromise phase space, very much like subtraction terms constructed to render real-radiation phase space integrals finite in NLO calculations. 
This means, that we need to accommodate for the momentum recoil when the momentum $r$ is not exactly at the collinear limit.
Fortunately, we are privileged compared to collinear factorization in that we have initial-state transverse momentum that can be used for the recoil, and we can simply write
%
%%%%%%%%%%%%%%%%%%%%%%%%%%%%%%%%%%%%%%%%
\begin{multline}
\justMStar\big(\xiPin\pP+\kperp,\kinBar\,;r,\{p_i\}_{i=1}^n\big)
\\
\longrightarrow
\frac{2\Nc}{\lop{\pP}{r}}\,\frac{\xiPin^2}{\xiP_r(\xiPin-\xiP_r)^2}
\,\justMStar\big((\xiPin-\xiP_r)\pP+\kperp-\rperp,(\xMin-\xM_r)\pM\,;\{p_i\}_{i=1}^n\big)
~.
\label{Eq:050}
\end{multline}
%%%%%%%%%%%%%%%%%%%%%%%%%%%%%%%%%%%%%%%%
%
Now, $\{p_i\}_{i=1}^n$ can be seen as identical on the left-hand side and the right-hand side for any momentum $r$, not only at the limit $r=\xiP_r\pP$.
We will make one adjustments to the formula, that will still give the correct limit, namely we will neglect $\xM_r$ on the right-hand side of \Equation{Eq:050}.

In the following, we will employ integration over initial-state variables, and prefer to write formulas in terms of $\xP$ and $\pPA$ already instead of $\xiP$ and $\pP$ (\Section{Sec:023}).
We will use the symbol
%
%%%%%%%%%%%%%%%%%%%%%%%%%%%%%%%%%%%%%%%%
\begin{equation}
 \sTot = 2\lop{\pPA}{\pM}
\end{equation}
%%%%%%%%%%%%%%%%%%%%%%%%%%%%%%%%%%%%%%%%
%
but it will appear only multiplied with an $\xP$ variable.
Notice that the factor in front of the \matrixelement\ on the right-hand side of \Equation{Eq:050} is invariant under rescaling of the $\xiP$ variables and $\pP$.

\begin{figure}
\begin{center}
\epsfig{figure=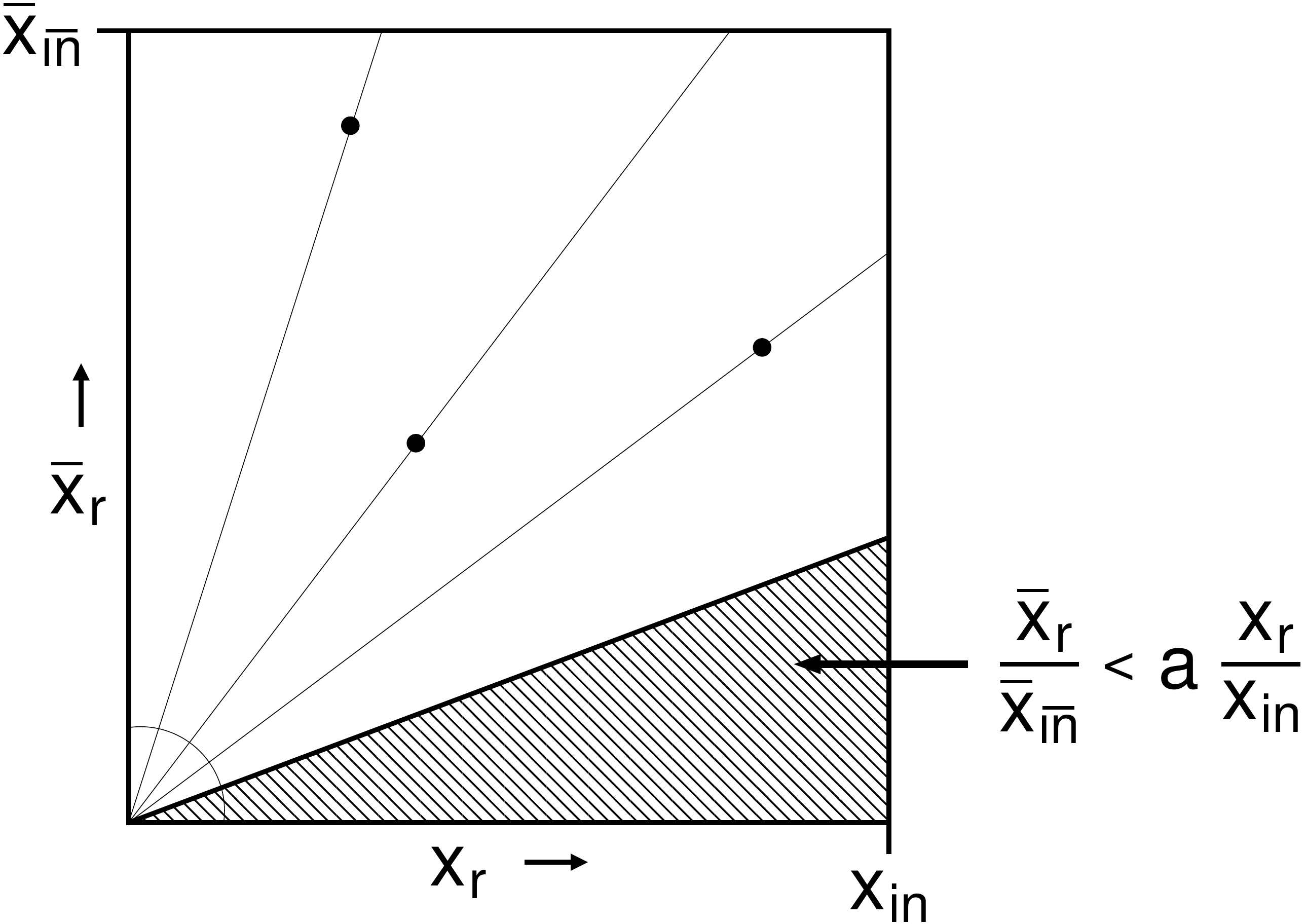,width=0.38\linewidth}
\caption{\label{Fig:04}Phase space for the collinear subtraction term. The lines with a dot correspond to final-state collinear singularities, and the vertical and horizontal axis correspond to initial-state collinear singularities. The region near $\xP_r=\xM_r=0$ is the soft region.}
\end{center}
\end{figure}

It must be stressed at this point that the ``subtraction term'' of \Equation{Eq:050} only produces the collinear limit correctly.
It can only produce some part of the soft limit of the \matrixelement.
We can however reduce the phase space on which it acts such that it still covers the whole relevant collinear region, but restricts the soft region to a part that can reasonably be associated with the particular collinear region.
This can be mimicked by demanding that $\xM_r/\xMin<\aM\xP_r/\xPin$ for some $\aM<1$, see \Figure{Fig:04}.
Although we are eventually only aiming at the true collinear contribution, we will be able to make contact with some of the findings in~\mycite{Nefedov:2020ecb} this way.

We can go one step further, and choose $\aM = |\kperp+\rperp|^2/(\sTot\xPin\xMin)$, leading, together with the on-shellness condition $|\rperp|^2=\sTot\xP_r\xM_r$, to the phase space restriction
%
%%%%%%%%%%%%%%%%%%%%%%%%%%%%%%%%%%%%%%%%
\begin{equation}
|\rperp| < |\kperp+\rperp|\,\frac{\xP_r}{\xPin}
~,
\end{equation}
%%%%%%%%%%%%%%%%%%%%%%%%%%%%%%%%%%%%%%%%
%
which is the complement of the phase space restriction imposed in \Equation{Eq:086} to cut out the collinear region.
So in that sense, we will now find what we wanted to keep inside the \familiar\ constribution and decided to cut out of the \unfamiliar\ contribution to avoid double counting.

We insert the formula, now in terms of $\xP$ variables and $\pPA$, into the cross section formula, to get (remember the flux factor)
%
%%%%%%%%%%%%%%%%%%%%%%%%%%%%%%%%%%%%%%%%
\begin{align}
&\dRfamColl\big(\kin,\kinBar\,;\{p_i\}_{i=1}^n\big)
\notag\\&\hspace{2ex}=
    \frac{\cNLO}{\piep\mu^\vep}
    \frac{1}{|\kperp|^2}
    \int d^{4+\vep}r\,\delta(r^2)
    \,\theta(\xP_r<\xPin)
    \,\theta\bigg(\frac{\xM_r}{\xMin}<\aM\,\frac{\xP_r}{\xPin}\bigg)
    \,2\Nc\,\frac{1}{\lop{\pPA}{r}}\,\frac{\xPin^2}{\xP_r(\xPin-\xP_r)^2}
\notag\\&\hspace{28ex}\times
    \,\frac{\xPin-\xP_r}{\xPin}
    \,d\PSTstr_{n}\big(\xPin-\xP_r,\kperp-\rperp,\xMin\,;\{p_i\}_{i=1}^{n}\big)
    \,\JetB\big(\{p_i\}_{i=1}^{n}\big)
%\notag\\&\hspace{2ex}=
%    \frac{2\cNLO\Nc}{\piep\mu^\vep}
%    \int d^{4+\vep}r\,\delta(r^2)\,\frac{2}{|\rperp|^2}
%    \,\theta(\xP_r<\xPin)
%    \,\theta\bigg(\frac{\xM_r}{\xMin}<\aM\,\frac{\xP_r}{\xPin}\bigg)
%    \,\frac{|\kperp-\rperp|^2}{|\kperp|^2}
%\notag\\&\hspace{28ex}\times
%    \frac{\xPin}{\xPin-\xP_r}
%    \,\frac{d\PSTstr_{n}\big(\xPin-\xP_r,\kperp-\rperp,\xMin\,;\{p_i\}_{i=1}^{n}\big)}
%           {|\kperp-\rperp|^2}
%    \,\JetB\big(\{p_i\}_{i=1}^{n}\big)
%\notag\\&\hspace{2ex}=
%    \frac{2\cNLO\Nc}{\piep\mu^\vep}
%    \int_0^{\xPin}\frac{d\xP_r}{\xP_r}
%    \int_0^{\aM\xMin\xP_r/\xPin} d\xM_r
%    \int\frac{d^{2+\vep}\rperp}{|\rperp|^2}
%    \,\frac{|\kperp-\rperp|^2}{|\kperp|^2}
%    \,\delta\bigg(\xM_r-\frac{|\rperp|^2}{\sTot\xP_r}\bigg)
%\notag\\&\hspace{34ex}\times
%    \frac{\xPin}{\xPin-\xP_r}
%    \,\dBstar\big(\xPin-\xP_r,\kperp-\rperp,\xMin\,;\{p_i\}_{i=1}^{n}\big)
\notag\\&\hspace{2ex}=
    \frac{2\cNLO\Nc}{\piep\mu^\vep}
    \int_0^{\xPin}\frac{d\xP_r}{\xP_r}
    \int\frac{d^{2+\vep}\rperp}{|\rperp|^2}
    \,\frac{|\kperp-\rperp|^2}{|\kperp|^2}
    \,\theta\bigg(|\rperp|<|\kperp+\rperp|\,\frac{\xP_r}{\xPin}\bigg)
\notag\\&\hspace{34ex}\times
    \frac{\xPin}{\xPin-\xP_r}
    \,\dBstar\big(\xPin-\xP_r,\kperp-\rperp,\xMin\,;\{p_i\}_{i=1}^{n}\big)
~.
\end{align}
%%%%%%%%%%%%%%%%%%%%%%%%%%%%%%%%%%%%%%%%
%
Now we put this underneath an integral over $\xPin$ and $\kperp$ with an extra function $F(\xPin,\kperp)$
%
%%%%%%%%%%%%%%%%%%%%%%%%%%%%%%%%%%%%%%%%
\begin{align}
&\int_0^1\frac{d\xPin}{\xPin}\int d^2\kperp\,F(\xPin,\kperp)\,\dRfamColl\big(\xPin\pPA+\kperp,\kinBar\,;\{p_i\}_{i=1}^n\big)
\notag\\&\hspace{2ex}=
    \frac{2\cNLO\Nc}{\piep\mu^\vep}
    \int_0^1\frac{dy}{y}
    \int d^2\kperp
    \int_y^1\frac{dz}{z(1-z)}
    \int\frac{d^{2+\vep}\rperp}{|\rperp|^2}\,\frac{|\kperp|^2}{|\kperp+\rperp|^2}
    \,\theta\big(|\rperp|<|\kperp|(1-z)\big)
\notag\\&\hspace{40ex}\times
    \,F\bigg(\frac{y}{z},\kperp+\rperp\bigg)
    \,\dBstar\big(y,\kperp,\xMin\,;\{p_i\}_{i=1}^{n}\big)
\notag\\&\hspace{2ex}=
    \int_0^1\frac{d\xPin}{\xPin}\,\int d^2\kperp\,\tilde{F}(\xPin,\kperp)
    \,\dBstar\big(\xPin,\kperp,\xMin\,;\{p_i\}_{i=1}^{n}\big)
\end{align}
%%%%%%%%%%%%%%%%%%%%%%%%%%%%%%%%%%%%%%%%
%
with
%
%%%%%%%%%%%%%%%%%%%%%%%%%%%%%%%%%%%%%%%%
\begin{equation}
\tilde{F}(\xPin,\kperp)
=
    \frac{2\cNLO\Nc}{\piep\mu^\vep}
    \int_{\xPin}^1\frac{dz}{z(1-z)}
    \int_0^{|\kperp|(1-z)}\frac{d^{2+\vep}\rperp}{|\rperp|^2}\,\frac{|\kperp|^2}{|\kperp+\rperp|^2}
    %\,\theta\bigg(|\rperp|^2<\frac{\sTot\aM\xPin^2(1-z)^2}{z^2}\bigg)
%\\\times
    \,
    F\bigg(\frac{\xPin}{z},\kperp+\rperp\bigg)
~,
\label{Eq:051}
\end{equation}
%%%%%%%%%%%%%%%%%%%%%%%%%%%%%%%%%%%%%%%%
%
where we introduced the notation
%
%
%%%%%%%%%%%%%%%%%%%%%%%%%%%%%%%%%%%%%%%%
\begin{equation}
\int_{0}^{U}d^{2+\vep}\rperp \equiv \int d^{2+\vep}\rperp\,\theta(0<|\rperp|<U) 
~.
\end{equation}
%%%%%%%%%%%%%%%%%%%%%%%%%%%%%%%%%%%%%%%%
%
%
Now we can compare this formula with Equation~(2.21) in~\cite{Nefedov:2020ecb}, which was interpreted in the context of multi-Regge kinematics evolution.
The factor $|\kperp|^2/|\kperp+\rperp|^2$ is missing in~\cite{Nefedov:2020ecb}, but we could also get rid of it by re-defining $F(\xPin,\kperp) \to |\kperp|^{2}\,F(\xPin,\kperp)$.
Notice that neglecting $\xM_r$ on the right-hand side of \Equation{Eq:051} is equivalent to crossing out $q_1^+$ in Equation~(2.17) in~\cite{Nefedov:2020ecb}.
In the following, we prefer to stick to the equation above, while realizing that
%
%%%%%%%%%%%%%%%%%%%%%%%%%%%%%%%%%%%%%%%%
\begin{equation}
F(\xP,0) = 0
~,
\end{equation}
%%%%%%%%%%%%%%%%%%%%%%%%%%%%%%%%%%%%%%%%
%
so there is no singularity at $\rperp=-\kperp$.
We would also like to mention that \Equation{Eq:051} looks similar to the real emission part of the CCFM equation~\mycite{Ciafaloni:1987ur,Catani:1989sg}.

% Observe that \Equation{Eq:051} does not conflict with the demand that $\xPin$ is small.
% %
% This demand can be stated as the condition that $F(\xP,\kperp)=0$ for $\xP>\xP_{\mathrm{max}}$ and some $\xP_{\mathrm{max}}\ll1$.
% %
% This means that the integral only contributes for $\xPin/z<\xP_{\mathrm{max}}$, that is for $z>\xPin/\xP_{\mathrm{max}}$.
% %
% Since $z<1$, this then implies that $\xPin$ must be smaller than $\xP_{\mathrm{max}}$ for the integral, and thus $\tilde{F}(\xPin,\kperp)$, not to vanish.
% %

In order to isolate all $1/\vep$ divergencies, we add and subtract identical terms in \Equation{Eq:051}, and write
%
%%%%%%%%%%%%%%%%%%%%%%%%%%%%%%%%%%%%%%%%
\begin{align}
\tilde{F}(\xPin,\kperp) &
   =\frac{2\cNLO\Nc}{\piep\mu^\vep}
    \int_{\xPin}^{1}\frac{dz}{z(1-z)}
    \int_{0}^{|\kperp|(1-z)}\!\!d^{2+\vep}\rperp
    \,\frac{|\kperp|^2}{|\rperp|^2|\rperp+\kperp|^2}
\notag\\&\hspace{24ex}
    \times\bigg[ F\bigg(\frac{\xPin}{z},\kperp+\rperp\bigg) 
     - \frac{|\rperp+\kperp|^2}{|\kperp|^2}\,F\bigg(\frac{\xPin}{z},\kperp\bigg)\bigg]
\notag\\&\hspace{0ex}
   + \tilde{F}^{\mathrm{div}}(\xPin,\kperp)
\label{Eq:058}
\end{align}
%%%%%%%%%%%%%%%%%%%%%%%%%%%%%%%%%%%%%%%%
%
The integral above is finite for $\vep\to0$, while al poles are contained in 
%
%%%%%%%%%%%%%%%%%%%%%%%%%%%%%%%%%%%%%%%%
\begin{align}
\tilde{F}^{\mathrm{div}}(\xPin,\kperp) &= 
\frac{2\cNLO\Nc}{\piep\mu^\vep}
    \int_{\xPin}^{1}\frac{dz}{z(1-z)}
    \int_{0}^{|\kperp|(1-z)}\frac{d^{2+\vep}\rperp}{|\rperp|^2}
    \,F\bigg(\frac{\xPin}{z},\kperp\bigg)
\notag\\&\hspace{0ex}=
\frac{4\cNLO\Nc}{\vep}
\bigg(\frac{|\kperp|}{\mu}\bigg)^{\vep}
    \int_{\xPin}^{1}\frac{dz}{z(1-z)^{1-\vep}}
     \,F\bigg(\frac{\xPin}{z},\kperp\bigg)
\notag\\&\hspace{0ex}=
  4\cNLO\Nc\bigg(\frac{|\kperp|}{\mu}\bigg)^{\vep}
  \Bigg\{
    \frac{F(\xPin,\kperp)}{\vep^2}
   +\frac{1}{\vep}\int_{\xPin}^{1}dz\bigg[\frac{1}{[1-z]_+}+\frac{1}{z}\bigg]
          F\bigg(\frac{\xPin}{z},\kperp\bigg)
  \Bigg\}
\\&\hspace{0ex}
   +4\cNLO\Nc\int_{\xPin}^{1}dz\bigg[\bigg[\frac{\ln(1-z)}{1-z}\bigg]_++\frac{\ln(1-z)}{z}\bigg]
          F\bigg(\frac{\xPin}{z},\kperp\bigg)
 + \Ord(\vep)
\notag~.
\end{align}
The $1/\vep^2$ term is part of the soft contribution which cancels against virtual contributions.
It is essentially the $1/\vep^2$ term that was removed from \Equation{Eq:093} to avoid double counting.
The remaining divergent, pure $1/\vep$, term gives \Equation{Eq:062} with $\vepv=-\vep/2$.

\section{Summary}

In the present paper we extended the auxiliary parton method for obtaining the impact factors in high energy factorization to the next-to-leading order. The main results are the promotion of the hybrid factorization formula to NLO, and a calculational scheme that allows to obtain the NLO impact factors from suitable helicity amplitudes with on-shell partons.  We also rederived NLO impact factors for the quark and the gluon obtained at first in Ref.\ \cite{Ciafaloni:1998hu}.

Applying the auxiliary parton method, we find the NLO hybrid factorization formula given in \Equation{Eq:094}.
The first line is free of divergencies, and involves the same TMD and PDF as at LO.
The individual virtual and real contributions are divergent, but we know that the divergencies cancel because their structure is exactly the same as if the space-like initial-state gluon were on-shell.
We named this contribution to the NLO cross section the (finite) {\em \familiar\ contribution}, and it can be calculated for arbitrary processes with methods known in literature.

The third line contains the collinear divergence $\Delta_{\collBar}$ given in \Equation{Eq:064}.
It cancels against a similar divergence in the PDF correction $f^{(1)}$ following the well-know mechanism in collinear factorization.
The second line contains a similar collinear-type divergence $\Delta^\star_{\coll}$, given in \Equation{Eq:065}, which must cancel against a TMD correction we denote $F^{(1)}$.
The latter is a condition on the TMD that we find must hold for the factorization to work.

Finally, the second line contains the divergencies $\Delta_{\unf}$ given in \Equation{Eq:047}.
These turn out to be identical to the impact factor corrections calculated in \mycite{Ciafaloni:1998hu} plus a contribution to the space-like Regge trajectory.
Consequently, we find {\it a posteriori} that the hybrid factorization formula must be cast in the form of high-energy factorization.

In the derivations presented here a bottom-up approach was taken, based on exact NLO expressions of general collinear partonic amplitudes in QCD in the high energy limit, \ie\ by taking the leading term in the $1/\Lambda$ power expansion, where the large parameter $\Lambda$ proportional to the large total energy squared $\sTot$ available in a particle collision. We found that the auxiliary parton method in the large $\Lambda$ limit and inclusion of NLO corrections do not commute. It means, that the auxiliary parton line cannot be factored out in the high energy limit before it is treated at NLO accuracy. In order to make this point explicit we introduced nomenclature motivated by the structure of the tree level expressions for scattering amplitudes in the high energy limit: contributions with the auxiliary parton treated at the LO approximation were named {\em \familiar}, while NLO contributions to the auxiliary parton impact factors were called {\em \unfamiliar}. 

The main guiding principle of the analysis described in the paper was following how the infrared divergencies emerging in the on-shell expressions are organized in the hybrid factorization scheme. 
% corrected - lm
To be more specific, we verified that the NLO expressions including the infrared singularities, are consistent with the Regge factorization: the singularities and the NLO corrections may be separated into an impact factor of the target  (in our case --- the auxiliary parton) and to the remaining contribution coming from the projectile  and from the in-between region, corresponding to an evolution kernel absorbing terms enhanced by a large $\ln\Lambda$.

Using this factorization one may proceed further and split off the evolution part from the \familiar\ contribution, and in this way obtain the projectile impact factor from the on-shell amplitudes. In particular, we showed that the \familiar\ contribution obtained in this procedure is universal w.r.t.\ the auxiliary parton flavor --- that is the same for an auxiliary quark and gluon. Hence, the projectile impact factor coming from the proposed scheme is universal as well. This universality holds for virtual and real NLO corrections independently.  

We analyzed cancellation of $\epsilon$-poles in the dimensional regularization, coming both from infrared and ultraviolet divergencies. The infrared singularities were found to cancel in the NLO target impact factor and the remaining part of NLO corrections separately when combined with the infrared singularities coming from the initial state parton lines, and the ultraviolet poles cancelled out after performing the renormalization. 
In the {\em \familiar} part of the amplitudes we found terms corresponding to one step of the evolution of transverse momentum dependent parton distribution function. The real emission part of this evolution equation is given by \Equation{Eq:051}. This integral kernel closely resembles the LO~BFKL kernel with a difference in the longitudinal momentum part: the $1/z$ factor present in the BFKL equation is replaced by $1/(z(1-z))$ in  \Equation{Eq:051}, and the transverse momentum integration is limited in \Equation{Eq:051}. This difference occurs because we did not isolate the leading $\ln\sTot$ approximation, but kept all the NLO contribution at the leading power of $1/S$. The $1/(1-z)$ factor in the evolution kernel (141) is a manifestation of the soft gluon corrections. 
Essentially the same equation was found in \cite{Nefedov:2020ecb}, where it was related to the last emission in a cascade of rapidity-ordered gluons. Of course the present results are still fixed order results and some more work is neccesary to obtain the evolution equation that would resum both BFKL and soft gluon logarithms in a rigorous manner.

The results obtained here show how to systematically extract the NLO impact factors for hybrid factorization from on-shell helicitiy amplitudes. In fact, the extraction of the virtual NLO contributions to the impact factors is straightforward and boils down to analytic subtractions of universal virtual terms described above from corresponding NLO virtual corrections to on-shell amplitudes. The computation of the real emission corrections to impact factors still remains to be done for each process separately, using the same scheme that was applied to compute the real radiation corrections to the auxiliary quark and gluon.\\

\subsection*{Acknowledgments}
The authors would like to thank Maxim Nefedov, Aleksander Kusina, and Alessandro Papa for enlightening discussions.
GZ is supported by grant no.\ 2019/35/B/ST2/03531 of the Polish National Science Centre.
AvH and LM gratefully acknowledge support of  the Polish National Science Center (NCN) grant No.\ 2017/27/B/ST2/02755.

%\bibliography{10-refs}{}\bibliographystyle{JHEP}
\providecommand{\href}[2]{#2}\begingroup\raggedright\endgroup

\begin{appendix}
\addtocontents{toc}{\protect\setcounter{tocdepth}{1}}
\section{\label{App:010}Spinors and transverse momenta }
Given a light-like momentum $p$, we define the spinors as four-component vectors with two vanishing components:
%
%%%%%%%%%%%%%%%%%%%%%%%%%%%%%%%%%%%%%%%%
\begin{align}
\Srght{p} &= \begin{pmatrix}L(p)\\\mathbf{0}\end{pmatrix}
\quad,\quad
L(p) = \frac{1}{\sqrt{|p_0+p_3|}}\begin{pmatrix}-p_1+\imag p_2\\ p_0+p_3\end{pmatrix}
\\
\Arght{p} &= \begin{pmatrix}\mathbf{0}\\ R(p)\end{pmatrix}
\quad,\quad
R(p) = \frac{\sqrt{|p_0+p_3|}}{p_0+p_3}\begin{pmatrix}p_0+p_3\\p_1+\imag p_2\end{pmatrix}
~.
\end{align}
%%%%%%%%%%%%%%%%%%%%%%%%%%%%%%%%%%%%%%%%
%
The ``dual'' spinors are defined as
%
%%%%%%%%%%%%%%%%%%%%%%%%%%%%%%%%%%%%%%%%
\begin{equation}
\leftS{p} = \big(\,(\EuScript{E}L(p))^T\,,\,\mathbf{0}\,\big)
\quad,\quad
\leftA{p} = \big(\,\mathbf{0}\,,\,(\EuScript{E}^TR(p))^T\,\big)
\quad,\quad\textrm{where}\quad
\EuScript{E} = \begin{pmatrix}0&1\\-1&0\end{pmatrix}
~.
\end{equation}
%%%%%%%%%%%%%%%%%%%%%%%%%%%%%%%%%%%%%%%%
%
Defined as such, their dyadic products satisfy the relation
%
%%%%%%%%%%%%%%%%%%%%%%%%%%%%%%%%%%%%%%%%
\begin{equation}
\Arght{p}\leftS{p} + \Srght{p}\leftA{p} = \slashp = \gamma_\mu p^\mu
~
\end{equation}
%%%%%%%%%%%%%%%%%%%%%%%%%%%%%%%%%%%%%%%%
%
where the $\gamma$-matrices are in the Weyl representation with
%
%%%%%%%%%%%%%%%%%%%%%%%%%%%%%%%%%%%%%%%%
\begin{equation}
\gamma^5 \equiv \imag\gamma^0\gamma^1\gamma^2\gamma^3
=\begin{pmatrix}-\mathbf{1}&\mathbf{0}\\\mathbf{0}&\mathbf{1}\end{pmatrix}
~.
\end{equation}
%%%%%%%%%%%%%%%%%%%%%%%%%%%%%%%%%%%%%%%%
%
For the spinor products, we also use the notation
%
%%%%%%%%%%%%%%%%%%%%%%%%%%%%%%%%%%%%%%%%
\begin{equation}
\leftA{p}\Arght{q} = \ANG{pq} = (pq)_-
\quad,\quad
\leftS{p}\Srght{q} = \SQR{pq} = (pq)_+
~.
\end{equation}
%%%%%%%%%%%%%%%%%%%%%%%%%%%%%%%%%%%%%%%%
%
The most important relations for the spinor products are antisymmetry
%
%%%%%%%%%%%%%%%%%%%%%%%%%%%%%%%%%%%%%%%%
\begin{equation}
\ANG{pq}=-\ANG{qp} \quad,\quad \SQR{pq}=-\SQR{qp}
\end{equation}
%%%%%%%%%%%%%%%%%%%%%%%%%%%%%%%%%%%%%%%%
%
and
%
%%%%%%%%%%%%%%%%%%%%%%%%%%%%%%%%%%%%%%%%
\begin{equation}
\ANG{pq}\SQR{qp} = (p+q)^2
~.
\end{equation}
%%%%%%%%%%%%%%%%%%%%%%%%%%%%%%%%%%%%%%%%
%
For a momentum $\rperp$ transverse to two light-like momenta $\pP$ and $\pM$ that are not collinear, we define
%
%%%%%%%%%%%%%%%%%%%%%%%%%%%%%%%%%%%%%%%%
\begin{equation}
\kstr_r = \frac{\brktAS{\pP|\slashr_\perp|\pM}}{\SQR{\pP\pM}} 
\quad,\quad
\kapp_r = \frac{\brktAS{\pM|\slashr_\perp|\pP}}{\ANG{\pM\pP}} 
~.
\end{equation}
%%%%%%%%%%%%%%%%%%%%%%%%%%%%%%%%%%%%%%%%
%
Realize that $\kapp_r,\kstr_r$ are invariant under replacement of $\pM$ with any other light-like momentum that is not collinear to $\pP$~\cite{vanHameren:2014iua}.
Using polariztion vectors
%
%%%%%%%%%%%%%%%%%%%%%%%%%%%%%%%%%%%%%%%%
\begin{equation}
e_{+}^\mu = \frac{\brktAS{\pM|\gamma^\mu|\pP}}{\sqrt{2}\ANG{\pM\pP}}
\quad,\quad
e_{-}^\mu = \frac{\brktAS{\pP|\gamma^\mu|\pM}}{\sqrt{2}\SQR{\pP\pM}} 
~,
\end{equation}
%%%%%%%%%%%%%%%%%%%%%%%%%%%%%%%%%%%%%%%%
%
satisfying $\lop{e_{-}}{e_{-}}=\lop{e_{+}}{e_{+}}=0$, $\lop{e_{-}}{e_{+}}=-1$, 
we can then write
%
%%%%%%%%%%%%%%%%%%%%%%%%%%%%%%%%%%%%%%%%
\begin{equation}
\rperp^\mu = -\frac{\kstr_r}{\sqrt{2}}\,e_{+}^\mu 
             -\frac{\kapp_r}{\sqrt{2}}\,e_{-}^\mu
~.
\end{equation}
%%%%%%%%%%%%%%%%%%%%%%%%%%%%%%%%%%%%%%%%
%
The inner product of two transverse moment is given by
%
%%%%%%%%%%%%%%%%%%%%%%%%%%%%%%%%%%%%%%%%
\begin{equation}
\lop{\rperp}{\qperp} = \frac{\lop{e_{-}}{e_{+}}}{2}(\kapp_r\kstr_q+\kstr_r\kapp_q)
=-\frac{\kapp_r\kstr_q+\kstr_r\kapp_q}{2}
~.
\end{equation}
%%%%%%%%%%%%%%%%%%%%%%%%%%%%%%%%%%%%%%%%
%

\section{\label{App:060}Color representation}

The relation of \Equation{Eq:007} holds as such after summation over color degrees of freedom of the external partons.
Before squaring and summation, the color-dependend amplitude has a color representation that comes natural with the auxiliary partons used.
In case of auxiliary quarks, the (anti-)fundamental color indices of these constitute the color of the space-like gluon in a color-flow-type of representation.
Let us write the color indices explicitly with an $i$ for the final-state auxiliary quark, and $j$ for the initial-state auxiliary quark.
We consider the situation with $n$ more gluons.
%
%All amplitudes considered in this section will be tree-level, and we do not explicitly label them as such.
%
In \mycite{vanHameren:2012if} it was shown that indeed (employing Einstein summation)
%
%%%%%%%%%%%%%%%%%%%%%%%%%%%%%%%%%%%%%%%%
\begin{equation}
\ColoredAmp^{\star a_1a_2\cdots a_n\,ji}_{\auxq}\delta_{ij} = 0
~,
\end{equation}
%%%%%%%%%%%%%%%%%%%%%%%%%%%%%%%%%%%%%%%%
%
as required for the color pair to represent a single gluon.
The auxiliary label only refers to the color content and regarding the kinematics the $\Lambda$ limit is already performed.
The amplitude can be decomposed in terms of partial amplitudes $\Amp^{\star}$ as
%
%%%%%%%%%%%%%%%%%%%%%%%%%%%%%%%%%%%%%%%%
\begin{equation}
\ColoredAmp^{\star a_1a_2\cdots a_n\,ji}_{\auxq}
=
2^{n/2}\sum_{\sigma\in S_n}\big(T^{a_{\sigma(1)}}T^{a_{\sigma(2)}}\cdots T^{a_{\sigma(n)}}\big)_{ji}
\,\Amp^{\star}(g^\star,\sigma(1),\sigma(2),\ldots,\sigma(n)\big)
~.
\end{equation}
%%%%%%%%%%%%%%%%%%%%%%%%%%%%%%%%%%%%%%%%
%
In~\mycite{vanHameren:2017hxx} it was shown that the partial amplitudes obtained with auxiliary quarks are the same as the ones obtained with auxiliary gluons, that is why we do not give them the auxiliary label.
We immediately see that
%
%%%%%%%%%%%%%%%%%%%%%%%%%%%%%%%%%%%%%%%%
\begin{align}
\ColoredAmp^{\star a_1a_2\cdots a_n\,ji}_{\auxq}\sqrt{2}T^{b}_{ij}
&=2^{(n+1)/2}\sum_{\sigma\in S_n}\Tr\big(T^bT^{a_{\sigma(1)}}T^{a_{\sigma(2)}}\cdots T^{a_{\sigma(n)}}\big)
\,\Amp^{\star}(g^\star,\sigma(1),\sigma(2),\ldots,\sigma(n)\big)
\notag\\&\hspace{0ex}
=\ColoredAmp^{\star ba_1a_2\cdots a_n}
\end{align}
%%%%%%%%%%%%%%%%%%%%%%%%%%%%%%%%%%%%%%%%
%
gives the color decomposition of $(n+1)$-gluon amplitude, one of which happens to be space-like with color index $b$.
The relation can be inverted as, suppressing the irrelevant $a_i$ indices,
%
%%%%%%%%%%%%%%%%%%%%%%%%%%%%%%%%%%%%%%%%
\begin{equation}
\ColoredAmp^{\star b}\sqrt{2}T^b_{kl}
=
\ColoredAmp_{\auxq}^{\star \,ji}2T^{b}_{ij}T^b_{kl}
=
\ColoredAmp^{\star \,ji}_{\auxq}\bigg(\delta_{il}\delta_{kj}-\frac{1}{\Nc}\delta_{ij}\delta_{kl}\bigg)
=
\ColoredAmp^{\star \,kl}_{\auxq}
~.
\end{equation}
%%%%%%%%%%%%%%%%%%%%%%%%%%%%%%%%%%%%%%%%

It must be mentioned that the partial amplitudes for auxiliary gluons are only the same as for auxiliary quarks in case the auxiliary gluons are adjacent.
It is however not difficult to understand that partial amplitudes for which the auxiliary gluons are not adjacent vanish in the large $\Lambda$ limit because they have an extra power of $\Lambda$ in the denominator.
Let the auxiliary gluons be number $n+1$ and $n+2$.
The full amplitude can be decomposed as
%
%%%%%%%%%%%%%%%%%%%%%%%%%%%%%%%%%%%%%%%%
\begin{align}
&\ColoredAmp_{\auxg}^{\star a_1a_2\cdots a_{n+2}} = 
\notag\\&\hspace{5ex}
2^{n/2+1}\sum_{\sigma\in S_{n}}\Tr\big(T^{a_{n+2}}T^{a_{\sigma(1)}}\cdots T^{a_{\sigma(n)}}T^{a_{n+1}}\big)
\Amp_{\auxg}^{\star}(n+2,\sigma(1),\ldots,\sigma(n),n+1\big)
\notag\\&\hspace{2.5ex}
+
2^{n/2+1}\sum_{\sigma\in S_{n}}\Tr\big(T^{a_{n+2}}T^{a_{n+1}}T^{a_{\sigma(1)}}\cdots T^{a_{\sigma(n)}}\big)
\Amp_{\auxg}^{\star}(n+2,n+1,\sigma(1),\ldots,\sigma(n)\big)
~.
\end{align}
%%%%%%%%%%%%%%%%%%%%%%%%%%%%%%%%%%%%%%%%
%
Usually, an $(n+2)$-gluon amplitude has $(n+1)!$ terms, but we excluded the permutations for which the auxiliary gluons are not adjacent, so $2(n!)$ remain.
Also, the partial amplitudes carry the auxiliary label.
In terms of the eikonal Feynman rules for the tree-level space-like amplitude, it is however not difficult to see that exchanging the role of the auxiliary partons effectively means replacing $\pP\leftrightarrow-\pP$, and leads to an overall minus sign.
Thus we can write
%
%%%%%%%%%%%%%%%%%%%%%%%%%%%%%%%%%%%%%%%%
\begin{align}
\ColoredAmp_{\auxg}^{\star a_1a_2\cdots a_{n+2}}
&=
2^{n/2+1}\sum_{\sigma\in S_{n}}\Tr\big(\big[T^{a_{n+1}},T^{a_{n+2}}\big]T^{a_{\sigma(1)}}\cdots T^{a_{\sigma(n)}}\big)
\,\Amp^{\star}(g^\star,\sigma(1),\ldots,\sigma(n)\big)
\notag\\&\hspace{0ex}
=2^{n/2+1}\imag f^{a_{n+1}a_{n+1}b}\sum_{\sigma\in S_{n}}\Tr\big(T^{b}T^{a_{\sigma(1)}}\cdots T^{a_{\sigma(n)}}\big)
\,\Amp^{\star}(g^\star,\sigma(1),\ldots,\sigma(n)\big)
~.
\end{align}
%%%%%%%%%%%%%%%%%%%%%%%%%%%%%%%%%%%%%%%%
%
We conclude that, suppressing irrelevant indices now,
%
%%%%%%%%%%%%%%%%%%%%%%%%%%%%%%%%%%%%%%%%
\begin{equation}
\ColoredAmp_{\auxq}^{\star\,ji}
=\sqrt{2}\,T^{b}_{ij}\,\ColoredAmp^{\star b}
\quad,\quad
\ColoredAmp_{\auxg}^{\star a_{n+1}a_{n+2}}
=\sqrt{2}\,\imag\,f^{a_{n+1}a_{n+2}b}\,\ColoredAmp^{\star b}
~.
\end{equation}
%%%%%%%%%%%%%%%%%%%%%%%%%%%%%%%%%%%%%%%%
%
Squaring and summing over color degrees of freedom, we get
%
%%%%%%%%%%%%%%%%%%%%%%%%%%%%%%%%%%%%%%%%
\begin{equation}
\sum_{\mathrm{color}}\big|\ColoredAmp_{\auxq}^{\star\,ji}\big|^2
=
2T^{b}_{ij}T^{c\dagger}_{ij}\sum_{\mathrm{color}}\ColoredAmp^{\star\,b}\ColoredAmp^{\star\,c\dagger}
=
\sum_{\mathrm{color}}\big|\ColoredAmp^{\star\,b}\big|^2
~,
\end{equation}
%%%%%%%%%%%%%%%%%%%%%%%%%%%%%%%%%%%%%%%%
%
where we used the $\dagger$ to indicate complex conjugation.
Averaging over initial-state color, we get
%
%%%%%%%%%%%%%%%%%%%%%%%%%%%%%%%%%%%%%%%%
\begin{equation}
\frac{1}{\Nc}\sum_{\mathrm{color}}\big|\ColoredAmp_{\auxq}^{\star\,ji}\big|^2
=
\frac{\Nc^2-1}{\Nc}\,\frac{1}{\Nc^2-1}\sum_{\mathrm{color}}\big|\ColoredAmp^{\star\,b}\big|^2
\end{equation}
%%%%%%%%%%%%%%%%%%%%%%%%%%%%%%%%%%%%%%%%
%
were we put the natural color average factor for the space-like gluon.
For the gluon case, we get
%
%%%%%%%%%%%%%%%%%%%%%%%%%%%%%%%%%%%%%%%%
\begin{equation}
\sum_{\mathrm{color}}\big|\ColoredAmp_{\auxg}^{\star\,ab}\big|^2
=
2f^{abe}f^{abd}\sum_{\mathrm{color}}\ColoredAmp^{\star\,e}\ColoredAmp^{\star\,d\dagger}
=
2\Nc\sum_{\mathrm{color}}\big|\ColoredAmp^{\star\,e}\big|^2
~,
\end{equation}
%%%%%%%%%%%%%%%%%%%%%%%%%%%%%%%%%%%%%%%%
%
so the same times $2\Nc$.
Averaging over initial-state color, we get
%
%
%%%%%%%%%%%%%%%%%%%%%%%%%%%%%%%%%%%%%%%%
\begin{equation}
\frac{1}{\Nc^2-1}\sum_{\mathrm{color}}\big|\ColoredAmp_{\auxg}^{\star\,ab}\big|^2
=
2\Nc\,\frac{1}{\Nc^2-1}\sum_{\mathrm{color}}\big|\ColoredAmp^{\star\,c}\big|^2
~.
\end{equation}
%%%%%%%%%%%%%%%%%%%%%%%%%%%%%%%%%%%%%%%%
%
We also need color-correlated \matrixelement{}s.
Highlighting only the color indices for the relevant parton, they are~\mycite{Catani:1996vz}
%
%%%%%%%%%%%%%%%%%%%%%%%%%%%%%%%%%%%%%%%%
\begin{equation}
\textrm{quark:}\;\ColoredAmp^{i\dagger}T^a_{i\bar{\imath}}\ColoredAmp^{\bar{\imath}}
\quad,\quad
\textrm{antiquark:}\;\ColoredAmp^{j\dagger}\big({-}T^a_{\bar{\jmath}j}\big)\ColoredAmp^{\bar{\jmath}}
\quad,\quad
\textrm{gluon:}\;\ColoredAmp^{b\dagger}\big(\imag f^{bac}\big)\ColoredAmp^{c}
~.
\end{equation}
%%%%%%%%%%%%%%%%%%%%%%%%%%%%%%%%%%%%%%%%
%
%
For the sum of \matrixelement{}s with color correlators for the auxiliary lines with equal coefficients (put to $1$ here), we get
%
%%%%%%%%%%%%%%%%%%%%%%%%%%%%%%%%%%%%%%%%
\begin{align}
\ColoredAmp_{\auxq}^{\star\,ji\,\dagger}T^{a}_{i\bar{\i}}\ColoredAmp_{\auxq}^{\star\,j\bar{\i}}
+
\ColoredAmp_{\auxq}^{\star\,ji\,\dagger}\big({-}T^{a}_{\bar{\jmath}j}\big)\ColoredAmp_{\auxq}^{\star\,\bar{\jmath}i}
&=
2\ColoredAmp^{\star\,b\dagger}T_{ij}^{b\dagger}T^{a}_{i\bar{\i}}T^{c}_{\bar{\imath}j}\ColoredAmp^{\star\,c}
-
2\ColoredAmp^{\star\,b\dagger}T^{b\dagger}_{ij}T^{a}_{\bar{\jmath}j}T^{c}_{i\bar{\jmath}}\ColoredAmp^{\star\,c}
% \notag\\&
% =
% 2\ColoredAmp^{\star\,b\dagger}\Tr\big(T^b[T^a,T^c]\big)\ColoredAmp^{\star\,c}
\notag\\&
=\ColoredAmp^{\star\,b\dagger}\big(\imag f^{bac}\big)\ColoredAmp^{\star\,c}
~,
\end{align}
%%%%%%%%%%%%%%%%%%%%%%%%%%%%%%%%%%%%%%%%
%
as desired.
The one term with both auxiliary quark correlators is given by
%
%%%%%%%%%%%%%%%%%%%%%%%%%%%%%%%%%%%%%%%%
\begin{align}
\ColoredAmp_{\auxq}^{\star\,ji\,\dagger}T^{a}_{i\bar{\i}}\big({-}T^{a}_{\bar{\jmath}j}\big)\ColoredAmp_{\auxq}^{\star\,\bar{\jmath}\bar{\i}}
&=
-2\ColoredAmp^{\star\,b\dagger}
  T_{ij}^{b\dagger}
%  T^{a}_{i\bar{\i}}T^{a}_{\bar{\jmath}j}
  \frac{1}{2}\bigg(\delta_{ij}\delta_{\bar{\imath}\bar{\jmath}}-\frac{1}{\Nc}\delta_{i\bar{\imath}}\delta_{j\bar{\jmath}}\bigg)
  T^{c}_{\bar{\imath}\bar{\jmath}}
  \ColoredAmp^{\star\,c}
% \notag\\&
% =\frac{1}{\Nc}\ColoredAmp^{\star\,b\dagger}
%   T^{b\dagger}_{\bar{\imath}j}
%   T^{c}_{\bar{\imath}j}
%   \ColoredAmp^{\star\,c}
=
\frac{1}{2\Nc}\ColoredAmp^{\star\,b\dagger}\ColoredAmp^{\star\,b}
~.
\end{align}
%%%%%%%%%%%%%%%%%%%%%%%%%%%%%%%%%%%%%%%%
%
%
With auxiliary gluons, we get, using the Jacobi identity,
%
%%%%%%%%%%%%%%%%%%%%%%%%%%%%%%%%%%%%%%%%
\begin{align}
&
\ColoredAmp_{\auxg}^{\star\,b_1b_2\,\dagger}\big(\imag f^{b_1ac_1}\big)\ColoredAmp_{\auxg}^{\star\,c_1b_2}
+
\ColoredAmp_{\auxg}^{\star\,b_1b_2\,\dagger}\big(\imag f^{b_2ac_2}\big)\ColoredAmp_{\auxg}^{\star\,b_1c_2}
\notag\\&\hspace{20ex}
=
2\imag\ColoredAmp^{\star\,d\,\dagger}f^{b_1b_2d}f^{b_1ac_1}f^{c_1b_2e}\ColoredAmp^{\star\,e}
+
2\imag\ColoredAmp^{\star\,d\,\dagger}f^{b_1b_2d}f^{b_2ac_2}f^{b_1c_2e}\ColoredAmp^{\star\,e}
\notag\\&\hspace{20ex}
% =
% 2\imag\ColoredAmp^{\star\,d\,\dagger}f^{b_1b_2d}\big(f^{b_1ac_1}f^{b_2ec_1}
% +
% f^{b_2ac_1}f^{eb_1c_1}\big)\ColoredAmp^{\star\,e}
% \notag\\&\hspace{20ex}
% =
% 2\imag\ColoredAmp^{\star\,d\,\dagger}f^{b_1b_2d}f^{eac_1}f^{b_2b_1c_1}\ColoredAmp^{\star\,e}
% \notag\\&\hspace{20ex}
% =
% -2\Nc\imag\ColoredAmp^{\star\,d\,\dagger}f^{ead}\ColoredAmp^{\star\,e}
=
2\Nc\,\ColoredAmp^{\star\,d\,\dagger}\big(\imag f^{dae}\big)\ColoredAmp^{\star\,e}
~,
\end{align}
%%%%%%%%%%%%%%%%%%%%%%%%%%%%%%%%%%%%%%%%
%
so again $2\Nc$ times the desired result, like for the non-correlated square.
The term with both correlators is
%
%
%%%%%%%%%%%%%%%%%%%%%%%%%%%%%%%%%%%%%%%%
\begin{align}
\ColoredAmp_{\auxg}^{\star\,b_1b_2\,\dagger}
\big(\imag f^{b_1ac_1}\big)
\big(\imag f^{b_2ac_2}\big)
\ColoredAmp_{\auxg}^{\star\,c_1c_2}
&=
-2\ColoredAmp^{\star\,d\,\dagger}
f^{b_1b_2d}f^{b_1ac_1}f^{b_2ac_2}f^{c_1c_2e}
\ColoredAmp^{\star\,e}
\notag\\&\hspace{0ex}
% =
% -2\ColoredAmp^{\star\,d\,\dagger}
% f^{b_2db_1}f^{b_1ac_1}f^{c_1ec_2}f^{c_2ab_2}
% \ColoredAmp^{\star\,e}
% \notag\\&\hspace{0ex}
% =
% -\Nc^2\ColoredAmp^{\star\,d\dagger}\ColoredAmp^{\star\,d}
=
2\Nc\,\frac{-\Nc}{2}\,\ColoredAmp^{\star\,d\dagger}\ColoredAmp^{\star\,d}
~.
\end{align}
%%%%%%%%%%%%%%%%%%%%%%%%%%%%%%%%%%%%%%%%

\section{\label{App:020}Collinear splitting functions}
The collinear limit of a momentum pair $p_i^\mu,p_j^\mu$ in an amplitude can be derived by factoring out the obvious singular part, and then replacing
%
%%%%%%%%%%%%%%%%%%%%%%%%%%%%%%%%%%%%%%%%
\begin{equation}
  p_i^\mu \to (1-z)p_{ij}^\mu
\quad,\quad
  p_j^\mu \to zp_{ij}^\mu
~.
\end{equation}
%%%%%%%%%%%%%%%%%%%%%%%%%%%%%%%%%%%%%%%%
%
Consider for example the MHV amplitude (we omit the overall imaginary unit)
%
%%%%%%%%%%%%%%%%%%%%%%%%%%%%%%%%%%%%%%%%
\begin{equation}
\Amp_n\big(1^-,2^+,3^-,4^+,\ldots,n^+\big)
=
\frac{\ANG{p_1p_3}^4}{\ANG{p_1p_2}\ANG{p_2p_3}\ANG{p_3p_4}\cdots\ANG{p_{n-1}p_n}\ANG{p_np_1}}
\end{equation}
%%%%%%%%%%%%%%%%%%%%%%%%%%%%%%%%%%%%%%%%
%
which is singular when $p_1^\mu,p_2^\mu$ become collinear.
Performing the procedure mentioned above, we get
%
%%%%%%%%%%%%%%%%%%%%%%%%%%%%%%%%%%%%%%%%
\begin{multline}
\Amp_n\big(1^-,2^+,3^-,4^+,\ldots,n^+\big)
\to
\frac{1}{\ANG{p_1p_2}}\,\frac{(1-z)^2\ANG{p_{12}p_3}^4}{\sqrt{z}\ANG{p_{12}p_3}\ANG{p_3p_4}\cdots\ANG{p_{n-1}p_n}\sqrt{1-z}\ANG{p_np_{12}}}
\\
= \frac{1}{\ANG{p_1p_2}}\,\frac{(1-z)^2}{\sqrt{z(1-z)}}\,\Amp_{n-1}\big((12)^-,3^-,4^+,\ldots,n^+\big)
\end{multline}
%%%%%%%%%%%%%%%%%%%%%%%%%%%%%%%%%%%%%%%%
%
which corresponds to Eq.(6.15)~in~\mycite{Mangano:1990by} (because the all-plus amplitude vanishes).

Let us consider a slightly more general procedure, in which we take the limits $p_i^\mu\to xp^\mu$, $p_j^\mu\to yp^\mu$, and then correct the factorized amplitude so it contains momentum $(x+y)p^\mu$.
For the example above we get
\begin{multline}
\Amp_n\big(1^-,2^+,3^-,4^+,\ldots,n^+\big)
\to
\frac{1}{\ANG{p_1p_2}}\,\frac{x^2}{\sqrt{xy}}\,\frac{\ANG{pp_3}^4}{\ANG{pp_3}\ANG{p_3p_4}\cdots\ANG{p_{n-1}p_n}\ANG{p_np}}
\\
= \frac{1}{\ANG{p_1p_2}}\,\frac{x^2}{\sqrt{xy}(x+y)}\,\frac{\ANG{(x+y)pp_3}^4}{\ANG{(x+y)pp_3}\ANG{p_3p_4}\cdots\ANG{p_{n-1}p_n}\ANG{p_n(x+y)p}}
\\
= \frac{1}{\ANG{p_1p_2}}\,\frac{x^2}{\sqrt{xy}(x+y)}\,\Amp_{n-1}\big((x+y)p^-,3^-,4^+,\ldots,n^+\big)
~.
\end{multline}
We get the previous formula by simply setting $x=1-z$ and $y=z$.
In our definition of the spinors, they scale with not-necessarily positive fractions as
%
%%%%%%%%%%%%%%%%%%%%%%%%%%%%%%%%%%%%%%%%
\begin{equation}
\Arght{xp} = \sqrt{|x|}\,\Arght{p}
\quad,\quad
\Srght{xp} = \frac{x}{\sqrt{|x|}}\,\Srght{p}
= \sgn(x)\sqrt{|x|}\,\Srght{p}
~,
\end{equation}
%%%%%%%%%%%%%%%%%%%%%%%%%%%%%%%%%%%%%%%%
%
so for the previous splitting function, the fractions $x$, $y$, and $(x+y)$ just need to be replaced by their absolute values.
Let us also consider the explicit example with an anti-MHV amplitude.
We get
\begin{multline}
\Amp_n\big(1^-,2^+,3^+,4^-,\ldots,n^-\big)
=
\frac{\SQR{p_3p_{2}}^4}{\SQR{p_{1}p_n}\SQR{p_np_{n-1}}\cdots\SQR{p_4p_3}\SQR{p_3p_{2}}\SQR{p_2p_1}}
\\
\to
\frac{1}{\SQR{p_2p_1}}\,\frac{\sgn(xy)y^2}{\sqrt{|xy|}}
\,\frac{\SQR{p_3p}^4}{\SQR{pp_n}\SQR{p_np_{n-1}}\cdots\SQR{p_4p_3}\SQR{p_3p}}
\\
=
\frac{-1}{\SQR{p_1p_2}}\,\frac{\sgn(xy)y^2}{\sqrt{|xy|}\,|x+y|}
\,\Amp_{n-1}\big((x+y)p^+,3^+,4^-,\ldots,n^-\big)
~.
\end{multline}
We can summarize
%
%%%%%%%%%%%%%%%%%%%%%%%%%%%%%%%%%%%%%%%%
\begin{align}
\mathrm{Split}^{g\to g_1g_2}\big((x+y)^-\to 1_x^-,2_y^+\big)
&=
\frac{1}{(12)_-}\,\frac{x^2}{\sqrt{|xy|}\,|x+y|}
\\
\mathrm{Split}^{g\to g_1g_2}\big((x+y)^+\to 1_x^-,2_y^+\big)
&=
\frac{-1}{(12)_+}\,\frac{\sgn(xy)\,y^2}{\sqrt{|xy|}\,|x+y|}
~,
\end{align}
%%%%%%%%%%%%%%%%%%%%%%%%%%%%%%%%%%%%%%%%
%
where the $(1_x^+,2_y^-)$ functions are obtained by exchanging $x^2\leftrightarrow y^2$ in the numerators.
Realize that here the splitting functions indicate the amplitude helicities, while in \mycite{Bern:1999ry} the collinear limit is indicated with the opposite helicity.
The other non-vanishing ones are
%
%%%%%%%%%%%%%%%%%%%%%%%%%%%%%%%%%%%%%%%%
\begin{align}
\mathrm{Split}^{g\to g_1g_2}\big((x+y)^+\to 1_x^+,2_y^+\big)
&=
\frac{1}{(12)_-}\,\frac{1}{\sqrt{|xy|}\,|x+y|}
\\
\mathrm{Split}^{g\to g_1g_2}\big((x+y)^-\to 1_x^-,2_y^-\big)
&=
\frac{-1}{(12)_+}\,\frac{\sgn(xy)}{\sqrt{|xy|}\,|x+y|}
~.
\end{align}
%%%%%%%%%%%%%%%%%%%%%%%%%%%%%%%%%%%%%%%%
%
For squared amplitudes summed over helicities, the gluon splitting function becomes
%
%%%%%%%%%%%%%%%%%%%%%%%%%%%%%%%%%%%%%%%%
\begin{equation}
\EuScript{Q}_{gg}(x,y)
= \frac{ x^4 +y^4 +(x+y)^4 }{|xy|(x+y)^2}
= \frac{ 1 +(y/x)^4 +(1+y/x)^4 }{|y/x|(1+y/x)^2}
=\frac{\EuScript{P}_{gg}\big(|1+y/x|\big)}{|1+y/x|}
~.
\label{Eq:035}
\end{equation}
%%%%%%%%%%%%%%%%%%%%%%%%%%%%%%%%%%%%%%%%
%
For the splitting of a quark or anti-quark into a quark or anti-quark and a gluon we with $p_{q/\bar{q}}\to xp_{q/\bar{q}}$ and $p_g\to yp_{q/\bar{q}}$ are given by
%
%%%%%%%%%%%%%%%%%%%%%%%%%%%%%%%%%%%%%%%%
\begin{align}
\mathrm{Split}(q_{x+y}^\mp\to q_x^\pm\,g_y^\pm) &= \sigma_\mp(y)\,\frac{\pm\sqrt{|x+y|}}{\sqrt{|y|}\,(p_qp_g)_\mp}
\\
\mathrm{Split}(q_{x+y}^\mp\to q_x^\pm\,g_y^\mp) &= \sigma_\pm(y)\,\frac{\mp|x|}{\sqrt{|y||x+y|}\,(p_qp_g)_\pm}
\\
\mathrm{Split}(\bar{q}_{x+y}^\pm\to \bar{q}_x^\mp\,g_y^\mp) &= \sigma_\pm(y)\,\frac{\mp\sqrt{|x+y|}}{\sqrt{|y|}\,(p_gp_{\bar{q}})_\pm}
\\
\mathrm{Split}(\bar{q}_{x+y}^\pm\to \bar{q}_x^\mp\,g_y^\pm) &= \sigma_\mp(y)\,\frac{\pm|x|}{\sqrt{|y||x+y|}\,(p_gp_{\bar{q}})_\mp}
\end{align}
%%%%%%%%%%%%%%%%%%%%%%%%%%%%%%%%%%%%%%%%
%
where
%%%%%%%%%%%%%%%%%%%%%%%%%%%%%%%%%%%%%%%%
\begin{equation}
\sigma_+(y) = 1 \quad,\quad \sigma_-(y) = \sgn(y)
~,
\end{equation}
%%%%%%%%%%%%%%%%%%%%%%%%%%%%%%%%%%%%%%%%
%
leading to the splitting function
%
%%%%%%%%%%%%%%%%%%%%%%%%%%%%%%%%%%%%%%%%
\begin{equation}
\EuScript{Q}_{qq}(x,y) 
= \frac{1}{|y|}\left(|x+y|+\frac{x^2}{|x+y|}\right)
= \frac{1+(1+y/x)^2}{|y/x||1+y/x|}
= \frac{\EuScript{P}_{qq}\big(|1+y/x|\big)}{|1+y/x|}
~.
\end{equation}
%%%%%%%%%%%%%%%%%%%%%%%%%%%%%%%%%%%%%%%%
%
The remaining splitting functions are given by
%
%%%%%%%%%%%%%%%%%%%%%%%%%%%%%%%%%%%%%%%%
\begin{align}
\mathrm{Split}(g_{x+y}^\mp\to \bar{q}_x^\pm\,q_y^\mp) &=
\frac{1}{(p_{\bar{q}}p_{q})_\mp}\,\frac{|y|}{|x+y|}
\\
\mathrm{Split}(g_{x+y}^\mp\to \bar{q}_x^\mp\,q_y^\pm) &=
\frac{1}{(p_{\bar{q}}p_{q})_\mp}\,\frac{|x|}{|x+y|}
~.
\end{align}
%%%%%%%%%%%%%%%%%%%%%%%%%%%%%%%%%%%%%%%%

\subsection{\label{App:030}Collinear limit with $\pP$ for space-like amplitudes}
We extract the splitting functions by condidering the example of MHV amplitudes.
For a space-like gluon with momentum $k=p_1+\kperp$ we have~\mycite{vanHameren:2014iua},
%
%%%%%%%%%%%%%%%%%%%%%%%%%%%%%%%%%%%%%%%%
\begin{equation}
\Amp_n^\star\big(p_1+\kperp,2^+,3^-,4^+,\ldots,n^+\big)
=
\frac{1}{\kstr}\frac{\ANG{p_1p_3}^4}{\ANG{p_1p_2}\ANG{p_2p_3}\ANG{p_3p_4}\cdots\ANG{p_{n-1}p_n}\ANG{p_np_1}}
~.
\end{equation}
%%%%%%%%%%%%%%%%%%%%%%%%%%%%%%%%%%%%%%%%
%
As such, the amplitude is singlular for $\kperp\to0$ and still requires a factor $|\kperp|$.
We, however, prefer to define amplitudes like this and not to break their analytic structure.
The necessary factor is included when constructing the \matrixelement.

We can easily extract the limit $p_1\to x\pP$ and $p_2\to y\pP$
%
%
%%%%%%%%%%%%%%%%%%%%%%%%%%%%%%%%%%%%%%%%
\begin{align}
&\Amp_n^\star\big(p_1+\kperp,2^+,3^-,4^+,\ldots,n^+\big)
\notag\\&
\overset{p_1\to x\pP,\,p_2\to y\pP}{\longrightarrow}
\frac{1}{\ANG{p_1p_2}}
\frac{1}{\kstr}\frac{x^2\ANG{\pP p_3}^4}{\sqrt{|y|}\ANG{\pP p_3}\ANG{p_3p_4}\cdots\ANG{p_{n-1}p_n}\sqrt{|x|}\ANG{p_n\pP}}
\notag\\&\hspace{4ex}
=\frac{1}{\ANG{p_1p_2}}\,\frac{x^2}{\sqrt{|xy|}|x+y|}
\,\frac{1}{\kstr}\frac{\ANG{(x+y)\pP|p_3}^4}{\ANG{(x+y)\pP|p_3}\ANG{p_3p_4}\cdots\ANG{p_{n-1}p_n}\ANG{p_n|(x+y)\pP}}
\notag\\&\hspace{4ex}
=\frac{1}{\ANG{\pP|p_2}}\,\frac{|x|}{\sqrt{|y|}|x+y|}
 \,\Amp_{n-1}^\star\big((x+y)\pP+\kperp,3^-,4^+,\ldots,n^+\big)
~.
\end{align}
%%%%%%%%%%%%%%%%%%%%%%%%%%%%%%%%%%%%%%%%
%
For the opposite helicity there is $\sqrt{|y|}/y$ instead of $1/\sqrt{|y|}$, and also $1/\SQR{p_2\pP}=-1/\SQR{\pP p_2}$.
Thus we find
%
%%%%%%%%%%%%%%%%%%%%%%%%%%%%%%%%%%%%%%%%
\begin{align}
\mathrm{Split}\big(\,g^\star((x+y)\pP+\kperp) \to g^\star(x\pP+\kperp),g^+(y\pP)\,\big)
&=
\frac{1}{(\pP r)_-}\,\frac{|x|}{\sqrt{|y|}\,|x+y|}
\\
\mathrm{Split}\big(\,g^\star((x+y)\pP+\kperp) \to g^\star(x\pP+\kperp),g^-(y\pP)\,\big)
&=
\frac{-1}{(\pP r)_+}\,\frac{\sgn(y)|x|}{\sqrt{|y|}\,|x+y|}
~.
\end{align}
%%%%%%%%%%%%%%%%%%%%%%%%%%%%%%%%%%%%%%%%
%
For \matrixelement{}s with a space-like initial-state gluon, this leads to the collinear limit for a final-state gluon with momentum $r$ (and taking $\gQCD=1$)
%
%%%%%%%%%%%%%%%%%%%%%%%%%%%%%%%%%%%%%%%%
\begin{equation}
\MtreeStar\big(\xP_k,\kperp,\xM\,;r,\{p_i\}_{i=1}^n\big)
\overset{r\to \xP_r\pP}{\longrightarrow}
2\Nc\,\frac{1}{\lop{\pP}{r}}\,\frac{\xP_k^2}{\xP_r(\xP_k-\xP_r)^2}
\,\MtreeStar\big((\xP_k-\xP_r),\kperp,\xM\,;\{p_i\}_{i=1}^n\big)
~.
\end{equation}
%%%%%%%%%%%%%%%%%%%%%%%%%%%%%%%%%%%%%%%%
%
At first glance it seems that these formulas cannot be obtained by taking the large-$\Lambda$ after the collinear limit, from the expressions of the previous section with $x\to\pm\Lambda$, seemingly only producing the factor $1/\sqrt{|y|}$.
Realize, however, that the definition of the space-like amplitudes requires an extra factor $x$ from the space-like gluon momentum $k=x\pP+\kperp$ (the factor $\sqrt{\xiPin^2}$ in \Equation{Eq:007}), resulting in the extra factor $|x|/|x+y|$.

\section{\label{App:050}The triple auxiliary parton limit}
Let us start with a simple example of the auxiliary parton method applied to amplitudes.
Consider the partial amplitude for a process involving a quark-antiquark pair and $3$ gluons.
%
%%%%%%%%%%%%%%%%%%%%%%%%%%%%%%%%%%%%%%%%
\begin{equation}
\Amp({q}^{-},g_1^{-},g_2^{+},g_3^{+},\bar{q}^{+})
=
\frac{\ANG{q1}^3\ANG{\bar{q}1}}{\ANG{\bar{q}q}\ANG{q1}\ANG{12}\ANG{23}\ANG{3\bar{q}}}
~.
\label{Eq:036}
\end{equation}
%%%%%%%%%%%%%%%%%%%%%%%%%%%%%%%%%%%%%%%%
%
The quark-antiquark pair take the role of auxiliary partons, which means they get the spinors
%
%%%%%%%%%%%%%%%%%%%%%%%%%%%%%%%%%%%%%%%%
\begin{equation}
\Arght{\bar{q}} = \sqrt{\Lambda+\xP}\,\Arght{P}
\quad,\quad
\Arght{q} = \sqrt{\Lambda}\,\Arght{\pP}
          + \frac{\kstr}{\sqrt{\Lambda}\,\ANG{\pP\pM}}\,\Arght{\pM}
~.
\end{equation}
%%%%%%%%%%%%%%%%%%%%%%%%%%%%%%%%%%%%%%%%
%
We shifted $\Lambda$ by an amount $x$ for convenience.
Contrary to the construction in~\mycite{Blanco:2020akb}, these spinors will give the sum of the quark and antiquark momenta a non-zero component for $\pM^\mu$, which however still vanishes for $\Lambda\to\infty$.
Important is that
%
%%%%%%%%%%%%%%%%%%%%%%%%%%%%%%%%%%%%%%%%
\begin{equation}
\ANG{\bar{q}q}=\kstr +\OrdL{-1}
\end{equation}
%%%%%%%%%%%%%%%%%%%%%%%%%%%%%%%%%%%%%%%%
%
does not vanish or become large for $\Lambda\to\infty$, and the amplitude scales as $\Lambda$ due to the occurrence of $\Arght{\bar{q}}$ and $\Arght{q}$ in other spinor products.
Inserting the spinors, we find
%
%
%%%%%%%%%%%%%%%%%%%%%%%%%%%%%%%%%%%%%%%%
\begin{align}
\Amp({q}^{-},g_1^{-},g_2^{+},g_3^{+},\bar{q}^{+})
&\to
\frac{\Lambda\,\ANG{\pP1}^4}{\kstr\ANG{\pP1}\ANG{12}\ANG{23}\ANG{3\pP}}
=
\Lambda\,\Amp^\star(g^\star,g_1^{-},g_2^{+},g_3^{+})
~.
\end{align}
%%%%%%%%%%%%%%%%%%%%%%%%%%%%%%%%%%%%%%%%
%
For the triple-parton limit, we distribute $\Lambda$ over the quark and a gluon with spinors
%
%%%%%%%%%%%%%%%%%%%%%%%%%%%%%%%%%%%%%%%%
\begin{align}
\Arght{q} = \sqrt{z\Lambda}\,\Arght{\pP}
          + \frac{\kstr_q}{\sqrt{z\Lambda}\,\ANG{\pP\pM}}\,\Arght{\pM}
\;\;,\;\;
\Arght{r} = \sqrt{(1-z)\Lambda}\,\Arght{\pP}
          + \frac{\kstr_r}{\sqrt{(1-z)\Lambda}\,\ANG{\pP\pM}}\,\Arght{\pM}
~.
\end{align}
%%%%%%%%%%%%%%%%%%%%%%%%%%%%%%%%%%%%%%%%
%
We see that if we choose $r$ to refer to $2$ in \Equation{Eq:036}, then it will produce a factor $\Lambda$ in the denominator and give a subleading contribution.
If we choose $1$, then the spinor products in the numerator will not produce powers of $\Lambda$.
In order to produce a leading contribution, it must refer to $3$, and inserting the spinors, we get
%
%%%%%%%%%%%%%%%%%%%%%%%%%%%%%%%%%%%%%%%%
\begin{align}
\Amp({q}^{-},g_1^{-},g_2^{+},g_r^{+},\bar{q}^{+})
&=\frac{\ANG{q1}^3\ANG{\bar{q}1}}{\ANG{\bar{q}q}\ANG{q1}\ANG{12}\ANG{2r}\ANG{r\bar{q}}}
\notag\\&\to
\frac{\big(\sqrt{z\Lambda}\,\ANG{\pP1}\big)^3\big(\sqrt{\Lambda}\,\ANG{\pP1}\big)}{\frac{\kstr_q}{\sqrt{z}}\big(\sqrt{z\Lambda}\ANG{\pP1}\big)\ANG{12}\big(\sqrt{(1-z)\Lambda}\,\ANG{2\pP}\big)\frac{-\kstr_r}{\sqrt{1-z}}}
\notag\\&=
\frac{-z^{3/2}\,\Lambda\,\ANG{\pP1}^4}{\kstr_q\kstr_r\ANG{\pP1}\ANG{12}\ANG{2\pP}}
=
-z^{3/2}\,\frac{\kstr_q+\kstr_r}{\kstr_q\kstr_r}\,\Lambda\,\Amp(g^\star,g_1^{-},g_2^{+})
~,
\end{align}
%%%%%%%%%%%%%%%%%%%%%%%%%%%%%%%%%%%%%%%%
%
where the factor $\kstr_q+\kstr_r$ occurs, because the space-like gluon in $A(g^\star,g_1^{-},g_2^{+})$ has transverse momentum $\qperp^\mu+\rperp^\mu$, and thus the amplitude a factor $\kstr_{q+r}=\kstr_q+\kstr_r$ in the denominator.

Below we give all possible ``splitting functions''.
The complete construction of momenta and spinors is
%
%%%%%%%%%%%%%%%%%%%%%%%%%%%%%%%%%%%%%%%%
\begin{equation}
K_{\bar{q}}^\mu = -(\xP+\Lambda)\pP^\mu
\;\;,\;\;
K_q^\mu = \zq\Lambda\pP^\mu + \qperp^\mu - \frac{\qperp^2}{s\zq\Lambda}\pM^\mu
\;\;,\;\;
K_r^\mu = \zr\Lambda\pP^\mu + \rperp^\mu - \frac{\rperp^2}{s\zr\Lambda}\pM^\mu
~,
\end{equation}
%%%%%%%%%%%%%%%%%%%%%%%%%%%%%%%%%%%%%%%%
%
with
%
%%%%%%%%%%%%%%%%%%%%%%%%%%%%%%%%%%%%%%%%
\begin{equation}
\zr+\zq=1
~.
\end{equation}
%%%%%%%%%%%%%%%%%%%%%%%%%%%%%%%%%%%%%%%%
%
So we have, to $\OrdL{-1}$, 
%
%%%%%%%%%%%%%%%%%%%%%%%%%%%%%%%%%%%%%%%%
\begin{equation}
K_{\bar{q}}^\mu + K_q^\mu + K_r^\mu = -x\pP^\mu + \qperp^\mu + \rperp^\mu 
\quad,\quad
(K_{\bar{q}} + K_q + K_r)^2 = (\qperp + \rperp)^2
\label{Eq:037}
\end{equation}
%%%%%%%%%%%%%%%%%%%%%%%%%%%%%%%%%%%%%%%%
%
and
%
%%%%%%%%%%%%%%%%%%%%%%%%%%%%%%%%%%%%%%%%
\begin{gather}
s_{\qL\qB} = (K_{\bar{q}} + K_q)^2 = \frac{\qperp^2}{\zq}
\;\;,\;\;
s_{\rL\qB} = (K_{\bar{q}} + K_r)^2 = \frac{\rperp^2}{\zr}
\\
s_{\rL\qL} = (K_{q} + K_r)^2 = (\qperp+\rperp)^2-\frac{\qperp^2}{\zq}-\frac{\rperp^2}{\zr}
\;\;.
\end{gather}
%%%%%%%%%%%%%%%%%%%%%%%%%%%%%%%%%%%%%%%%
%
It is important to observe that these three invariants do not depend on $\Lambda$ and do not produce a factor of $\Lambda$ in propagator denominators.
The spinors of the momenta are give by
%
%%%%%%%%%%%%%%%%%%%%%%%%%%%%%%%%%%%%%%%%
\begin{align}
\Arght{\bar{q}} &= \sqrt{\Lambda+x}\,\Arght{\pP}
  \hspace{20ex}
  \Srght{\bar{q}} =-\sqrt{\Lambda+x}\,\Srght{\pP}
\\
\Arght{q} &= \sqrt{\zq\Lambda}\,\Arght{\pP} + \frac{\kstr_q}{\sqrt{\zq\Lambda}\ANG{\pP\pM}}\,\Arght{\pM}
  \hspace{4.9ex}
  \Srght{q} = \sqrt{\zq\Lambda}\,\Srght{\pP} + \frac{\kapp_q}{\sqrt{\zq\Lambda}\SQR{\pM\pP}}\,\Srght{\pM}
\\
\Arght{r} &= \sqrt{\zr\Lambda}\,\Arght{\pP} + \frac{\kstr_r}{\sqrt{\zr\Lambda}\ANG{\pP\pM}}\,\Arght{\pM}
  \hspace{5.7ex}
  \Srght{r} = \sqrt{\zr\Lambda}\,\Srght{\pP} + \frac{\kapp_r}{\sqrt{\zr\Lambda}\SQR{\pM\pP}}\,\Srght{\pM}
\end{align}
%%%%%%%%%%%%%%%%%%%%%%%%%%%%%%%%%%%%%%%%
%
with products
%
%%%%%%%%%%%%%%%%%%%%%%%%%%%%%%%%%%%%%%%%
\begin{align}
\ANG{q\bar{q}} = -\frac{\kstr_q}{\sqrt{\zq}}
\quad&,\quad
\SQR{\bar{q}q} = \frac{\kapp_q}{\sqrt{\zq}}
\notag\\
\ANG{r\bar{q}} = -\frac{\kstr_r}{\sqrt{\zr}}
\quad&,\quad
\SQR{\bar{q}r} = \frac{\kapp_r}{\sqrt{\zr}}
\label{Eq:038}\\
\ANG{rq} = \sqrt{\frac{\zr}{\zq}}\,\kstr_q - \sqrt{\frac{\zq}{\zr}}\,\kstr_r
\quad&,\quad
\SQR{qr} = \sqrt{\frac{\zr}{\zq}}\,\kapp_q - \sqrt{\frac{\zq}{\zr}}\,\kapp_r
\notag
\end{align}
%%%%%%%%%%%%%%%%%%%%%%%%%%%%%%%%%%%%%%%%
%
It is convenient to introducing the functions
%
%%%%%%%%%%%%%%%%%%%%%%%%%%%%%%%%%%%%%%%%
\begin{equation}
\Sqr(\qB,\qL,\rL) = \frac{\kapp_q+\kappa_r}{\SQR{\qB\qL}\SQR{\qL\rL}\SQR{\rL\qB}}
\quad,\quad
\Ang(\qB,\qL,\rL) = \frac{\kstr_q+\kstr_r}{\ANG{\qB\qL}\ANG{\qL\rL}\ANG{\rL\qB}}
~.
\end{equation}
%%%%%%%%%%%%%%%%%%%%%%%%%%%%%%%%%%%%%%%%
%

\subsection{Auxiliary quark-antiquark pair plus a gluon}
We have
%
%%%%%%%%%%%%%%%%%%%%%%%%%%%%%%%%%%%%%%%%
\begin{align}
\Amp(\qL,g_\rL,\ldots,\qB) &\to \Lambda\,\Splt(\qB,\qL,g_\rL)\,\Amp^\star(g^\star,\ldots)
\\
\Amp(\qL,\ldots,g_\rL,\qB) &\to \Lambda\,\Splt(g_\rL,\qB,\qL)\,\Amp^\star(g^\star,\ldots)
\end{align}
%%%%%%%%%%%%%%%%%%%%%%%%%%%%%%%%%%%%%%%%
%
The logic in the arguments of $\Splt$ is that $g$ is next to $\qB$ or next to $\qL$ while keeping the cyclic order.
The function $\Splt$ for non-vanishing helicities are
%
%%%%%%%%%%%%%%%%%%%%%%%%%%%%%%%%%%%%%%%%
\begin{align}
 \Splt(\qB^+,\qL^-,g_\rL^-) &=
  \Sqr(\qB,\qL,\rL)
  \,\frac{\zq^{1/2}\SQR{\rL\qB}}{\sqrt{\zr}}
\quad,\quad\hspace{1.0ex}
 \Splt(\qB^-,\qL^+,g_\rL^+) =
  \Ang(\qB,\qL,\rL)
  \,\frac{\zq^{1/2}\ANG{\rL\qB}}{\sqrt{\zr}}
\quad,
\\
 \Splt(\qB^+,\qL^-,g_\rL^+) &=
  \Ang(\qB,\qL,\rL)
  \,\frac{\zq^{3/2}\ANG{\rL\qB}}{\sqrt{\zr}}
\quad,\quad\hspace{0.0ex}
 \Splt(\qB^-,\qL^+,g_\rL^-) =
  \Sqr(\qB,\qL,\rL)
  \,\frac{\zq^{3/2}\SQR{\rL\qB}}{\sqrt{\zr}}
\quad,
\\
 \Splt(g_\rL^-,\qB^+,\qL^-) &=
  \Sqr(\qB,\qL,\rL)
  \,\frac{-\SQR{\qL\rL}}{\sqrt{\zr}}
\quad,\quad\hspace{2.9ex}
 \Splt(g_\rL^+,\qB^-,\qL^+) =
  \Ang(\qB,\qL,\rL)
  \,\frac{\ANG{\qL\rL}}{\sqrt{\zr}}
\quad,
\\
 \Splt(g_\rL^+,\qB^+,\qL^-) &=
  \Ang(\qB,\qL,\rL)
  \,\frac{\zq\ANG{\qL\rL}}{\sqrt{\zr}}
\quad,\quad\hspace{1.2ex}
 \Splt(g_\rL^-,\qB^-,\qL^+) =
  \Sqr(\qB,\qL,\rL)
  \,\frac{-\zq\SQR{\qL\rL}}{\sqrt{\zr}}
\quad.
\end{align}
%%%%%%%%%%%%%%%%%%%%%%%%%%%%%%%%%%%%%%%%
%

In general an amplitude for $(n+1)$ gluons and a quark-antiquark pair can for any given helicity configuration be decomposed as
%
%%%%%%%%%%%%%%%%%%%%%%%%%%%%%%%%%%%%%%%%
\begin{equation}
\ColoredAmp^{a_1 a_2\cdots a_{n+1}\,ji}_{\Tree}
= 2^{\frac{n+1}{2}}\sum_{\sigma\in S_{n+1}}
  \big(T^{a_{\sigma(1)}}T^{a_{\sigma(2)}}\cdots T^{a_{\sigma(n+1)}}\big)_{ji}
  \,\Amp(\qL,g_{\sigma(1)},g_{\sigma(2)},\ldots,g_{\sigma(n+1)},\qB)
\end{equation}
%%%%%%%%%%%%%%%%%%%%%%%%%%%%%%%%%%%%%%%%
%
where $\Amp$ without color indices is the color-ordered partial amplitude.
Now, gluon number $n+1$ is the radiative one, and only when it is the first or the last in the argument list, then the partial amplitude contributes in the joint large-$\Lambda$ limit for the radiative gluon and the auxiliary pair:
%
%%%%%%%%%%%%%%%%%%%%%%%%%%%%%%%%%%%%%%%%
\begin{equation}
\ColoredAmp^{a_1 a_2\cdots a_{n} a_\rL\,ji}_{\Tree}
\to 
        \Lambda\,\Splt(\qB,\qL,g_r)\,\sqrt{2}\big(T^{a_\rL}\Omega^{\star}\big)_{ji}
       +\Lambda\,\Splt(g_r,\qB,\qL)\,\sqrt{2}\big(\Omega^{\star}T^{a_\rL}\big)_{ji}
~,
\label{Eq:039}
\end{equation}
%%%%%%%%%%%%%%%%%%%%%%%%%%%%%%%%%%%%%%%%
%
where
%
%%%%%%%%%%%%%%%%%%%%%%%%%%%%%%%%%%%%%%%%
\begin{equation}
\Omega^{\star}_{lk} = 
2^{\frac{n}{2}}\sum_{\sigma\in S_{n}}
  \big(T^{a_{\sigma(1)}}T^{a_{\sigma(2)}}\cdots T^{a_{\sigma(n)}}\big)_{lk}
  \,\Amp(g^\star,g_{\sigma(1)},g_{\sigma(2)},\ldots,g_{\sigma(n)})
\label{Eq:049}
\end{equation}
%%%%%%%%%%%%%%%%%%%%%%%%%%%%%%%%%%%%%%%%
%
is the full colored amplitude with $1$ space-like and $n$ on-shell gluons.
The right-hand side of \Equation{Eq:039} must be squared and summed over color and helicities to obtain the formula for \matrixelement{}s.
We find
%
%
%%%%%%%%%%%%%%%%%%%%%%%%%%%%%%%%%%%%%%%%
%
%%%%%%%%%%%%%%%%%%%%%%%%%%%%%%%%%%%%%%%%
\begin{multline}
\Mtree\big(\Lambda+x,\xM\,;K_r,K_q,\{p_i\}\big)
\\
\overset{\Lambda\to\infty}{\longrightarrow}\;
\Caux\,\Paux(\zq,\qperp,\zr,\rperp)
\,\frac{\Lambda^2\MtreeStar\big(x,-\qperp-\rperp,\xM\,;\{p_i\}\big)}{x^2|\qperp+\rperp|^2}
~,
\label{Eq:041}
\end{multline}
%%%%%%%%%%%%%%%%%%%%%%%%%%%%%%%%%%%%%%%%
%
with
%
%%%%%%%%%%%%%%%%%%%%%%%%%%%%%%%%%%%%%%%%
\begin{equation}
\Paux(\zq,\qperp,\zr,\rperp)
=
 \EuScript{P}_{\mathrm{aux}}(\zq,\zr)\,|\qperp+\rperp|^2\left(
  \frac{\Cqbar}{|s_{\rL\qB}||s_{\qL\qB}|}
+ \frac{\Cq\,\zq}{|s_{\rL\qL}||s_{\qL\qB}|}
+ \frac{\Cr\,\zr}{|s_{\rL\qL}||s_{\rL\qB}|}
\right)
~,
\label{Eq:042}
\end{equation}
%%%%%%%%%%%%%%%%%%%%%%%%%%%
%
and
%
%%%%%%%%%%%%%%%%%%%%%%%%%%%%%%%%%%%%%%%%
\begin{equation}
\Caux = \frac{\Nc^2-1}{\Nc}
\quad,\quad
\Cqbar = \Cq = \Nc \quad,\quad \Cr =-\frac{1}{\Nc}
~,
\end{equation}
%%%%%%%%%%%%%%%%%%%%%%%%%%%%%%%%%%%%%%%%
%
and collinear splitting function
%
%%%%%%%%%%%%%%%%%%%%%%%%%%%%%%%%%%%%%%%%
\begin{equation}
\EuScript{P}_{\mathrm{aux}}(\zq,\zr)=\EuScript{P}_{qq}(\zq,\zr) = \frac{1+\zq^2-\vepv\zr^2}{\zr}
~.
\end{equation}
%%%%%%%%%%%%%%%%%%%%%%%%%%%%%%%%%%%%%%%%
%
The term proportional to $\vepv$ from dimensional regularization was inferred from a cross-check using the explicit expressions for relevant \matrixelement{}s in \mycite{Ellis:1985er}.

\subsection{Auxiliary gluon pair plus a gluon}
We keep the labels $\qB$ and $\qL$ for the auxiliary pair, and have
%
%%%%%%%%%%%%%%%%%%%%%%%%%%%%%%%%%%%%%%%%
\begin{align}
\Amp(g_\qL,g_\rL,\ldots,g_\qB) &\to \Lambda\,\Splt(g_\qB,g_\qL,g_\rL)\,\Amp^\star(g^\star,\ldots)
\\
\Amp(g_\qL,\ldots,g_\rL,g_\qB) &\to \Lambda\,\Splt(g_\rL,g_\qB,g_\qL)\,\Amp^\star(g^\star,\ldots)
\end{align}
%%%%%%%%%%%%%%%%%%%%%%%%%%%%%%%%%%%%%%%%
%
with
%
%%%%%%%%%%%%%%%%%%%%%%%%%%%%%%%%%%%%%%%%
\begin{align}
\Splt(g_\qB^+,g_\qL^-,g_\rL^-) &=
 \Sqr(\qB,\qL,\rL)\,
 \frac{-\zq^{1/2}\SQR{\rL\qB}}{\sqrt{\zq\zr}}
\\
\Splt(g_\qB^+,g_\qL^-,g_\rL^+) &=
 \Ang(\qB,\qL,\rL)\,
 \frac{\zq^{5/2}\ANG{\rL\qB}}{\sqrt{\zq\zr}}
\\
\Splt(g_\qB^+,g_\qL^+,g_\rL^-) &=
 \Ang(\qB,\qL,\rL)\,
 \frac{\zq^{1/2}\zr^{2}\ANG{\rL\qB}}{\sqrt{\zq\zr}}
\\
\Splt(g_\rL^-,g_\qB^+,g_\qL^-) &=
 \Sqr(\qB,\qL,\rL)\,
 \frac{\SQR{\qL\rL}}{\sqrt{\zq\zr}}
\\
\Splt(g_\rL^+,g_\qB^+,g_\qL^-) &=
 \Ang(\qB,\qL,\rL)\,
 \frac{\zq^{2}\ANG{\qL\rL}}{\sqrt{\zq\zr}}
\\
\Splt(g_\rL^-,g_\qB^+,g_\qL^+) &=
 \Ang(\qB,\qL,\rL)\,
 \frac{\zr^{2}\ANG{\qL\rL}}{\sqrt{\zq\zr}}
%\\
%\Splt(g_\qL^-,g_\rL^-,g_\qB^+) &=
% \Sqr(\qB,\qL,\rL)\,
% \frac{-\zr^{1/2}\SQR{\qL\qB}}{\sqrt{\zq\zr}}
%\\
%\Splt(g_\qL^+,g_{\rL}^-,g_\qB^+) &=
% \Ang(\qB,\qL,\rL)\,
% \frac{\zr^{5/2}\ANG{\qL\qB}}{\sqrt{\zq\zr}}
%\\
%\Splt(g_\qL^-,g_{\rL}^+,g_\qB^+) &=
% \Ang(\qB,\qL,\rL)\,
% \frac{\zr^{1/2}\zq^{2}\ANG{\qL\qB}}{\sqrt{\zq\zr}}
\end{align}
%%%%%%%%%%%%%%%%%%%%%%%%%%%%%%%%%%%%%%%%
%
For \matrixelement{}s, we find \Equation{Eq:041} and \Equation{Eq:042} again, but with
%
%%%%%%%%%%%%%%%%%%%%%%%%%%%%%%%%%%%%%%%%
\begin{equation}
\Caux = 2\Nc
\quad,\quad
\Cqbar = \Cq = \Cr = \Nc
~,
\end{equation}
%%%%%%%%%%%%%%%%%%%%%%%%%%%%%%%%%%%%%%%%
%
and collinear splitting function
%
%%%%%%%%%%%%%%%%%%%%%%%%%%%%%%%%%%%%%%%%
\begin{equation}
\EuScript{P}_{gg}(\zq,\zr) = \frac{1+\zr^4+\zq^4}{\zr\zq}
~.
\end{equation}
%%%%%%%%%%%%%%%%%%%%%%%%%%%%%%%%%%%%%%%%
%

\subsection{Initial-state gluon with final-state quark-antiquark pair}
For the triple limit, we also need to include this case, which has no double limit equivalent.
We will still use the label $\rL$ for the gluon, but realize that now for $\Srght{\rL}$ there is a minus sign $\Srght{\rL}\to-\Lambda\Srght{P}$, and a momenum fraction $z_\qB$ will be associated with the final-state anti-quark.
\begin{align}
\Splt_g(\qB^+,\qL^-,g_\rL^-)&=
  \Sqr(\qB,\qL,\rL)\,
  \big(-\zq^{1/2}\zqbar\SQR{\rL\qB}\big)
\\
\Splt_g(\qB^+,\qL^-,g_\rL^+)&=
  \Ang(\qB,\qL,\rL)\,
  \big(\zq^{3/2}\ANG{\rL\qB}\big)
\\
\Splt_g(g_\rL^-,\qB^+,\qL^-)&=
  \Sqr(\qB,\qL,\rL)\,
  \big(-\zqbar^{3/2}\SQR{\qL\rL}\big)
\\
\Splt_g(g_\rL^+,\qB^+,\qL^-)&=
  \Ang(\qB,\qL,\rL)\,
  \big(\zq\zqbar^{1/2}\ANG{\qL\rL}\big)
\\
\Splt_g(\qB^-,\qL^+,g_\rL^+)&=
  \Ang(\qB,\qL,\rL)\,
  \big(\zq^{1/2}\zqbar\ANG{\rL\qB}\big)
\\
\Splt_g(\qB^-,\qL^+,g_\rL^-)&=
  \Sqr(\qB,\qL,\rL)\,
  \big(-\zq^{3/2}\SQR{\rL\qB}\big)
\\
\Splt_g(g_\rL^+,\qB^-,\qL^+)&=
  \Ang(\qB,\qL,\rL)\,
  \big(\zqbar^{3/2}\ANG{\qL\rL}\big)
\\
\Splt_g(g_\rL^-,\qB^-,\qL^+)&=
  \Sqr(\qB,\qL,\rL)\,
  \big(-\zq\zqbar^{1/2}\SQR{\qL\rL}\big)
\end{align}
%%%%%%%%%%%%%%%%%%%%%%%%%%%%%%%%%%%%%%%%
%
The functions $\Ang$ and $\Sqr$ now contain $\kapp_{\qB},\kstr_{\qB}$ instead of $\kapp_{\rL},\kstr_{\rL}$.
For \matrixelement{}s, we find \Equation{Eq:041} with
%
%%%%%%%%%%%%%%%%%%%%%%%%%%%%%%%%%%%%%%%%
\begin{equation}
\EuScript{P}_{\aux}(\zq,\qperp,\zqbar,\qbarperp)
=
 \EuScript{P}_{ab}(\zq,\zr)\,|\qperp+\qbarperp|^2\left(
  \frac{\Cr}{|s_{\qB\rL}||s_{\qL\rL}|}
+ \frac{\Cq\,\zq}{|s_{\qB\qL}||s_{\qL\rL}|}
+ \frac{\Cqbar\,\zqbar}{|s_{\qB\qL}||s_{\qB\rL}|}
\right)
~,
\label{Eq:043}
\end{equation}
%%%%%%%%%%%%%%%%%%%%%%%%%%%
%
and
%
%%%%%%%%%%%%%%%%%%%%%%%%%%%%%%%%%%%%%%%%
\begin{equation}
\Caux = \frac{\Nc^2-1}{\Nc}
\quad,\quad
\Cqbar = \Cq = \frac{1}{2} \quad,\quad \Cr =-\frac{1}{2\Nc}
~,
\end{equation}
%%%%%%%%%%%%%%%%%%%%%%%%%%%%%%%%%%%%%%%%
%
and collinear splitting function
%
%%%%%%%%%%%%%%%%%%%%%%%%%%%%%%%%%%%%%%%%
\begin{equation}
\EuScript{P}_{qg}(\zq,\zr) = 1 - \frac{\zq\zqbar}{1-\vepv}
~.
\end{equation}
%%%%%%%%%%%%%%%%%%%%%%%%%%%%%%%%%%%%%%%%
%

\section{\label{App:040}Some integrals}
The following integrals appear in this paper
%
%%%%%%%%%%%%%%%%%%%%%%%%%%%%%%%%%%%%%%%%
\begin{align}
I_A 
=
\int d^{2+\vep}\rperp\,\frac{\theta\big(0<\rperp^2<\infty\big)}{\rperp^2|\rperp+\kperp|^2}
=
\frac{\piep|\kperp|^{\vep}}{|\kperp|^2}\,\frac{4}{\vep}
+\Ord(\vep^2)
~,
\label{Eq:052}
\end{align}
%%%%%%%%%%%%%%%%%%%%%%%%%%%%%%%%%%%%%%%%
%
and
%
%%%%%%%%%%%%%%%%%%%%%%%%%%%%%%%%%%%%%%%%
\begin{align}
I_B 
&=
\int d^{2+\vep}\rperp\,\frac{\theta\big(0<\rperp^2<\rupp^2\big)}{\rperp^2|\rperp+\kperp|^2}
\quad,\quad
\frac{|\kperp|^2}{\rupp^2}<1
\notag\\&=
\frac{\piep|\kperp|^{\vep}}{|\kperp|^2}\bigg\{
  \frac{4}{\vep} + \ln\bigg(1-\frac{|\kperp|^2}{\rupp^2}\bigg)
 \bigg\}
+\Ord(\vep)
\notag\\&=
\frac{\piep\rupp^{\vep}}{|\kperp|^2}\bigg\{
  \frac{4}{\vep} + 2\ln\frac{|\kperp|^2}{\rupp^2} + \ln\bigg(1-\frac{|\kperp|^2}{\rupp^2}\bigg)
 \bigg\}
+\Ord(\vep)
~,
\label{Eq:053}
\end{align}
%%%%%%%%%%%%%%%%%%%%%%%%%%%%%%%%%%%%%%%%
%
and
%
%%%%%%%%%%%%%%%%%%%%%%%%%%%%%%%%%%%%%%%%
\begin{align}
I_C 
&=
\int_0^{\upper}\frac{dx}{x}\int d^{2+\vep}\rperp\,\frac{\theta\big(0<\rperp^2<\rupp x\big)}{\rperp^2|\rperp+\kperp|^2}
\quad,\quad
\frac{|\kperp|^2}{\upper\rupp}<1
\notag\\&=
 \frac{\piep|\kperp|^\vep}{|\kperp|^2}
    \bigg\{\frac{4}{\vep^2}-\frac{4}{\vep}\ln\frac{|\kperp|^2}{\upper\rupp}
   +\mathrm{Li}_2\bigg(\frac{|\kperp|^2}{\upper\rupp}\bigg)
    \bigg\}
+\Ord(\vep)
~.
\label{Eq:054}
\end{align}
%%%%%%%%%%%%%%%%%%%%%%%%%%%%%%%%%%%%%%%%
and
%
%%%%%%%%%%%%%%%%%%%%%%%%%%%%%%%%%%%%%%%%
\begin{align}
I_{C2} 
&=
\int_0^{\upper}\frac{dx}{x}\int d^{2+\vep}\rperp\,\frac{\theta\big(0<\rperp^2<\rupp x^2\big)}{\rperp^2|\rperp+\kperp|^2}
\quad,\quad
\frac{|\kperp|^2}{\upper^2\rupp}<1
\notag\\&=
\int_0^{\upper^2}\frac{d\sqrt{x}}{\sqrt{x}}\int d^{2+\vep}\rperp\,\frac{\theta\big(0<\rperp^2<\rupp x\big)}{\rperp^2|\rperp+\kperp|^2}
=
\frac{1}{2}I_C\big(\upper\to\upper^2\big)
~.
\label{Eq:078}
\end{align}
%%%%%%%%%%%%%%%%%%%%%%%%%%%%%%%%%%%%%%%%
and
%
%%%%%%%%%%%%%%%%%%%%%%%%%%%%%%%%%%%%%%%%
\begin{align}
I_D 
&=
\int_0^{1}\frac{dx}{x}\int d^{2+\vep}\rperp\,\frac{1}{\rperp^2|\rperp+\kperp|^2}
\,\theta\bigg(\frac{|\rperp|}{v\sqrt{\Lambda}}<x<\frac{|\rperp|}{|\rperp+\kperp|}\bigg)
\notag\\&=
  \frac{2}{|\kperp|^2}
  \frac{\piep|\kperp|^{\vep}}{\vep}\ln\frac{\rupp^2\Lambda}{|\kperp|^2}
  + \Ord\big(\vep\big) + \Ord\big(\Lambda^{-1}\big)
~.
\label{Eq:079}
\end{align}
\end{appendix}

\end{document}